\documentclass[sw,crcready]{iosart2x}

\usepackage{algorithm} 
\usepackage[noend]{algpseudocode}

\usepackage{xspace,amstext,url}
\usepackage{enumerate}

\usepackage{framed}

\usepackage{comment}

\usepackage{caption}
\usepackage{subcaption}

\usepackage{stmaryrd}

\usepackage{amssymb}

\newcommand{\semp}[2]{\llbracket #1\rrbracket_{#2}}

\usepackage{tikz}
\usetikzlibrary{automata, graphs,positioning,chains,arrows,decorations.pathmorphing}
\usetikzlibrary{calc}

\usepackage{balance}

\newcommand{\nc}{\newcommand}

\nc{\src}{\ensuremath{\mathit{src}}}
\nc{\tgt}{\ensuremath{\mathit{tgt}}}
\nc{\lab}{\ensuremath{\mathit{lab}}}
\nc{\len}{\ensuremath{\mathit{len}}}
\nc{\paths}{\ensuremath{\mathit{Paths}}}

\newcommand{\sysneo}{\textsc{Neo4j}\xspace}
\newcommand{\sysjena}{\textsc{Jena}\xspace}
\newcommand{\sysblaze}{\textsc{BlazeGraph}\xspace}
\newcommand{\sysvirtuoso}{\textsc{Virtuoso}\xspace}
\newcommand{\mdb}{\textsc{PathFinder}\xspace}
\newcommand{\sysnebula}{\textsc{Nebula}\xspace}
\newcommand{\sysmem}{\textsc{Memgraph}\xspace}
\newcommand{\syskuzu}{\textsc{Kuzu}\xspace}
\newcommand{\sysduck}{\textsc{DuckPGQ}\xspace}
\newcommand{\mdbbfs}{\textsc{PathFinder-BFS}\xspace}
\newcommand{\mdbdfs}{\textsc{PathFinder-DFS}\xspace}

\newcommand{\millennium}{\textsc{MillenniumDB}\xspace}

\usepackage{xcolor}
\usepackage[normalem]{ulem}

\usetikzlibrary{arrows,positioning,backgrounds} 

\usepackage{tikz-uml}
\usepackage{kg-macros}

\def\checkmark{\tikz\fill[scale=0.4](0,.35) -- (.25,0) -- (1,.7) -- (.25,.15) -- cycle;}

\nc{\SIMPLE}{\ensuremath{\mathit{SIMPLE}}\xspace}
\nc{\TRAIL}{\ensuremath{\mathit{TRAIL}}\xspace}
\nc{\TRAILS}{\ensuremath{\mathit{TRAILS}}\xspace}
\nc{\ANY}{\ensuremath{\mathit{ANY}}\xspace}
\nc{\ALL}{\ensuremath{\mathit{ALL}}\xspace}
\nc{\SHORTEST}{\ensuremath{\mathit{SHORTEST}}\xspace}
\nc{\WALK}{\ensuremath{\mathit{WALK}}\xspace}
\nc{\WALKS}{\ensuremath{\mathit{WALKS}}\xspace}
\nc{\ACYCLIC}{\ensuremath{\mathit{ACYCLIC}}\xspace}
\nc{\GROUPS}{\ensuremath{\mathit{GROUPS}}\xspace}

\nc{\kshortest}{\ensuremath{k\text{-}\mathit{ShortestLength}}}
\nc{\nodes}{\ensuremath{\mathit{Nodes}}}
\nc{\edges}{\ensuremath{\mathit{Edges}}}
\nc{\labels}{\ensuremath{\mathit{Lab}}}
\nc{\variables}{\ensuremath{\mathit{Var}}}

\nc{\sel}{\ensuremath{\mathit{sel}}}
\nc{\res}{\ensuremath{\mathit{res}}}
\nc{\rgx}{\ensuremath{\mathit{rgx}}}

\nc{\cA}{\ensuremath{\mathcal{A}}}
\nc{\cL}{\ensuremath{\mathcal{L}}}

\usepackage{amsthm}
\newtheorem{theorem}{Theorem}
\newtheorem{definition}{Definition}
\newtheorem{example}{Example}

\pubyear{0000}
\volume{0}
\firstpage{1}
\lastpage{1}

\begin{document}

\begin{frontmatter}

\title{PathFinder: Returning Paths in Graph Queries}
\runtitle{PathFinder: Returning Paths in Graph Queries}


\begin{aug}
\author[A,C]{\inits{B.}\fnms{Benjam\'in} \snm{Far\'ias}\ead[label=e1]{bffarias@uc.cl}}
\author[B]{\inits{W.}\fnms{Wim} \snm{Martens}\ead[label=e2]{wim.martens@uni-bayreuth.de}}
\author[C]{\inits{C.}\fnms{Carlos} \snm{Rojas}\ead[label=e3]{cirojas6@uc.cl}}
\author[A,C]{\inits{D.}\fnms{Domagoj} \snm{Vrgo\v{c}}\ead[label=ed]{vrdomagoj@uc.cl}
\thanks{Corresponding author. \printead{ed}.}
}
\address[A]{\orgname{Pontificia Universidad Cat\'olica de Chile}, \cny{Chile}}
\address[B]{\orgname{University or Bayreuth}, \cny{Germany}}
\address[C]{\orgname{Millennium Institute for Foundational Research on Data (IMFD)}, Santiago, \cny{Chile}\printead[presep={\\}]{e1,e2,e3,ed}}
\end{aug}

%


\begin{abstract}
Path queries are a central feature of all modern graph query languages and standards, such as SPARQL, Cypher, SQL/PGQ, and GQL. While SPARQL returns \emph{endpoints} of path queries, it is possible in Cypher, SQL/PGQ, and GQL to return \emph{entire paths}. In this paper, we present the first framework for returning paths that match regular path queries under all twenty seven modes supported by the SQL/PGQ and GQL standards. At the core of our approach is the product graph construction combined with a way to compactly represent a huge number of results that can match a path query. We describe the approach on a conceptual level and provide runtime guarantees for evaluating path queries. We also show how it can be incorporated into the SPARQL query language, thus extending RDF engines with the ability to return witnesses to property path queries (under the GQL semantics). To show the feasibility of our methods in practice, we develop a reference implementation on top of an existing open-source graph processing engine MillenniumDB, and perform a detailed analysis of path querying over several synthetic and real-world datasets. In particular, we test our implementation over Wikidata using property path queries posted by users, extending them with the ability to return paths. We obtain order-of-magnitude speedups over modern graph database engines and remarkably stable performance, even for theoretically intractable queries.
\end{abstract}


\end{frontmatter}

\section{Introduction}
\label{sec:intro}


Graph databases~\cite{AnglesABBFGLPPS18,SakrBVIAAAABBDV-cacm21} have gained significant popularity and are used in areas such as Knowledge Graph management~\cite{HoganBC21}, Semantic Web applications~\cite{AnglesABHRV17}, and Biology~\cite{Alphafold}. There are roughly two generations of query languages for handling graph data. The first is SPARQL~\cite{HarrisS13}, which is being developed by the W3C since 2004 and became an official W3C standard in 2008. The second generation originated with Neo4j's Cypher~\cite{FrancisGGLLMPRS18}, which was initially released in 2011. Cypher heavily inspired SQL/PGQ and GQL, which were standardized by ISO in 2023 and 2024, respectively.

All these languages have \emph{path queries} as a core feature. In SPARQL, these are supported through \emph{property paths}, which are a variant of \emph{regular path queries} (\emph{RPQs})
~\cite{CalvaneseGLV02,MendelzonW89,Baeza13,CruzMW87}. Intuitively, an RPQ is an expression $(?x,\texttt{regex},?y)$, where \texttt{regex} is a regular expression and $?x,?y$ are variables. When evaluated over an edge-labeled graph $G$ (e.g., RDF or property graph), it extracts all node pairs $(n_1, n_2)$ such that there is a path in $G$ from $n_1$ to $n_2$, whose edge labels form a word that matches \texttt{regex}. SPARQL property paths therefore return \emph{endpoints} of paths. In important applications such as money laundering detection \cite{laundering} or in investigative journalism~\cite{GQL}, retrieving the \emph{entire paths that match the expression} is of crucial importance. To accommodate this use case, Cypher, GQL, and SQL/PGQ all extend RPQs with a mechanism for returning paths.

\begin{figure}
\begin{tikzpicture}[->,>=stealth',auto, thick, scale = 1.0,initial text= {},    node/.style={ square,inner sep=2pt    }]

\tikzstyle{every state}=[draw,thick,rectangle,rounded corners,fill=white!85!black,minimum size=5mm, text=black, font=\ttfamily, inner sep=3pt]
            \tikzstyle{every node}=[font=\ttfamily]

        \node [state] at (0,0) (n0) {\small John};
        \node [state] at (4,0) (n1) {\small Joe};
        \node [state] at (8,1.3) (n2) {\small Paul};
        \node [state] at (8,-1.3) (n3) {\small Lily};	  
        \node [state] at (12,0) (n4) {\small Jane};
        \node [state] at (0,3.5) (n5) {\small Rome};
        \node [state] at (12,3.5) (n6) {\small ENS Paris};
        \node [state] at (5,3.5) (n7) {\small Anne};      
  
        \path[->] (n0) edge[bend left]   node [above] {\small e1:follows} (n1);
        \path[->] (n1) edge[bend left]   node [below] {\small e2:follows} (n0);		
        \path[->] (n1) edge   node [above, sloped] {\small  e3:follows} (n2);
        \path[->] (n1) edge   node [below, sloped] {\small  e4:follows} (n3);
        \path[->] (n2) edge   node [above, sloped] {\small  e6:follows} (n4);
        \path[->] (n3) edge   node [below, sloped] {\small  e7:follows} (n4);
        \path[->] (n0) edge   node [above, sloped] {\small  e8:lives} (n5);
        \path[->] (n4) edge   node [below, sloped] {\small  e10:works} (n6);        
        \path[->] (n7) edge   node [above, sloped] {\small  e9:lives} (n5); 
        \path[->] (n2) edge   node [above, sloped] {\small  e5:follows} (n7); 
        \path[->] (n7) edge   node [above, sloped] {\small  e11:works} (n6);
        \path[->] (n3) edge   node [right] {\small e12:follows} (n2);
        
		 \end{tikzpicture}
\caption{An example graph database.}
\label{fig:introNew}
\end{figure}

\paragraph{Path Modes By Example.}
We illustrate how these languages allow returning paths. Consider the social network example in Figure~\ref{fig:introNew}. Here, we have node identifiers (such as \texttt{Joe} or \texttt{Rome}), edge identifiers (such as \texttt{e1}), and edge labels (for instance, edge \texttt{e7} has the label \texttt{follows}).\footnote{Throughout the paper we use a simplified model of property graphs which only considers nodes, edges, and edge labels. This is done since the queries we consider only utilize these graph elements. Additionally, this will allow to easily transfer our results to RDF graphs.} 
A natural task would be to explore the network that influences \texttt{Joe}, i.e.,  finding the people that \texttt{Joe} follows, then people they follow, and so on, transitively traversing \texttt{follows}-edges. This can be written as an RPQ of the form $(\texttt{Joe}, \texttt{follows}^+,?x)$, signaling that we wish to start at \texttt{Joe}, and traverse any nonzero number of \texttt{follows}-edges. In this case, the query returns all the people in the database. If we also wish to return  paths witnessing these connections, we might run into some issues. Most notably, given the cycle formed by the edges \texttt{e1} and \texttt{e2}, there is an infinite number of paths from \texttt{Joe} to \texttt{John}. To avoid having to return an infinite number of paths, GQL and SQL/PGQ let the user specify restrictions on the type of paths to be returned through so-called \emph{path modes}.

We give a few examples of path modes and provide comprehensive details later. Two common path modes are \SIMPLE and \TRAIL. Whereas \SIMPLE requires the paths to not repeat any node (apart from the first and last one), \TRAIL, requires to not repeat any edge. For instance, our example has one path satisfying \SIMPLE from \texttt{Joe} to \texttt{Lily} (only taking edge \texttt{e4}), but two paths satisfying \TRAIL (the additional one being \texttt{e2} $\rightarrow$ \texttt{e1} $\rightarrow$ \texttt{e4}).
 
Another common way to force the number of paths to be finite is by selecting only the shortest paths. The \ANY \SHORTEST mode in GQL and SQL/PGQ will return a single shortest path for each different pair of endpoints in the answer, whereas \ALL \SHORTEST will return all such shortest paths. For instance, there are five paths matching the RPQ $(\texttt{Joe},\texttt{follows}^+\cdot \texttt{works},?x)$. Incidentally they are all from \texttt{Joe} to \texttt{ENS PARIS}. The \ALL \SHORTEST mode returns the three shortest ones, which are $\texttt{e3}\rightarrow \texttt{e5}\rightarrow \texttt{e11}$, $\texttt{e4}\rightarrow \texttt{e7}\rightarrow \texttt{e10}$, and $\texttt{e3}\rightarrow \texttt{e6}\rightarrow \texttt{e10}$). \ANY \SHORTEST would return an arbitrary path among these. We note that these modes select the shortest paths \emph{among all paths that match}. For instance, \ALL \SHORTEST $(\texttt{Joe},\texttt{follows}\cdot\texttt{follows}\cdot\texttt{follows},?x)$ returns the path 
$\texttt{e2}\rightarrow \texttt{e1}\rightarrow \texttt{e4}$ from \texttt{Joe} to \texttt{Lily} and $\texttt{e2}\rightarrow \texttt{e1}\rightarrow \texttt{e3}$ from \texttt{Joe} to \texttt{Paul}, even though shorter paths (that, crucially, do not match the RPQ) between these end-nodes exist. The standard supports a total of twenty seven path modes, most of which use the ingredients we outlined here, thus giving the user fine-grained control over which paths should be returned as path of the query answer.

\paragraph{Support for Path Modes is Lacking.} While Cypher, GQL, and SQL/PGQ all recognize the need to return paths that match regular path queries, the support for these features in modern graph databases is still lacking. As already stated, in SPARQL engines, RPQ matching is supported for all regular languages, but paths cannot be returned. In property graph engines that implement (fragments of) GQL or SQL/PGQ, only a handful of path modes is supported, and no property graph engine to date supports all regular path queries like SPARQL engines do. For reference, we summarize the support for path features in several popular SPARQL and property graph engines in Table~\ref{tab:intro} below. Here, ``RPQ'' means the support for all regular languages in RPQs, while the other columns signal which types of paths can be returned. 
\begin{table}
\begin{center}
\resizebox{0.7\columnwidth}{!}{
    \begin{tabular}{c|ccccccc}
       & RPQ & \WALK & \TRAIL & \SIMPLE & \ACYCLIC & \SHORTEST & \GROUPS\\
      \hline
      \sysblaze~\cite{ThompsonPC14} & \checkmark & -- & -- & -- & -- & -- & --\\
      \sysjena~\cite{JenaTDB} & \checkmark & -- & -- & -- & -- & -- & --\\    
      \sysvirtuoso~\cite{Erling12} & \checkmark & -- & -- & -- & -- & -- & --\\    \sysneo~\cite{Webber12} & partial & \checkmark & \checkmark & -- & -- & \checkmark & --\\  
      \sysnebula~\cite{nebula} & partial & partial & \checkmark & -- & \checkmark & \checkmark & --\\  
      \sysmem~\cite{memgraph} & partial & \checkmark & \checkmark & -- & -- & \checkmark & --\\ 
      \syskuzu~\cite{kuzu} & partial & \checkmark & -- & -- & -- & \checkmark & --\\
      \sysduck~\cite{duckDB} & partial & \checkmark & -- & -- & -- & \checkmark  & --\\ 
      \mdb & \checkmark & \checkmark & \checkmark & \checkmark & \checkmark & \checkmark      & \checkmark      
    \end{tabular}
}    
\end{center}
\caption{Support for path queries in popular graph database engines.}
\label{tab:intro}
\end{table}

The overall picture tells us that SPARQL engines support all RPQs, but do not return paths. On the other hand, while several property graph engines can return some (but not all) types of paths, they only support a limited number of regular expressions in RPQs. \mdb is the \emph{first engine that can return paths for every RPQ.}

In addition, both SPARQL engines and property graph engines are known to exhibit slow execution times when dealing with path queries~\cite{ReutterSV21,MillDB}. We believe that this is the case because of scalability. In particular, \emph{the number of paths that match an RPQ can be exponential in the size of the graph}~\cite{pmr} and existing engines don't have a principled method to tackle this issue.
In this paper, we tackle these problems and show that:
\begin{enumerate}
    \item Regular path queries can be evaluated efficiently while at the same time returning paths under all the path modes in the GQL standard at the point of writing this paper.
    \item Adding these implementations on top of an existing property graph engine can be done with ease.
    \item One can extend these algorithms to the RDF/SPARQL context and extend existing SPARQL engines with the ability to return paths.
\end{enumerate}

\smallskip
\noindent \textbf{Our Contribution.} To achieve the three points above, we introduce  \mdb, a unified framework for returning paths in graph query answers. Our concrete contributions can be summarized as follows:
\begin{itemize}
    \item We describe implementable algorithms for evaluating RPQs and returning paths under each GQL path mode.
    \item We show how all these algorithms can share a common core and can build on the same fundamental principle: the \emph{product graph}~\cite{MendelzonW89,pmr}. Indeed, each of the algorithms is a modification of classical graph search on the product graph, with extra bookkeeping. As such, our algorithms always produce a \emph{path multiset representation}~\cite{pmr}, a compact representation of all paths that need to be returned. We show how this compact representation can be constructed lazily, such that paths can be returned as soon as they are detected, allowing for seamless incorporation into a database pipeline.
    \item We provide a full implementation of \mdb, which is the first property graph engine to  support returning entire paths in \emph{all 27 modes} in the GQL and SQL/PGQ standards, for \emph{all RPQs}. 
    \item \textcolor{black}{\mdb is the first graph database engine that supports returning paths for all regular languages.}
    \item We extend the SPARQL syntax with the ability to return paths and implement this extension on top of an existing RDF engine, showing that our approach is a feasible extension for existing triplestores.
    \item We perform a detailed experimental evaluation of our solution, using both synthetic and real-world data and queries. Most notably, we test our approach over Wikidata~\cite{VrandecicK14} using real world queries~\cite{MalyshevKGGB18} and show remarkably scalable performance, even for worst-case intractable queries. 
\end{itemize}
Our work shows that it is possible to extend SPARQL with path returning features, while keeping  query evaluation performant. Importantly for a clean design, however, is that the semantics of path returning queries is consistent with standard constructions in automata theory~\cite{PODSgems}. We discuss this in more depth in Section~\ref{sec:concl}. 

\medskip
\noindent \textbf{Remark.} This manuscript is an extension of Farias et al.~\cite{FariasMRV24}, which appeared in the International Semantic Web Conference 2024. Compared to the conference version, we extend the work to also cover group modes of GQL, thus covering all path modes that are prescribed by the standard. Additionally, we provide more details to the algorithmic solutions and additional examples illustrating how these operate. We also discuss the complexity of our algorithms and extend the experimental setup to include more use cases and to illustrate finer points about our algorithms.

\medskip
\noindent \textbf{Code Availability.} All of the code and experimental data is available in our repository~\cite{ANONrepo}. 

\medskip
\noindent \textbf{Organization.} We define GQL path queries in Section~\ref{sec:prelim}. \mdb and the way it handles arbitrary paths is presented in Section~\ref{sec:walk}. Paths satisfying the \TRAIL and \SIMPLE modes are treated in Section~\ref{sec:simple}. Implementation details are presented in Section~\ref{sec:implementation}. Experimental evaluation is performed in Section~\ref{sec:exp}. Related work is discussed in Section~\ref{sec:related}. We conclude in Section~\ref{sec:concl}. Additional details, code, and experiments can be found in our repository~\cite{ANONrepo}.

\section{Graph Databases and Path Queries}
\label{sec:prelim}
We define graph databases and path queries as  supported by the GQL and SQL/PGQ standard.

\medskip
\noindent \textbf{Graph Databases.}
Let $\nodes$ be a set of node identifiers and $\edges$ be a set of edge identifiers, with $\nodes$ and $\edges$ being disjoint. Additionally, let $\labels$ be a set of labels.  Following \cite{AnglesABHRV17,GQL,GQLdigest,pmr}, we define graph databases as follows.

\begin{definition}
A \emph{graph database} $G$ is a tuple $(V,E,\rho,\lambda)$, where
\begin{itemize}
\item $V\subseteq \nodes$ is a finite set of nodes,
\item $E\subseteq \edges$ is a finite set of edges,
\item $\rho:E\rightarrow(V\times V)$ is a total function, and 
\item $\lambda:E\rightarrow \labels$ is a total function assigning a label to each edge.
\end{itemize}
\end{definition}
Intuitively, $\rho(e)=(v_1,v_2)$ means that $e$ is a directed edge going from $v_1$ to $v_2$.
We use a simplified version of property graphs that omits node and edge properties (with their associated values). This is done since the type of queries we consider only use nodes, edges, and edge labels. However, all of our results transfer to the full version of property graphs. We also remark that our results apply directly to RDF graphs \cite{PerezAG09} and edge-labeled graphs \cite{MendelzonW89,Baeza13}, which do not use explicit edge identifiers.

\medskip
\noindent \textbf{Paths.} A sequence
$p = v_0 e_1 v_1 e_2 v_2 \cdots e_n v_n$ is called a \emph{path} in a graph database $G=(V,E,\rho,\lambda)$, if $n \geq 0$, $e_i \in E$, and $\rho(e_i) = (v_{i-1},v_i)$ for $i = 1,\ldots ,n$. If $p$ is a path in $G$, we  write $\lab(p)$ for the sequence of labels $\lambda(e_1) \cdots \lambda(e_n)$ occurring on the edges of $p$. We write $\src(p)$ for the starting node $v_0$ of $p$, and $\tgt(p)$ for the end node $v_n$ of $p$. The length of a path $p$, denoted
$\len(p)$, is defined as the number $n$ of edges it uses. 
We say that a path $p$ is 
\begin{itemize}
    \item a \emph{walk}, for every $p$ (as per GQL and SQL/PGQ terminology);
    \item a \emph{trail}, if $p$ does not repeat an edge (that is, $e_i\neq e_j$ for every $i\neq j$);   
    \item \emph{acyclic}, if $p$ does not repeat a node (that is, $v_i\neq v_j$ for every $i\neq j$); and     
    \item \emph{simple}, if $p$ does not repeat a node, except that possibly $\src(p) = \tgt(p)$. 
\end{itemize}
In examples, we will often use the lighter notation $e_1 \rightarrow \cdots \rightarrow e_n$ for paths. Given a set of paths $P$ over a graph database $G$, we say that $p\in P$ is a \emph{shortest path in $P$}, if $\len(p)\leq \len(p')$ for each $p'\in P$. We use $\paths(G)$ to denote the (possibly infinite) set of all paths in a graph database $G$. By $\paths(G,\mathit{walk})$, $\paths(G,\mathit{trail})$, $\paths(G,\mathit{acyclic})$, and $\paths(G,\mathit{simple})$, we denote the subsets of $\paths(G)$ that are walks, trails, acyclic paths, and simple paths, respectively. Notice that $\paths(G,\mathit{walk}) = \paths(G)$. 

\medskip
\noindent\textbf{Path Queries in GQL and SQL/PGQ.} To define path queries, we assume the existence of a set  $\variables$ of variables, disjoint from $\nodes$, $\edges$, and $\labels$. 
Following the GQL~\cite{gql-standard} and SQL/PGQ~\cite{sql-pgq-standard} standards, we define \emph{path queries} as expressions of the form
$$\sel?\  \res\  (X,\rgx,Y)$$
where $\sel$ is called a \emph{selector}, $\res$ is called a \emph{restrictor}, and $\rgx$ is a regular expression. We define all these ingredients next. Furthermore, \mbox{$\sel$?} means that the selector is optional. 

In this paper, regular expressions are inductively defined as follows. Each element of $\labels$ is a regular expression, $\varepsilon$ is a regular expression, and if $r$, $r_1$, and $r_2$ are regular expressions, then so are their disjunction $(r_1 + r_2)$, their concatenation $(r_1 \cdot r_2)$, and the Kleene closure $r^*$. By standard convention, we omit brackets and (usually) omit the sign $\cdot$ to simplify notation. We use $r?$ to abbreviate $(r+\varepsilon)$ and $r^+$ to abbreviate $(r \cdot r^*)$. The \emph{language} $\cL(r)$ of the regular expression $r$ is inductively defined as $\cL(\varepsilon) = \{\varepsilon\}$ and $\cL(a) = \{a\}$ for every $a \in \labels$. Furthermore, $\cL(r_1 + r_2) = \cL(r_1) \cup \cL(r_2)$, $\cL(r_1 \cdot r_2) = \cL(r_1) \cdot \cL(r_2)$, and $\cL(r^*) = \cup_{k=0}^\infty \cL(r)^k$. Here, for a set of words $S$, we denote by $S^k$ the $k$-fold concatenation of $S$. The \emph{size} of a regular expression, which we write as $|r|$, is the number of occurrences of symbols of $\labels \cup \{\varepsilon\}$ it contains. 
We will refer to the subexpression  $(X,\rgx,Y)$ as a \emph{regular path query} or \emph{RPQ}~\cite{Baeza13}.
In regular path queries, the symbols $X$ and $Y$ represent variables or nodes in the graph. To distinguish these cases, we will always write variables using question marks (e.g., $?x$, $?y$) and nodes without question marks (e.g., $v,v',n,n'$).

Selectors and restrictors are used to specify which paths are to be returned. Their grammar is: 
 \begin{align*}
     \sel \quad ::= \quad & \ \ANY \ |\ \ANY\ \SHORTEST \ | \ \ALL\ \SHORTEST \ | \ 
        \ANY\ k \ | \ 
        \SHORTEST\ k \ | \ 
        \SHORTEST\ k\ \GROUPS\\
     \res \quad ::= \quad & \ \WALK \ |\ \TRAIL \ |\  \SIMPLE \ |\ \ACYCLIC
 \end{align*}
Here, $k \in \{1,2,3,\ldots\}$ is a natural number.
 
A \emph{path mode} is either a restrictor or a combination of a selector and a restrictor. This gives us 28 path modes: 6 selectors times 4 restrictors, plus the 4 restrictors without selector. However, \WALK needs to be preceded by a selector~\cite{GQL} to avoid the need to return infinitely many paths in some cases, which gives 27 relevant modes. 


{ 

\paragraph{Semantics of Path Queries.} 
Let $G$ be a graph database. We first consider path queries $q$ of the form
$$\res\ (?x,\rgx,?y)\;,$$
that is, we omit the optional selector $\sel$ and assume that the RPQ uses two different variables $?x$ and $?y$.  
By slight abuse of notation, we use $\paths(G,\res)$ to denote the set of all paths in $G$ that are valid according to  $\res$. That is, we also use $\paths(G,\TRAIL)$ to denote $\paths(G,\mathit{trail})$, etc. We then define the semantics of $q$ over $G$, denoted $\semp{q}{G}$, as
$$\semp{\res\  (?x,\rgx,?y)}{G} = \{ (v,v',p)  \ \mid\  p\in \paths(G,\res),\ \src(p) = v,\ \tgt(p) = v',\ \lab(p)\in \cL(\rgx)\} \ .$$
For example, the semantics of ``$\TRAIL \ (?x,\allowbreak \rgx,?y)$'' over $G$ is the set of triples $(v,v',p)$ such that $p$ is a trail in $G$ from $v$ to $v'$ and $\lab(p)\in \cL(\rgx)$. If $G$ has cycles and the restrictor is \WALK, this set can be infinite.

\medskip
In order to define the semantics of selectors, we need some additional notation. We write 
\begin{itemize}
\item $\mathit{EndPoints}(\semp{q}{G})$ for the set $\{(v,v') \mid (v,v',p) \in \semp{q}{G}\}$, 
\item $\mathit{Paths}(\semp{q}{G},v,v')$ for the set $\{p \mid (v,v',p) \in \semp{q}{G}\}$,
\item for a set $P$ of paths, 
\begin{itemize}
    \item $\mathit{Shortest}(P)$ for the set $\{p \in P \mid \nexists p' \in P$ such that $\len(p') < \len(p)\}$, and
    \item $\kshortest(P)$ for the $k$-th smallest element of the set $\{\len(p) \mid p \in P\}$.
\end{itemize}
\end{itemize}
We now define the semantics of selectors and use $q$ to denote the selector-free query $\res\  (?x,\rgx,?y)$. Let $S \subseteq \semp{q}{G}$, which means that $S$ only contains triples of the form $(v,v',p)$ such that $p$ is a path from $v$ to $v'$ that matches $\rgx$. For nodes $v$ and $v'$, we denote by $S|_{v,v'}$ the subset $\{(u,u',p) \in S \mid u = v$ and $u' = v'\}$ of $S$.
We define when $S$ satisfies $q$ under $\sel$ in $G$, denoted as $S \models (\sel\ q)_{G}$, on a case-by-case basis:
\begin{itemize}
    \item $S \models (\ANY\ q)_{G}$ if for each pair $(v,v') \in \mathit{Endpoints}(\semp{q}{G})$ the set $S|_{v,v'}$ contains a single tuple $(v,v',p)$.
    \item $S \models (\ANY\ \SHORTEST\ q)_{G}$ if  for each pair $(v,v') \in \mathit{Endpoints}(\semp{q}{G})$ the set $S|_{v,v'}$ contains exactly one tuple $(v,v',p)$ and, additionally, $p \in \mathit{Shortest}(\mathit{Paths}(\semp{q}{G},v,v'))$.

    \item $S \models (\ALL\ \SHORTEST\ q)_{G}$ if, for each pair $(v,v') \in \mathit{Endpoints}(\semp{q}{G})$, we have 
    $S|_{v,v'} = \{(v,v',p) \mid p \in \mathit{Shortest}(\mathit{Paths}(\semp{q}{G},v,v'))\}$.

    \item $S \models (\ANY\ k\ q)_{G}$ if, for each  pair $(v,v') \in \mathit{Endpoints}(\semp{q}{G})$, either
    \begin{itemize}
    \item $S|_{v,v'} = \{(v,v',p) \mid p \in \mathit{Paths}(\semp{q}{G},v,v')\}$, if $|\mathit{Paths}(\semp{q}{G},v,v')| \leq k$; or
    \item $S$ has exactly $k$ tuples of the form $(v,v',p)$, otherwise. 
    \end{itemize}


    \item $S \models (\SHORTEST\ k\ q)_{G}$ if, for each  pair $(v,v') \in \mathit{Endpoints}(\semp{q}{G})$, either
    \begin{itemize}
    \item $S|_{v,v'} = \{(v,v',p) \mid p \in \mathit{Paths}(\semp{q}{G},v,v')\}$, if $|\mathit{Paths}(\semp{q}{G},v,v')| \leq k$; or
    \item $S$ has exactly $k$ tuples of the form $(v,v',p)$, all of which have $|\{p'\in \mathit{Paths}(\semp{q}{G},v,v') \mid \len(p') < \len(p)\}| < k$, otherwise. 
    \end{itemize}


    \item $S \models (\SHORTEST\ k\ \GROUPS\ q)_{G}$ if, for each  pair $(v,v') \in \mathit{Endpoints}(\semp{q}{G})$, we have that
        $S|_{v,v'} = \{(v,v',p) \mid \len(p) \leq \kshortest(\mathit{Paths}(\semp{q}{G},v,v'))\}$.

\end{itemize}

} 

Notice that the \SHORTEST modes select shortest paths \emph{grouped by endpoints}. That is, the selected paths do not need to be the shortest in the entire output, they only need to be the shortest matching paths from $v$ to $v'$. The rationale for \SHORTEST $k$ \GROUPS is similar. 

Finally, if $X$ is a node $u$ in $G$, then the semantics is defined exactly as above, but requires all answers $(v,v',p)$ to have  $v = u$. (The case where $Y$ is a node in $G$ is analogous.) As such, one can also use queries of the form
$\ALL\ \TRAILS\ (u,\texttt{a}^+,?y)$ or $\ALL\ \TRAILS\ (u,\texttt{a}^+,u')$, where $u$ and $u'$ are elements of \nodes.

Notice that the semantics of \ANY, \ANY \SHORTEST, \ANY $k$, and \SHORTEST $k$ is non-deterministic in the sense that there can be multiple subsets $S$ of $\semp{q}{G}$ that satisfy the condition. Terminologically, we deal with this as follows. By slight abuse of notation, we say that $S = \semp{\sel\ q}{G}$ if $S$ is the only subset of $\semp{q}{G}$ such that $S \models (\sel\ q)_G$. Otherwise, we say that $\semp{\sel\ q}{G}$ \emph{can be} $S$. We use analogous terminology for describing $S$, for instance, we can say that $S$ \emph{can contain} a given answer.

\begin{example}
\label{ex:semantics}
To illustrate different path modes, we return to the graph $G$ of Figure~\ref{fig:introNew} and consider the RPQ
$$(\texttt{Joe},\ \texttt{follows}^*\cdot \texttt{works},\ ?x)\;.$$
While $(\texttt{Joe},\texttt{ENS}\; \texttt{Paris})$ is the only pair of nodes connected by a path matching this regular expression, depending on different configuration of selectors and restrictors, the set of matching paths can differ significantly. For instance:
\begin{itemize}
    \item $\semp{\ANY\ \WALK\ (\texttt{Joe},\texttt{follows}^*\cdot \texttt{works},?x)}{G}$ \emph{can contain} the triple $(\texttt{Joe},\texttt{ENS}\; \texttt{Paris},p)$, where $p$ is the 
    path $\texttt{e2}\rightarrow \texttt{e1}\rightarrow \texttt{e3}\rightarrow\texttt{e5}\rightarrow\texttt{e11}$, which loops back to \texttt{Joe};
    \item $\semp{\ALL\ \SHORTEST\ \WALKS\ (\texttt{Joe},\texttt{follows}^*\cdot \texttt{works},?x)}{G}$ contains \texttt{Joe} and \texttt{ENS}\; \texttt{Paris} together with the three shortest paths linking them; namely:
    \begin{enumerate}
        \item $\texttt{e3}\rightarrow \texttt{e5}\rightarrow \texttt{e11}$;
        \item $\texttt{e3}\rightarrow \texttt{e6}\rightarrow \texttt{e10}$; and
        \item $\texttt{e4}\rightarrow \texttt{e7}\rightarrow \texttt{e10}$.
    \end{enumerate}
    \item $\semp{\ANY\ \SHORTEST\ \WALK\ (\texttt{Joe},\texttt{follows}^*\cdot \texttt{works},?x)}{G}$ contains a single triple $(\texttt{Joe},\texttt{ENS}\; \texttt{Paris},p)$, where $p$ is one of the three path in the previous bullet.
    \item $\semp{\SHORTEST\ 6\ \WALKS\ (\texttt{Joe},\texttt{follows}^*\cdot \texttt{works},?x)}{G}$ can contain  \texttt{Joe} and \texttt{ENS}\; \texttt{Paris} together with the following six paths:
    \begin{enumerate}
        \item $\texttt{e3}\rightarrow \texttt{e5}\rightarrow \texttt{e11}$;
        \item $\texttt{e3}\rightarrow \texttt{e6}\rightarrow \texttt{e10}$;
        \item $\texttt{e4}\rightarrow \texttt{e7}\rightarrow \texttt{e10}$;
        \item $\texttt{e4}\rightarrow \texttt{e12}\rightarrow\texttt{e5}\rightarrow \texttt{e11}$;
        \item $\texttt{e4}\rightarrow \texttt{e12}\rightarrow\texttt{e6}\rightarrow \texttt{e10}$; and
        \item $\texttt{e2}\rightarrow \texttt{e1}\rightarrow \texttt{e3}\rightarrow \texttt{e6}\rightarrow \texttt{e10}$.
    \end{enumerate}
    Notice here that the last three paths are longer than the shortest paths from \texttt{Joe} to \texttt{ENS}\; \texttt{Paris}, but are among the top six if we order them by increasing length. There are two other possible answers for this query, namely those where path 6 is replaced with one of the other two paths of length five from \texttt{Joe} to \texttt{ENS}\; \texttt{Paris}.
    If the query would state \SIMPLE instead of \WALKS, then path 6 would not be in the answer, since are only five simple paths from \texttt{Joe} to \texttt{ENS}\; \texttt{Paris}.
    \item $\semp{\ANY\ 4\ \TRAIL
    \ (\texttt{Joe},\texttt{follows}^*\cdot \texttt{works},?x)}{G}$ can contain \texttt{Joe} and \texttt{ENS}\; \texttt{Paris} together with the following four paths:
    \begin{enumerate}
        \item $\texttt{e3}\rightarrow \texttt{e5}\rightarrow \texttt{e11}$;
        \item $\texttt{e3}\rightarrow \texttt{e6}\rightarrow \texttt{e10}$;
        \item $\texttt{e4}\rightarrow \texttt{e7}\rightarrow \texttt{e10}$; and
        \item $\texttt{e2}\rightarrow \texttt{e1}\rightarrow \texttt{e3}\rightarrow \texttt{e6}\rightarrow \texttt{e10}$.
    \end{enumerate}
    Replacing \TRAIL with \SIMPLE would force the fourth path to be replaced by a simple path, e.g., $\texttt{e4}\rightarrow \texttt{e12}\rightarrow\texttt{e5}\rightarrow \texttt{e11}$.
    \item $\semp{\SHORTEST\ $2$\ \GROUPS\ \WALKS\ (\texttt{Joe},\texttt{follows}^*\cdot \texttt{works},?x)}{G}$ \emph{ contains}  \texttt{Joe} and \texttt{ENS}\; \texttt{Paris} together with the following two groups of paths:
    \begin{enumerate}
        \item Group 1 (paths of length three): 
        \begin{enumerate}
        \item $\texttt{e3}\rightarrow \texttt{e5}\rightarrow \texttt{e11}$;
        \item $\texttt{e3}\rightarrow \texttt{e6}\rightarrow \texttt{e10}$; and 
        \item $\texttt{e4}\rightarrow \texttt{e7}\rightarrow \texttt{e10}$.
        \end{enumerate}
        \item Group 2 (paths of length four): 
        \begin{enumerate}
        \item $\texttt{e4}\rightarrow \texttt{e12}\rightarrow\texttt{e5}\rightarrow \texttt{e11}$; and
        \item $\texttt{e4}\rightarrow \texttt{e12}\rightarrow\texttt{e6}\rightarrow \texttt{e10}$. \qed
        \end{enumerate}
    \end{enumerate}
\end{itemize}
\end{example}

\medskip
\noindent \textbf{Remark.} While we define the semantics of queries as \emph{sets} of answers, the GQL specification requires answers to be returned grouped by endpoints $(v,v')$. For instance, when answering a query of the form \ALL\ \SHORTEST\ $q$, if there are multiple matching paths from $v$ to $v'$, these should be returned one after the other before moving to another pair of nodes. All algorithms we present in the paper enumerate the results in such a fashion.

\medskip
\noindent \textbf{Output-Linear Delay.} 
Returning paths can result in a large number of outputs. Therefore, to measure the efficiency of our algorithms, we will use the paradigm of \emph{enumeration algorithms} \cite{bagan2006mso,LosemannM14,Segoufin13,FlorenzanoRUVV20,pmr,MartensT-icdt18}. Such algorithms work in two phases: a \emph{pre-processing phase}, which constructs a data structure allowing us to output the solutions; and the \emph{enumeration phase}, which enumerates these solutions \emph{without repetitions}. The efficiency of an enumeration algorithm is measured by the complexity of the pre-processing phase, and the  delay  between any two outputs produced during the enumeration phase. We will say that an enumeration algorithm works with \emph{output-linear delay}, when the delay is linear in the size of each output element. This means that the time needed to output a single path is linear in the number of nodes in the path, after which we immediately start to output the next path. Notice that this is optimal in a sense that to return a path to the user, we have to at least write down each element of the path.

\medskip
\noindent \textbf{Regular Expressions and Automata.} We assume basic familiarity with regular expressions and finite state automata~\cite{sakarovitch2009}. 
We use $\mathcal{A} = (Q,\Sigma,\delta,q_0,F)$ to denote a non-deterministic finite automaton (NFA). Here, $Q$ is a set of states, $\Sigma$ a finite alphabet of edge labels, $\delta$  the transition relation over $Q\times \Sigma\times Q$, $q_0$ the initial  state, and $F$ the set of final states, respectively. An NFA  is deterministic (or a DFA for short) if $\delta$ is a function. A regular expression $\rgx$ can be converted into an equivalent $\varepsilon$-free NFA of size quadratic in $|\rgx|$ (see~\cite{sakarovitch2009}). Here, the quadratic factor only depends on the size of the alphabet $\Sigma$. There are several standard ways of converting an expression to an automaton, for instance using Thompson's or Glushkov's construction \cite{sakarovitch2009}.  In this paper, we assume that the automaton has a single initial state and that no $\varepsilon$-transitions are present. An NFA is called \emph{unambiguous} if it has at most one accepting run for every word. Every DFA is unambiguous, but the converse is not necessarily true.

\section{Backbone of \mdb and the WALK Semantics}
\label{sec:walk}
We describe the main idea behind \mdb and show how it can be used to
 evaluate RPQs under the \WALK semantics. That is, we show how to treat queries of the form
$$\sel\ \WALK\ (v,\rgx,?x)\;.$$
Recall that the selector $\sel$ for such queries is obligatory in GQL, since the set of all walks can be infinite. For simplicity of exposition, we assume for now that the starting point of the RPQ is fixed (i.e., we start from a concrete node $v \in \nodes$) and discuss the other cases for endpoints later. 
Our approach relies heavily on the product automaton construction~\cite{HopcroftUllman}, but we apply it to an automaton for the RPQ and the graph database, as in \cite{MendelzonW89}. We call the resulting structure the \emph{product graph}. We will present  how the product graph is built in its entirety, but in practice it is important to construct it \emph{lazily}, i.e., only construct the part of it that is needed to evaluate the query. 

\medskip
\noindent \textbf{Product Graph.}  Given a graph database $G = (V,E,\rho,\lambda)$ and an expression of the form $q = (v,\rgx,?x)$,
the \emph{product graph} is constructed by first converting the regular expression $\rgx$ into an equivalent non-deterministic finite automaton $\cA = (Q,\Sigma,\delta,q_0,F)$. The automaton consists of a set of states $Q$, the finite subset $\Sigma$ of $\labels$ that is used in $\rgx$, the transition relation $\delta \subseteq Q\times \Sigma\times Q$, the initial  state $q_0$, and the set of final states $F$. 
The product graph $G_\times(\cA,G)$, which is obtained from $\cA$ and $G$ is then defined as the graph database $(V_\times, E_\times, \rho_\times, \lambda_\times)$, where
\begin{itemize}
    \item $V_\times = V\times Q$;
    \item $E_\times = \{(e,(q_1,a,q_2)) \in E\times \delta \mid  \lambda(e) = a\}$;
    \item $\rho_\times((e,d)) = ((x,q_1),(y,q_2))$ if:
    \begin{itemize}
        \item $d = (q_1,a,q_2)$ and
        \item $\rho(e)=(x,y)$;
    \end{itemize}
    \item $\lambda_\times((e,d)) = \lambda(e)$.
\end{itemize}
We sometimes write $G_\times$ when $\cA$ and $G$ are clear from the context.
Each node of the form $(u,q)$ in $G_\times$ corresponds to the node $u$ in $G$ and, furthermore, each path $P$ of the form $(v,q_0) (e_1,(q_0,a_1,q_1)) (v_1,q_1) \cdots (e_n,(q_{n-1},a_n,q_n)) (v_n,q_n)$ in $G_\times$ corresponds to a path $p = v e_1 v_1 \cdots e_n v_n$ in $G$ that (a) has the same length as $P$ and (b) brings the automaton from state $q_0$ to $q_n$.
As such, when $q_n \in F$, then the path $p$ in $G$ matches $\rgx$. In other words, all nodes $v'$ that can be reached from $v$ by a path that matches $\rgx$ can be found by using standard graph search algorithms (e.g., BFS/DFS) on $G_\times$ starting in the node $(v,q_0)$. 

While this approach allows finding pairs $(v,v')$ in answers to an RPQ, we show next how it can be used to also find paths. 
Our approach is rooted in this fairly standard construction in automata theory and has important advantages: (a) the product graph can indeed be explored lazily \emph{for all 27 path modes}, (b) paths can be always returned \emph{on-the-fly}, allowing pipelined execution, and (c) the approach is highly efficient (Section~\ref{sec:exp}). We point out in which cases subtleties such as \emph{unambiguity of the automaton} need to be taken into account to achieve a correct algorithm.

\subsection{ANY (SHORTEST) WALK}
\label{ss:anyWalk}
We first treat the \WALK restrictor combined with selectors \ANY and \ANY \SHORTEST, that is, queries of the form:
\begin{gather}
\label{lab:anywalk}
    Q = \ANY\ (\SHORTEST)?\ \WALK\ (v,\rgx,?x)
\end{gather}
Let $\cA = (S,\Sigma,\delta,q_0,F)$ be an NFA for $\rgx$. The idea is that, for queries $Q$ of the form \eqref{lab:anywalk}, we can perform a classical graph search algorithm such as BFS or DFS starting at the node $(v,q_0)$ of the product graph $G_\times(\cA,G)$. Since both BFS and DFS support reconstructing a single (shortest in the case of BFS) path to any reached node, we obtain the desired semantics for queries of the form (\ref{lab:anywalk}). Query evaluation is presented in Algorithm \ref{alg:any}. The algorithm is a straightforward adaptation of BFS and DFS on a lazily constructed $G_\times$ with modifications that eliminate multiple paths reaching the same node and multiple accepting runs in an automaton. We present the algorithm in detail since we extend the approach to support different semantics later.

The basic object we manipulate is a \emph{search state}, i.e., a quadruple of the form $(n,q,e,prev)$, where $n$ is the node of $G$ we are currently exploring, $q$ is the current state of $\cA$, while $e$ is the edge of $G$ we used to reach $n$, and $prev$ is a pointer to the search state we used to reach $(n,q)$ in $G_\times$. Intuitively, the $(n,q)$-part of the search state allows us to track the node of $G_\times$ we are traversing, while $e$, together with $prev$ allows to reconstruct the path from $(v,q_0)$ that we used to reach $(n,q)$. The algorithm uses four data structures: 
\begin{itemize}
    \item \emph{Open}, which is a queue (in case of BFS), or stack (in case of DFS) of search states, with usual push() and pop() methods. 
    \item \emph{Visited}, which is a dictionary of search states we have already visited in our traversal. It prevents us from entering an infinite loop. We assume that $(n,q)$ can be used as a search key to check if some $(n,q,e,prev)\in \textit{Visited}$. We remark that $prev$ always points to a state stored in \textit{Visited}.
    \item \textit{Solutions}, which is a set containing (pointers to) search states in \textit{Visited} that encode a solution path to be returned; and
    \item \textit{ReachedFinal}, a set containing nodes we already returned as query answers, in case we re-discover them via a different end state (recall that an NFA can have several end states).
\end{itemize}

\begin{algorithm}
\caption{Evaluation of $Q = \ANY\ (\SHORTEST)? \WALK\ (v,\rgx,?x)$}
\label{alg:any}
\begin{algorithmic}[1]
\Function{AnyWalk}{$G,Q$}
    \State $\cA \gets$ Automaton($\rgx$) \Comment{\emph{with initial state $q_0$ and final states $F$}}
    \State \textit{Open}.init(); \textit{Visited}.init(); \textit{ReachedFinal}.init() 
    \State startState $\gets$ ($v,q_0,null,\bot$)
    \State \textit{Visited}.push(startState); 
           \textit{Open}.push(startState)
    \While{\textit{Open} $\neq \varnothing$}
        \State current $\gets$ \textit{Open}.pop() \Comment{\emph{with current = ($n,q,\mathit{edge},\mathit{prev}$)}}
        \If{$q \in F$ and $n \notin$ \textit{ReachedFinal}} 
            \State \textit{ReachedFinal}.add($n$)
            \State \textit{Solutions}.add(current)
        \EndIf
 	\For{\textbf{each} ($n',q',\mathit{edge}'$) $\in$ Neighbors(current$,G,\cA$)}
            \If{($n',q',*,*$) $\notin$ \textit{Visited}} 
                \State newState $\gets$ ($n',q',\mathit{edge}',$ current)
                \State \textit{Visited}.push(newState); 
                       \textit{Open}.push(newState)
            \EndIf
 	\EndFor 
    \EndWhile
\EndFunction
\end{algorithmic}
\end{algorithm}


The algorithm explores the product $G_\times(\cA,G)$ using either BFS (if \textit{Open} is a queue) or DFS (if \textit{Open} is a stack), starting from $(v,q_0)$. It starts by initializing the data structures and setting up the start node in $G_\times$ (lines 2--5). The main loop of line 6 is the classical BFS/DFS algorithm that pops an element $(n,q,e,prev)$ from \textit{Open} (line 7) and starts exploring its neighbors in $G_\times$ (lines 11--14). When exploring $(n,q,e,prev)$, we scan all the transitions $(q,a,q')$ of $\cA$ that originate from state $q$, and look for neighbors of $n$ in $G$ reachable by an $a$-labeled edge (line 11). The test $(n',q',\mathit{edge}')\in$ Neighbors$((n,q,\mathit{edge},\mathit{prev}),G,\cA)$ in line 11 checks whether $\rho(\mathit{edge}')=(n,n')$ in $G$, and whether $(q,\lambda(\mathit{edge}'),q')$ is a transition of $\cA$. If the pair $(n',q')$ has not been visited yet, we add it to $\textit{Visited}$ and $\textit{Open}$ (lines 12--14), which allows it to be expanded later on in the algorithm. When popping from \textit{Open} in line 7, we also check if $q$ is a final state and that $n$ has not been reached by a solution path yet (line 8). In this case we found a new solution; i.e., a walk from $v$ to $n$ whose label is in the language of \rgx, so we add it to \textit{Solutions} (line 10) and record it as reached (line 9). The $\textit{ReachedFinal}$ set is used to ensure that each solution is returned only once. 
Basically, since our automaton is non-deterministic and can have  multiple final states, two things can happen:
\begin{itemize}
    \item[(i)] The automaton might be ambiguous, that is, it could have multiple accepting runs on the same word. This, in turn, could result in the same path being returned twice, which is incorrect.
    \item[(ii)] There could be two different paths $p$ and $p'$ in $G$ from $v$ to some $n$, such that both $\lab(p)\in \mathcal{L}(\rgx)$ and $\lab(p')\in \mathcal{L}(\rgx)$, but the accepting runs of $\cA$ on these two words end up in different end states of $\cA$. Again, this could result with $n$ and a path to it being returned twice.
\end{itemize}
Both issues are handled using the \textit{ReachedFinal} set, which allows a solution to be recorded only once (lines 8--10). So, for each node that is a query answer, a single path is returned, without any restrictions on the automaton used for modeling the query.  

The procedure continues until $\textit{Open}$ is empty, meaning that there are no more states to expand and all reachable nodes have been found with a walk from the starting node $v$. We note that \textit{Solutions} stores the pointers to states in \textit{Visited}, which define a solution path. So, for each tuple $(n,q,e,prev)$ in \textit{Solutions} after running the algorithm, a path from $v$ to $n$ can be reconstructed using the $prev$ part of search states stored in \textit{Visited}. Furthermore, solutions can be enumerated once the algorithm terminates, or returned as soon as they are detected (line 10). This approach allows for a pipelined execution of the algorithm. Using BFS guarantees that the returned path is indeed shortest.

\begin{example}\label{ex:anywalk}
    Consider again the graph $G$ in Figure~\ref{fig:introNew}, and let $Q$ be the query
    $$\ANY\ \SHORTEST\ \WALK\ (\texttt{John},\ \texttt{follows}^+\cdot \texttt{lives},\ ?x).$$
So, we wish to find places where people that John (transitively) follows live. Considering Figure~\ref{fig:introNew}, we see that Rome is such a place, and the shortest path reaching it starts with \texttt{John}, and loops back to the same node using the edges \texttt{e1} and \texttt{e2}, before reaching \texttt{Rome} (via \texttt{e8}), as required. To compute the answer, Algorithm~\ref{alg:any} first converts the regular expression $\texttt{follows}^+\cdot \texttt{lives}$ into the following automaton:
\begin{center}
    \begin{tikzpicture}[->,>=stealth',auto, thick, scale = 1.0,initial text= {},    a/.style={ circle,inner sep=2pt    }]
		  \node [state,initial] at (0,0) (q0) {$q_0$};
		  \node [state] at (3,0) (q1) {$q_1$};
        \node [state,accepting] at (6,0) (qF) {$q_F$};

		  \path[->] (q0) edge node[above] {\texttt{follows}} (q1);
		  \path[->] (q1) edge[loop above] node[above] {\texttt{follows}} (q1);    
		  \path[->] (q1) edge node[above] {\texttt{lives}} (qF);
\end{tikzpicture}
\end{center}

\begin{figure}
    \centering
    \resizebox{0.6\columnwidth}{!}{
\begin{tikzpicture}[->,>=stealth',auto, thick, scale = 1.0,initial text= {},    a/.style={ circle,inner sep=2pt    }]
        \node [a] at (0,0) (John1) {$(\texttt{John},q_0)$};
        \node [a] at (3,0) (Joe) {$(\texttt{Joe},q_1)$};
        \node [a] at (6,1) (John2) {$(\texttt{John},q_1)$};
        \node [a] at (6,0) (Paul) {$(\texttt{Paul},q_1)$};
        \node [a] at (6,-1) (Lily) {$(\texttt{Lily},q_1)$};
        \node [a] at (9,1) (Rome) {$(\texttt{Rome},q_F)$};
        \node [a] at (9,0) (Anne) {$(\texttt{Anne},q_1)$};
        \node [a] at (9,-1) (Jane) {$(\texttt{Jane},q_1)$};

		  \path[->] (Joe) edge node [above, sloped] {\texttt{e1}} (John1);	
        \path[->] (John2.west) edge node [above, sloped] {\texttt{e2}} (Joe);
        \path[->] (Paul) edge node [above, sloped] {\texttt{e3}} (Joe);
        \path[->] (Lily.west) edge node [above, sloped] {\texttt{e4}} (Joe);
        \path[->] (Rome) edge node [above, sloped] {\texttt{e8}} (John2);
        \path[->] (Anne) edge node [above, sloped] {\texttt{e5}} (Paul);
        \path[->] (Jane.west) edge node [above, sloped] {\texttt{e6}} (Paul);
\end{tikzpicture}
}
    \vspace{-15pt}
    \caption{\textit{Visited} after running Algorithm~\ref{alg:any} in Example~\ref{ex:anywalk}.}
    \label{fig:visitedany}
\end{figure}
To find shortest paths, we use the BFS version of Algorithm~\ref{alg:any} and explore the product graph starting at $(\texttt{John},q_0)$. The algorithm then explores the only reachable neighbor $(\texttt{Joe},q_1)$, and continues by visiting $(\texttt{John},q_1), (\texttt{Paul},q_1)$ and $(\texttt{Lily},q_1)$. When expanding $(\texttt{John},q_1)$ the first solution, \texttt{Rome}, is found and recorded in \textit{Solutions}. The algorithm continues by reaching $(\texttt{Anne},q_1)$ and $(\texttt{Jane},q_1)$ from $(\texttt{Paul},q_1)$. When the $(\texttt{Lily},q_1)$ node is then expanded, it would revisit $(\texttt{Jane},q_1)$ \textcolor{black}{and $(\texttt{Paul},q_1)$}, which are blocked in line \textcolor{black}{12}. Expanding $(\texttt{Anne},q_1)$ would try to revisit \texttt{Rome}, but since this solution was already returned, we ignore it. The structure of \textit{Visited} upon executing the algorithm is illustrated in Figure~\ref{fig:visitedany}. Here we represent the pointer $prev$ as an arrow to other search states in \textit{Visited}, and annotate the arrow with the edge witnessing the connection. Notice that we can revisit a node of $G$ (e.g. \texttt{John}), but not a node of $G_\times$ (e.g. $(\texttt{Jane},q_1)$). Since \textit{Solutions} contains only $(\texttt{Rome},q_F)$, we enumerate a single path, namely $\texttt{e1}\rightarrow \texttt{e2}\rightarrow \texttt{e8}$. \qed
\end{example}

\noindent \textbf{Enumerating the Results.} Conceptually, paths can be enumerated by following backward edges from any state in the set \textit{Solutions} created by Algorithm~\ref{alg:any}. For completeness, we now provide an enumeration procedure that can be applied to the set \textit{Solutions} once Algorithm~\ref{alg:any} had finished executing.

\bigskip

\begin{algorithmic}[1]
\Function{PrintSolutions}{\textit{Solutions}}
    \While{\textit{Solutions} $\neq \emptyset$} \Comment{\emph{enumerate \textit{Solutions}}}
        \State sol = \textit{Solutions}.pop()
        \State print(\textsc{getPath}(sol, [])) \Comment{\emph{retrieve paths}}
    \EndWhile    
\EndFunction
\item[]
\Function{getPath}{($n,q,\mathit{edge},\mathit{prev}$), output}
    \If{$\mathit{prev} = \bot$} 
        \Return [$v$] + output \Comment{\emph{we are in the first search state}}
    \Else\  
        \Return \textsc{getPath}($prev$, [$\mathit{edge},n$] + output) \Comment{\emph{recursive backtracking}}
    \EndIf
\EndFunction
\end{algorithmic}
We note that solutions can be returned at the moment they are found (line 7), but we first construct the entire solution set for simplicity of presentation.

\medskip

Algorithm~\ref{alg:any} is fairly simple, but shows in detail how to manipulate the product graph in an efficient manner and how to handle arbitrary RPQs while finding a single (shortest) path between each pair of nodes in the result set. In addition, it  eliminates duplicate results in the sense that, for every node $n$ of $G$, it only computes a single tuple of the form $(n,q,e,prev)$ in \textit{Solutions} if $q$ is accepting.
In the following sections, we show how this approach can be extended for more complex path modes, starting with finding \emph{all} shortest paths.

\medskip

\noindent \textbf{Runtime.} Given that Algorithm~\ref{alg:any} runs the standard BFS/DFS algorithm on the product graph to construct the set \textit{Solutions} which allows enumerating the output paths, it runtime is $O(|\cA|\cdot |G|)$. Notice that, since we only return a single answer, this also means that the algorithm runs in output-linear delay.

\subsection{ALL SHORTEST WALKS}
\label{ss:allWalk}

We next show how to evaluate queries of the form
\begin{gather}
\label{lab:allwalk}
    Q = \textit{ALL SHORTEST WALK}\ (v,\rgx,?x).
\end{gather}
For this, we extend the BFS version of Algorithm~\ref{alg:any} in order to find \emph{all shortest paths} between pairs $(v,v')$ of nodes, instead of a single path --- see Algorithm~\ref{alg:allshortest}. The intuition simple: to obtain all shortest paths, upon reaching $v'$ from $v$ by a path conforming to $\rgx$ \emph{for the first time}, BFS always does so using a shortest path. The length of this path can then be recorded together with $v'$. When a new path reaches the same node $v'$ later, it is also an answer to our query $Q$ if its length equals the recorded length for $v'$.

\begin{algorithm}[t]
\caption{Evaluation of $Q= \textit{ALL SHORTEST WALK}\ (v,\rgx,?x)$.}
\label{alg:allshortest}
\begin{algorithmic}[1]
\Function{AllShortestWalk}{$G,Q$}
    \State $\cA \gets$ UnambiguousAutomaton($\rgx$) \Comment{\emph{with initial state $q_0$ and final states $F$}}
    \State \textit{Open}.init(); \textit{Visited}.init(); \textit{ReachedFinal}.init() 
        \State startState $\gets$ $(v,q_0,0,\bot)$
        \State \textit{Visited}.push(startState);
               \textit{Open}.push(startState)
    \While{\textit{Open} $\neq \varnothing$}
        \State current $\gets$ \textit{Open}.pop() 
        \Comment{\emph{with current = $(n,q,\mathit{depth},\mathit{prevList})$}}
        \If{$q \in F$} 
            \If{$n\notin$ \textit{ReachedFinal}}
               \State \textit{ReachedFinal}.add($\langle n, \mathit{depth}\rangle$)
                \State \textit{Solutions}.add(current)
            \ElsIf{\textit{ReachedFinal}.get($n$).depth $= \mathit{depth}$}
                \State \textit{Solutions}.add(current)
            \EndIf
            
        \EndIf
 	\For{$(n',q',\mathit{edge}') \in$ Neighbors(current$,G,\cA$)}
            \If{$(n',q',*,*) \in$ \textit{Visited}} 
                \State $(n',q',\mathit{depth}',\mathit{prevList}') \gets$ \textit{Visited}.get($n',q'$)
                \If{$\mathit{depth}+1 = \mathit{depth}'$} \Comment{new shortest path to $(n',q')$}
                    \State $\mathit{prevList}'$.add($\langle$current$,\mathit{edge}'\rangle$)
	    	  \EndIf
            \Else
            \State \textit{prevList}.init() 
            \State \textit{prevList}.add($\langle$current$,\mathit{edge}'\rangle$)
            \State newState $\gets$ ($n',q',\mathit{depth}+1,\mathit{prevList}$)
            \State \textit{Visited}.push(newState);
                 \textit{Open}.push(newState)
            \EndIf
 	\EndFor 
    \EndWhile
\EndFunction
\end{algorithmic}
\end{algorithm}

As before, we use $\cA$ to denote an NFA for $\rgx$. We will additionally require that $\cA$ is \emph{unambiguous} (it has at most one accepting run for every word), which we need to ensure that we do not return the same path twice. This condition is easy to enforce for real-world RPQs~\cite{BonifatiMT-www19,BonifatiMT20}. Indeed, a recent study of almost 150 million RPQs, found in over 900 million real-world SPARQL queries discovered that essentially all RPQs allow linear-time translations to unambiguous NFAs~\cite{HammererM25}, although the worst-case blow-up is exponential~\cite{KimelfeldMN25,Leung05}. The main difference to Algorithm~\ref{alg:any} is in the \emph{search state} structure. A search state is now a quadruple of the form $(n,q,\mathit{depth},\mathit{prevList})$, where $n$ is a node of $G$ and $q$ a state of $\cA$, the length of the shortest path from $(v,q_0)$ to $(n,q)$ is $\mathit{depth}$, and $\mathit{prevList}$ is a list of pointers to a previous search state that allows us to reach $n$ via a shortest path from $v$.
We assume that $\mathit{prevList}$ is a linked list, which we initialize empty, and to which we insert pairs of the form $\langle \mathit{searchState},\mathit{edge}\rangle$ using add(). Intuitively, $\mathit{prevList}$ allows us to reconstruct \emph{all} shortest paths that reach a node.

When adding a pair $\langle \mathit{searchState},\mathit{edge}\rangle$, we assume $\mathit{searchState}$ to be a pointer to a previous search state, and $\mathit{edge}$ will be used to reconstruct the path passing through the node in this previous search state.  Again, \textit{Visited} is a dictionary of search states with search key $(n,q)$. So, for each $(n,q)$, there can be at most one tuple $(n,q,\mathit{depth},\mathit{prevList})$ in \textit{Visited}. This tuple will be returned by calling \textit{Visited}.get($n,q$) if it exists.

Algorithm~\ref{alg:allshortest} explores the product graph of $G$ and $\cA$ using BFS, so \textit{Open} is a queue. The main difference to Algorithm \ref{alg:any} is as follows: if a node $(n',q')$ of the product graph $G_\times$ has already been visited (line 15), we do not discard the new path, but keep it if and only if it is shortest (line 17). In this case, the $\mathit{prevList}$ for $(n',q')$ is extended by adding the new path (line 18). 
If a new pair $(n',q')$ is discovered for the first time, a fresh $\mathit{prevList}$ is created (lines 20--23). As in Algorithm~\ref{alg:any}, we check for solutions after a state has been removed from $\textit{Open}$ (lines 8--13). Basically, when a state is popped from the queue, the structure of the BFS algorithm assures that we already explored all shortest paths to this state. \emph{Notice that solutions can actually be returned before (i.e., after line 16 we can test if $q'\in F$).} Returning answers when popping from the queue in lines 8--13 has the benefit that the paths are grouped for a pair $(v,n)$ of connected nodes, which is required by the GQL standard. Notice that this is also a natural place to perform enumeration of paths to achieve pipelined execution instead of enumerating them after the entire graph had been explored. As in Algorithm~\ref{alg:any}, we need to make sure that reaching the same node $n$ via different accepting states of $\cA$ is permitted only when this results in a shortest path. To achieve this, \textit{ReachedFinal} is now a dictionary storing pairs $\langle n, \mathit{depth}\rangle$, where $n$ is a solution node and $\mathit{depth}$ the length of a shortest path from $v$ to $n$. We assume $n$ to be the dictionary key. If a solution is reached the first time (lines 9--11) we record this information. When reaching the same node with another accepting path (lines 12--13), we record the solution only if the length of the path is shortest, that is, the same as the length already recorded. Finally, we can enumerate the solutions by a depth-first traversal the DAG stored in \textit{Visited}, starting from the nodes in \textit{Solutions}. 

\begin{example}\label{ex:allshortest}
    We reconsider the social network graph $G$ in Figure~\ref{fig:introNew} and let $Q$ be the query
    $$\textit{ALL SHORTEST WALK}\ (\texttt{Joe},\texttt{follows}^*\cdot \texttt{works},?x).$$
As we can see, there are three shortest paths from \texttt{Joe} to \texttt{ENS Paris} matching the query. To compute these, Algorithm~\ref{alg:allshortest}  first converts the expression $\texttt{follows}^*\cdot \texttt{works}$ into the following automaton $\cA$:
    \begin{center}
\resizebox{0.3\columnwidth}{!}{
    \begin{tikzpicture}[->,>=stealth',auto, thick, scale = 1.0,initial text= {},    a/.style={ circle,inner sep=2pt    }]
		  \node [state,initial] at (0,0) (q0) {$q_0$};
		  \node [state,accepting] at (3,0) (qF) {$q_F$};

		  \path[->] (q0) edge [loop above] node[above] {\texttt{follows}} (q0);
		  \path[->] (q0) edge node[above] {\texttt{works}} (qF);
\end{tikzpicture}
}
\end{center}
Notice that the automaton is unambiguous, that is, it has exactly one accepting run for each word in its language.
Algorithm~\ref{alg:allshortest} then starts traversing the product graph $G_\times(\cA,G)$ from the node $(\texttt{Joe},q_0)$. Upon executing the algorithm, the structure of \textit{Visited} is as follows:
\begin{center}
\resizebox{0.75\columnwidth}{!}{
\begin{tikzpicture}[->,>=stealth',auto, thick, scale = 1.0,initial text= {},    a/.style={ circle,inner sep=2pt    }]
        	\node [a] at (0,0) (Joe) {$(\texttt{Joe},q_0,0)$};
        	\node [a] at (4,1.1) (Paul) {$(\texttt{Paul},q_0,1)$};
            \node [a] at (4,-1.1) (John) {$(\texttt{John},q_0,1)$};
        	\node [a] at (4,0) (Lily) {$(\texttt{Lily},q_0,1)$};    
        	\node [a] at (8,1.1) (Anne) {$(\texttt{Anne},q_0,2)$};         
        	\node [a] at (8,0) (Jane) {$(\texttt{Jane},q_0,2)$};
        	\node [a] at (12,0) (ENS) {$(\texttt{ENS}\;\texttt{Paris},q_F,3)$};

		  \path[->] (Paul.west) edge node [above, sloped] {\texttt{e3}} (Joe);	
		  \path[->] (Lily) edge node [above, sloped] {\texttt{e4}} (Joe);	
		  \path[->] (Anne) edge node [above, sloped] {\texttt{e5}} (Paul);	
		  \path[->] (Jane) edge node [above, sloped] {\texttt{e6}} (Paul);	
		  \path[->] (Jane) edge node [above, sloped] {\texttt{e7}} (Lily);	
		  \path[->] (ENS) edge node [above, sloped] {\texttt{e11}} (Anne);	    
		  \path[->] (ENS) edge node [above, sloped] {\texttt{e10}} (Jane);	
		  \path[->] (John.west) edge node [above, sloped] {\texttt{e2}} (Joe);    
    
\end{tikzpicture}
}
\end{center}

Here we represent $\mathit{prevList}$ as a series of arrows to other  states in \textit{Visited}, and only draw $(n,q,\mathit{depth})$ in each node. For instance, $(\texttt{Jane},q_0,2)$ has two outgoing edges, representing two pointers in its $\mathit{prevList}$. The arrow is also annotated with the edge witnessing the connection (as stored in $\mathit{prevList}$). 

To build this structure, Algorithm~\ref{alg:allshortest} explores the neighbors of $(\texttt{Joe},q_0)$; namely,  $(\texttt{Paul},q_0), (\texttt{Lily},q_0)$ and $(\texttt{John},q_0)$ and puts them to  \textit{Visited} and \textit{Open}, with $\mathit{depth}=1$. The algorithm proceeds by visiting $(\texttt{Anne},q_0)$ from $(\texttt{Paul},q_0)$ and $(\texttt{Jane},q_0)$ from $(\texttt{Paul},q_0)$. The interesting thing happens in the next step when $(\texttt{Lily},q_0)$ is the node being expanded to its neighbor $(\texttt{Jane},q_0)$, which is already present in \textit{Visited}. Here we trigger lines 15--18 of the algorithm for the first time, and update the $\mathit{prevList}$ for  $(\texttt{Jane},q_0)$, instead of ignoring this path. \textcolor{black}{We then trigger these lines again for $(\texttt{Paul},q_0)$, but do not add it because the depth test in line 17 fails (we visited \texttt{Paul} already with a path of length 1). When we explore neighbors of $(\texttt{John},q_0)$, we revisit $(\texttt{Joe},q_0)$, and again do not add it because the depth test fails.} We then explore the node $(\texttt{ENS}\;\texttt{Paris},q_F)$ in $G_\times$ by traversing the neighbors of $(\texttt{Anne},q_0)$. Finally, $(\texttt{ENS}\;\texttt{Paris},q_F)$ will be revisited as a neighbor of $(\texttt{Jane},q_0)$ on a previously unexplored shortest path. We can then enumerate all the paths from $(\texttt{ENS}\;\texttt{Paris},q_F)$, by following the edges.\qed
\end{example}

\medskip
Interestingly, Algorithm~\ref{alg:all} \emph{explores precisely the same portion of the product graph $G_\times$ as Algorithm~\ref{alg:any}}. Intuitively, the reason is that even if we only need to find one witnessing path for each $(v,v')$ where $v$ is fixed and $v'$ variable, neither algorithm can predict whether the next step is going to arrive at a new or already seen node $v'$ without taking a look. (Indeed, even when answering queries with \ANY \SHORTEST, we need to continue searching for other nodes $v'' \neq v'$ that are reachable from $v$ with a matching path.)



\medskip

\noindent \textbf{Enumerating the Results.} Similarly as in Algorithm~\ref{alg:any}, results can be enumerated by following backward edges from any state in the set \textit{Solutions}. However, we need some extra steps, since Algorithm~\ref{alg:allshortest} records multiple paths between a pair of nodes. The full enumeration procedure is as follows:
\begin{algorithmic}[1]
\Function{EnumeratePaths}{\textit{Solutions}}
        \While{\textit{Solutions} $\neq \emptyset$} \Comment{\emph{enumerate Solutions}}
        \State sol = \textit{Solutions}.pop()
        \State \textsc{getAllPaths}(sol, [])\Comment{\emph{retrieve paths}}
    \EndWhile 
\EndFunction
\item[]
\Function{getAllPaths}{state = $(n,q,\mathit{depth},\mathit{prevList})$, output}
    \If{$\mathit{prevList} = \bot$} 
        print($[v]$ + output) \Comment{\emph{we are in the first search state}}
    \EndIf
    \For{prev = $(\mathit{prevState}, \mathit{prevEdge}) \in \mathit{prevList}$} \Comment{\emph{recursive backtracking}}
        \State \textsc{getAllPaths}($\mathit{prevState}, [\mathit{prevEdge},n]$ + output)
    \EndFor 
\EndFunction
\end{algorithmic}
Notice that here a single search state in \textit{Solutions} can produce multiple output paths. This is handled in lines 7--8 of \textsc{GetAllPaths} by traversing each node's $\mathit{prevList}$.

\medskip

\noindent \textbf{Runtime.} For Algorithm~\ref{alg:allshortest} to work correctly, we need $\cA$ to be unambiguous. This can be achieved by a standard determinization procedure, but any algorithm that produces an unambiguous automaton is fine.
Concerning run-time, Algorithm~\ref{alg:allshortest} indeed explores precisely the same subgraph of $G_\times$ as Algorithm~\ref{alg:any}, since the nodes of $G_\times$ that we revisit do not get added to \textit{Open} again (lines 15--18). However, we can potentially add extra edges to \textit{Visited}, so $\semp{Q}{G}$ can be exponentially larger~\cite{pmr,pathChallenge}. Additionally, when enumerating the output paths, each path is returned in time proportional to its length, which is known as \emph{output-linear delay}~\cite{FlorenzanoRUVV20}. It is optimal in the sense that, to return a path we need to at least write it down. In short, we can conclude that Algorithm~\ref{alg:allshortest} runs with $O(|\cA|\cdot |G|)$ pre-processing time, after which the results can be enumerated in time proportional to their total length.

\subsection{ANY $k$ WALKS and SHORTEST $k$ WALKS}\label{ss:shortestk}

We now present an algorithm that returns $k$ paths in increasing length. Given that \textit{ANY} $k$ is automatically solved by the \textit{SHORTEST} $k$ case, we will focus on the queries of the form:
\begin{gather}
\label{lab:shortestwalk}
    Q = \textit{SHORTEST}\ k\ \textit{WALK}\ (v,\rgx,?x).
\end{gather}
The code is described in Algorithm~\ref{alg:allshortestkpaths}. We idea is similar as for \textit{ALL SHORTEST WALKS}, but instead of keeping track of only the shortest walks from $(v,q_0)$ to every pair $(n,q)$ in the product graph $G_\times$, we keep track of all paths of increasing lengths reaching each $(n,q)$, until we find at least $k$ paths for $(n,q)$. Intuitively, this is similar to running Algorithm~\ref{alg:allshortest} (that is, we use BFS), but not discarding new paths when they are not shortest. To achieve this, our \emph{search state} is going to be the tuple $(n,q,\mathit{depth},\mathit{numPaths},\mathit{prevList})$, where $n$ is a node in the graph, $q$ a state of the automaton for our query, $\mathit{depth}$ the length of the paths reaching $(n,q)$ in the product graph $G_\times$, while $\mathit{numPaths}$ is the total number of paths of length $\mathit{depth}$ from $(v,q_0)$ to $(n,q)$ and $\mathit{prevList}$ the list of pointers to predecessors of $(n,q)$ on any such path. 
The parameter $\mathit{numPaths}$ is used to keep count of up to $k$ paths (to each $(n,q)$) and to ensure termination when loops are present.

\begin{algorithm}[t]
\caption{Evaluation of $Q= \textit{SHORTEST}\ k\ \textit{WALK}\ (v,\rgx,?x)$.}
\label{alg:allshortestkpaths}
\begin{algorithmic}[1]
\Function{K-ShortestWalk}{$G,Q,k$}
    \State $\cA \gets$ UnambiguousAutomaton($\rgx$) \Comment{\emph{with initial state $q_0$ and final states $F$}}
    \State \textit{Open}.init(); \textit{Visited}.init();
    \State \textit{Solutions}.init() \Comment{\emph{dictionary with search key $n$}}
    \State \textit{CountSolutions}.init() \Comment{\emph{dictionary with search key $n$}}
    \State \textit{CountPaths}.init() \Comment{\emph{dictionary with with search key $(n,q)$}}
        \State startState $\gets$ $(v,q_0,0,1,\bot)$
        \State \textit{Visited}.push(startState); \textit{Open}.push(startState)
        \State \textit{CountPaths}[$(v,q_0)$] = 1
    \While{\textit{Open} $\neq \varnothing$}
        \State current $\gets$ \textit{Open}.pop() 
        \Comment{\emph{with current = $(n,q,\mathit{depth},\mathit{numPaths},\mathit{prevList})$}}
        \If{$q \in F$} 
        \If{\textit{CountSolutions}[$n$]  $ < k$}
               \State \textit{Solutions}[$n$].append(current)
               \State \textit{CountSolutions}[$n$] += $\mathit{numPaths}$                \Comment{counts accepting runs to $n$}
            \EndIf
            \EndIf
            
 	\For{$(n',q',\mathit{edge}') \in$ Neighbors(current$,G,\cA$)}
\If{\textit{CountPaths}[$(n',q')$] $ < k$} \Comment{we never need more than $k$ paths}
            \If{$(n',q',\mathit{depth}+1,*,*) \in$ \textit{Visited}} 
                \State $(n',q',\mathit{depth}+1,\mathit{numPaths}',\mathit{prevList}') \gets$ \textit{Visited}.get($n',q',\mathit{depth}+1$)
                    \State $\mathit{prevList}'$.add($\langle$current$,\mathit{edge}'\rangle$)
                    \State $\mathit{numPaths}'$ += $\mathit{numPaths}$ 
                    \State \textit{CountPaths}[$(n',q')$] += $\mathit{numPaths}$ \Comment{update the count}
            \Else \Comment{first time visiting $(n,q)$ at $\mathit{depth}+1$}
            \State $\mathit{prevList}$.init() 
            \State $\mathit{prevList}$.add($\langle$current$,\mathit{edge}'\rangle$)
            \State newState $\gets$ ($n',q',\mathit{depth}+1,\mathit{numPaths},\mathit{prevList}$)
 \State \textit{CountPaths}[$(n',q')$] += $\mathit{numPaths}$ \Comment{update the count}
        
            \State \textit{Visited}.push(newState);
                 \textit{Open}.push(newState)
                 \EndIf
            \EndIf
 	\EndFor 
    \EndWhile

\EndFunction



\end{algorithmic}
\end{algorithm}

To achieve the correct behavior, Algorithm~\ref{alg:allshortestkpaths} uses \textit{Open} as a queue and \textit{Visited} as a dictionary of search states. We now use \textit{Visited} so that its search key is the triple $(n,q,\mathit{depth})$, which will allow to track all the paths reaching $(n,q)$ of particular length. In addition, we will need the following three dictionaries which will allow us to keep track of encountered paths and tracking their total number:
\begin{itemize}
    \item \textit{CountPaths} is a dictionary searchable by the pair $(n,q)$, which stores the total number of paths reaching $(n,q)$ from $(v,q_0)$ in the product graph. (These paths can be longer than the shortest length.) This number will be used to prevent revisiting any  $(n,q)$ with more than $k$ paths (where $k$ is specified by the query) in order to ensure that the algorithm always terminates.
    \item \textit{CountSolutions} is a dictionary searchable by a node $n$ of our graph. \textit{CountSolutions}[$n$] stores the total number of paths reaching this node with a final state of the automaton.
    \item \textit{Solutions}, a dictionary searchable by $n$ which contains an ordered list of (pointers to) search states that are used to enumerate the solutions that reach $n$. Order is needed so that shortest paths are returned before longer ones. We use the \textit{Solutions}[$n$].append(current) to denote that \textit{current} is added to the end of the list in \textit{Solutions}[$n$].
\end{itemize}

We already gave the main idea: Algorithm~\ref{alg:allshortestkpaths} explores the product graph from $(v,q_0)$ using BFS, but keeps track of paths in increasing length until we find at least $k$ of them. In detail, when a node $(n',q')$ of the product graph is revisited by path of any length, this is recorded as long as the total number of paths to $(n',q')$ is less than $k$ (lines 17--28). Here, we consider two cases. First, when $(n',q')$ has already been visited by a path of length $\mathit{depth}+1$ (lines 18--22),  we increase the \textit{CountPaths}[$(n',q')$] by the total number of paths reaching $(n,q)$. Basically, by adding the single edge $\mathit{edge}'$, we can traverse every such path, since we can reach $(n,q)$. We also update the count of paths to $(n',q')$ of length $\mathit{depth} + 1$ by updating the $\mathit{numPaths}'$ parameter of the search state, and we add $\langle \textit{current}, \mathit{edge}'\rangle$ to  $\mathit{prevList}'$ thus updating its value in \textit{Visited}. The second case, when a path to $(n',q')$ of length $\mathit{depth}+1$ is considered for the first time (lines 24--28) is similar, but this time we also update \textit{Visited} and \textit{Open} with the newly created search state in order to continue looking for more solutions. Notice that Algorithm~\ref{alg:allshortestkpaths} always terminates, since in line 17 we check that no more than $k$ paths to each pair $(n',q')$ in the product graph are added. In particular, this means that no edge in the product graph can be traversed more than $k$ times. In lines 22 and 27 we can end up with more than $k$ paths to $(n',q')$ due to the existence of many paths of the same length. When enumerating solutions, we need to keep track of this as explained below.

The solutions are recorded in lines 12--15. We ensure that, when $k$ solution paths to a particular node $n$ have already been discovered, no new solutions are recorded for $n$ (line 13). Given that we run BFS, and assuming that \textit{Solutions}[$n$].append(current) creates an ordered list, paths reaching node $n$ will be given in increasing length. To achieve pipelinining, we could check whether \textit{CountSolutions[$n$] $ > k$} after line 15 and immediately return $k$ paths to $n$. Notice that since more than $k$ paths might be recorded (if CountSolutions[$n$] + $\mathit{numPaths}$ $ > k$ in line 15),  when enumerating the paths to any particular $n$ we should return $k$ path if more than $k$ are available. It is also important to notice how this might require extra computation since we cannot return the first batch of shortest paths before finding all $k$ shortest paths. Of course, returning the paths as they are encountered in line 12 could potentially result in the results not being grouped by endpoints of the paths, as required by the GQL standard. Next we illustrate how Algorithm~\ref{alg:allshortestkpaths} works by an example.

\begin{example}
    \label{ex:shortestk}
    Consider again the graph of Figure~\ref{fig:introNew} and the following query:
   $$Q = \SHORTEST\ 5\ \WALK\ (\texttt{Joe},\ \texttt{follows}^*\cdot \texttt{works},\ ?x).$$    

Algorithm~\ref{alg:allshortestkpaths} creates the structure in Figure~\ref{fig:allshortestk}. Here, each search state is depicted as a tuple $(n,q,\mathit{depth},\mathit{numPaths})$ with the $\mathit{prevList}$ being represented by outgoing edges. Above each search state, we show the total number of paths reaching the node $(n,q)$ in the product graph at this stage of the execution; namely, the content of \textit{CountPaths}[$(n,q)$]. 

\begin{figure}
\begin{center}
\resizebox{\columnwidth}{!}{
\begin{tikzpicture}[->,>=stealth',auto, thick, scale = 1.0,initial text= {},    a/.style={ circle,inner sep=2pt }, label distance=-25pt]
        \node [a] at (0,0) (Joe) {$(\texttt{Joe},q_0,0,1)$};
        \node [a] at (4,1.1) (Paul) {$(\texttt{Paul},q_0,1,1)$};
        \node [a] at (4,-2.2) (John) {$(\texttt{John},q_0,1,1)$};         
        \node [a] at (4,0) (Lily) {$(\texttt{Lily},q_0,1,1)$};    
        \node [a] at (8,1.1) (Anne) {$(\texttt{Anne},q_0,2,1)$};         
        \node [a] at (8,0) (Jane) {$(\texttt{Jane},q_0,2,2)$};
        \node [a] at (12,1.1) (ENS) {$(\texttt{ENS}\;\texttt{Paris},q_F,3,3)$};

        \node at ($(Joe)+(0,.4)$) {\textit{\textcolor{black}{1}}};
        \node at ($(Paul)+(0,.4)$) {\textit{\textcolor{black}{1}}};
        \node at ($(John)+(0,.4)$) {\textit{\textcolor{black}{1}}};
        \node at ($(Lily)+(0,.4)$) {\textit{\textcolor{black}{1}}};
        \node at ($(Anne)+(0,.4)$) {\textit{\textcolor{black}{1}}};
        \node at ($(Jane)+(0,.4)$) {\textit{\textcolor{black}{2}}};
        \node at ($(ENS)+(0,.4)$) {\textit{\textcolor{black}{3}}};

        \node [a] at (8,-1.1) (PPaul) {$(\texttt{Paul},q_0,2,1)$};
        \node [a] at ($(PPaul)+(4,1.1)$) (AAnne) {$(\texttt{Anne},q_0,3,1)$};
        \node [a] at ($(PPaul)+(4,0)$) (JJane) {$(\texttt{Jane},q_0,3,1)$};
        \node [a] at ($(AAnne)+(4,0)$) (EENS) {$(\texttt{ENS}\;\texttt{Paris},q_F,4,2)$};

        \node at ($(PPaul)+(0,.4)$) {\textit{\textcolor{black}{2}}};
        \node at ($(AAnne)+(0,.4)$) {\textit{\textcolor{black}{2}}};
        \node at ($(JJane)+(0,.4)$) {\textit{\textcolor{black}{3}}};
        \node at ($(EENS)+(0,.4)$) {\textit{\textcolor{black}{5}}};

        \node [a] at (8,-2.2) (Joe2) {$(\texttt{Joe},q_0,2,1)$};
        \node [a] at ($(Joe2)+(4,0.0)$) (Paul2) {$(\texttt{Paul},q_0,3,1)$};
        \node [a] at ($(Joe2)+(4,-1.1)$) (Lily2) {$(\texttt{Lily},q_0,3,1)$};
        \node [a] at ($(Joe2)+(4,-3.3)$) (John2) {$(\texttt{John},q_0,3,1)$};

        \node [a] at ($(Lily2)+(4,-1.1)$) (PPPaul) {$(\texttt{Paul},q_0,4,1)$};
        \node [a] at ($(PPPaul)+(4,1.1)$) (AAAnne) {$(\texttt{Anne},q_0,5,1)$};

        \node at ($(Paul2)+(0,.4)$) {\textit{\textcolor{black}{3}}};
        \node at ($(Lily2)+(0,.4)$) {\textit{\textcolor{black}{2}}};
        \node at ($(John2)+(0,.4)$) {\textit{\textcolor{black}{2}}};
        \node at ($(PPPaul)+(0,.4)$) {\textit{\textcolor{black}{4}}};
        \node at ($(AAAnne)+(0,.4)$) {\textit{\textcolor{black}{4}}};

        \node at ($(Joe2)+(0,.4)$) {\textit{\textcolor{black}{2}}};

        \node [a] at ($(Paul2)+(4,0)$) (Anne2) {$(\texttt{Anne},q_0,4,1)$};
        \node [a] at ($(Paul2)+(4,-1.1)$) (Jane3) {$(\texttt{Jane},q_0,4,2)$};
        \node [a] at ($(John2)+(4,0)$) (Joe3) {$(\texttt{Joe},q_0,4,1)$};            
        \node [a] at ($(Joe3)+(4,0)$) (Paul3) {$(\texttt{Paul},q_0,5,1)$};        	

        \node [a] at ($(Joe3)+(4,-1.1)$) (Lily3) {$(\texttt{Lily},q_0,5,1)$};        	
        \node [a] at (21.5,-6.6) (aaa) {$\textit{...}$};    

        \node [a] at 
        ($(Joe3)+(4,-2.2)$) (John3) {$(\texttt{John},q_0,5,1)$};        	
        \node [a] at (21.5,-7.7) (aaa) {$\textit{...}$};                

        \node at ($(Anne2)+(0,.4)$) {\textit{\textcolor{black}{3}}};
        \node at ($(Jane3)+(0,.4)$) {\textit{\textcolor{black}{5}}};
        \node at ($(Joe3)+(0,.4)$) {\textit{\textcolor{black}{3}}};
        \node at ($(Paul3)+(0,.4)$) {\textit{\textcolor{black}{5}}};
        \node at ($(Lily3)+(0,.4)$) {\textit{\textcolor{black}{3}}};
        \node at ($(John3)+(0,.4)$) {\textit{\textcolor{black}{3}}};

        	        	
		  \path[->] (Paul.west) edge node [above, sloped] {\texttt{e3}} (Joe);	
		  \path[->] (Lily) edge node [above, sloped] {\texttt{e4}} (Joe);	
		  \path[->] (Anne) edge node [above, sloped] {\texttt{e6}} (Paul);	
		  \path[->] (Jane) edge node [above, sloped] {\texttt{e7}} (Paul);	
		  \path[->] (Jane) edge node [above, sloped] {\texttt{e8}} (Lily);	
		  \path[->] (ENS) edge node [above, sloped] {\texttt{e12}} (Anne.east);	    
		  \path[->] (ENS) edge node [above, sloped] {\texttt{e11}} (Jane);	
		  \path[->] (John.west) edge node [above, sloped] {\texttt{e2}} (Joe);
          
    	  \path[->] (Anne2) edge node [above, sloped] {\texttt{e6}} (Paul2);	
		  \path[->] (Jane3) edge node [above, sloped] {\texttt{e7}} (Paul2);	
		  \path[->] (Jane3) edge node [above, sloped] {\texttt{e8}} (Lily2);	
          
		  \path[->] 
            (AAnne) edge node [above, sloped] {\texttt{e5}} (PPaul)
            (PPaul) edge node [above, sloped] {\texttt{e12}} (Lily)
            (JJane) edge node [above, sloped] {\texttt{e6}} (PPaul)
            (EENS) edge node [above, sloped] {\texttt{e10}} (JJane)
            (EENS) edge node [above, sloped] {\texttt{e11}} (AAnne)
            (PPPaul) edge node [above,sloped] {\texttt{e12}} (Lily2)
            (AAAnne) edge node [above,sloped] {\texttt{e5}} (PPPaul)
        ;

        \path[->] (Paul2.west) edge node [above, sloped] {\texttt{e3}} (Joe2);	
		  \path[->] (Lily2) edge node [above, sloped] {\texttt{e4}} (Joe2);	
		  \path[->] (John2.west) edge[bend left] node [above, sloped] {\texttt{e2}} ($(Joe2.south)+(0,0.5)$);
        \path[->] (Joe2) edge node [above, sloped] {\texttt{e1}} (John);
        \path[->] (Joe3) edge node [above, sloped] {\texttt{e1}} (John2);


\path[->] (Paul3) edge node [above, sloped] {\texttt{e3}} (Joe3);
\path[->] (Lily3) edge node [above, sloped] {\texttt{e4}} (Joe3);

\path[->] (John3.west) edge[bend left] node [above, sloped] {\texttt{e2}} ($(Joe3.south)+(0,0.5)$);
    
\end{tikzpicture}
}
\end{center}
    \vspace{-15pt}
    \caption{\textit{Visited} when running Algorithm~\ref{alg:allshortestkpaths} in Example~\ref{ex:shortestk}. The number above each search state is the content of \textit{CountPaths} for the $(n,q)$ portion of the search state.}
    \label{fig:allshortestk}
\end{figure}

\textcolor{black}{
The algorithm starts in $(\texttt{Joe},q_0)$ and records that there is a single path of length zero to this node. Then,
three paths of length one are discovered, going to $(\texttt{Paul},q_0)$, $(\texttt{Lily},q_0)$ and $(\texttt{John},q_0)$, respectively. 
}

\textcolor{black}{
Expanding $(\texttt{Paul},q_0)$ now visits $(\texttt{Anne},q_0)$ and $(\texttt{Jane},q_0)$ with paths of length two. If we expand $(\texttt{Lily},q_0)$, we find a second (shortest) path to $(\texttt{Jane},q_0)$, and a second (non-shortest) path to $(\texttt{Paul},q_0)$. Since this is a new path to this node of the product graph, the $\mathit{numPaths}$ component of $(\texttt{Lily},q_0,1,1)$, that is, 1, gets added to \textit{CountPaths}[$(\texttt{Paul},q_0)$], correctly tracking the total number of paths to be two: one of length one and another of length two. Next, we expand $(\texttt{John},q_0)$ to $(\texttt{Joe},q_0)$, but now with a path of length two. Again, this is a new path to this node of the product graph. Analogously to $(\texttt{Paul},q_0)$, we add 1 to \textit{CountPaths}[$(\texttt{Joe},q_0)$], for which we now have two paths: one of length zero and another of length two. 
}

\textcolor{black}{
Next, we explore paths of length three. Here $(\texttt{ENS}\;\texttt{Paris},q_F)$ is reached with three paths, so \textit{CountSolutions}[\texttt{ENS}\;\texttt{Paris}] also gets updated to three. From $(\texttt{Paul},q_0,2,1)$ and from $(\texttt{Joe},q_0,2,1)$, the algorithm does analogous steps as for $(\texttt{Paul},q_0,1,1)$ and $(\texttt{Joe},q_0,0,1)$, respectively, resulting in new values for \texttt{Anne}, \texttt{Jane}, \texttt{Paul}, \texttt{Lily}, and \texttt{John}. However, since we still haven't found five paths required by the query in any of these nodes, the exploration continues. 
}

\textcolor{black}{
For paths of length four, the same nodes as before are being explored, but now with paths of longer lengths. When we reach $(\texttt{ENS}\;\texttt{Paris},q_F,4,2)$ we actually found five solutions for \texttt{ENS}\;\texttt{Paris}. Notice that the dictionary \textit{Solutions[ENS]} contains two search states allowing for enumeration: $(\texttt{ENS}\;\texttt{Paris},q_F,3,3)$ and $(\texttt{ENS}\;\texttt{Paris},q_F,4,2)$. We can output the three paths for the first element of the list (paths of length three) and only the two paths of length four. (If there were more results of length four, we would only need to output two of them.)
}

\textcolor{black}{
Despite that we already have all the results, the algorithm will continue until each node of the product graph is visited at least five times or no further expansion is possible. Indeed, in principle, it is possible that there are other nodes we will still reach in state $q_F$. For example, if there were two more nodes in the graph: \texttt{Ted} and \texttt{Berkeley}, with edges stating that \texttt{Anne} follows \texttt{Ted} and that \texttt{Ted} works in \texttt{Berkeley}, then we would need to continue this exploration until a sufficient number of paths from \texttt{Joe} to \texttt{Berkeley} had been found.
}

\textcolor{black}{
However, when we would expand $(\texttt{Anne},q_0,4,1)$, we discover that we only reach $(\texttt{ENS}\;\texttt{Paris})$, for which we already found the required five paths, so the exploration stops there. For the same reason (but with \texttt{Jane} instead of \texttt{ENS} \texttt{Paris}, the expansion no longer follows \texttt{e6} out of $(\texttt{Paul},q_0,4,1)$ to add $(\texttt{Jane},q_0,5,1)$.
\qed}
\end{example}

\medskip

\noindent\textbf{Enumeration and Runtime.} The enumeration of results is similar as in Section~\ref{ss:allWalk}. The main difference is that now \textit{Solutions} is a dictionary which contains, for each solution node $n$, a search state that needs to be enumerated using the same procedure as in Section~\ref{ss:allWalk} when generating all shortest walks. Since the search states in \textit{Solutions}$[n]$ are added in order of increasing length, these will even be enumerated as required by the GQL standard. Since the algorithm sometimes discovers a number of paths of the same length in bulk, the algorithm sometimes finds more than the required number (for instance in lines 22 and 27 of Algorithm~\ref{alg:allshortestkpaths}). Therefore, when returning paths, we simply need to keep count of the exact number we return. 
Enumeration with output-linear delay is also very similar. There is a small change in the pre-processing phase, which can now run in time $O(k\cdot |\cA|\cdot |G|)$, since each node in the product graph is allowed to be revisited up to $k$ times (line 17 of Algorithm~\ref{alg:allshortestkpaths}, and as illustrated in Example~\ref{ex:shortestk}).

\subsection{SHORTEST $k$ GROUPS}
\label{ss:shortestGroups}

We now explain how to evaluate queries of the form
\begin{gather}
\label{lab:shortestgroups}
    Q = \textit{SHORTEST}\ k\ \textit{GROUPS WALK}\ (v,\rgx,?x)\;.
\end{gather}
The task is to return, for each node $v'$ such that the $k$-th smallest length of paths from $v$ to $v'$ that match $\rgx$ is $\ell$ (see Section~\ref{sec:prelim} for a formal definition), all tuples $(v,v',p)$ such that $p$ is matching path from $v$ to $v'$ of length at most $\ell$.


The solution is presented in Algorithm~\ref{alg:allshortestkgroupsTermination}. It follows a similar logic as our solution for \textit{SHORTEST} $k$ \textit{WALK}, but now it tracks how many different lengths of paths reach each pair $(n,q)$ in the product graph. To do this, we need to make some changes to search states. Now they are tuples of the form $(n,q,\mathit{depth},\mathit{prevList})$, where $n$ is the graph node, $q$ a state of the automaton $\cA$, and $\mathit{depth}$ the length of the path in the product graph that starts in $(v,q_0)$ and ends in $(n,q)$ (where $q_0$ is the start state of $\cA$). As before, $\mathit{prevList}$ is a list of pointers to search states that are immediate predecessors of $(n,q)$ on this path. As in Section~\ref{ss:shortestk}, we assume that \textit{Visited} is a dictionary of search states that has $(n,q,\mathit{depth})$ as the search key. We also use the following three data structures, which we use to track the number of groups and solutions encountered during the execution of the algorithm:
\begin{itemize}
    \item \textit{NumGroups}, a dictionary searchable by $(n,q)$ contains a pair $(\mathit{numGrps}, \mathit{lastDepth})$, where $\mathit{numPaths}$ is
    the number of different path groups that have visited $(n,q)$ thus far, while $\mathit{lastDepth}$ is the length of paths in the last group that visited $(n,q)$ at this point. We use the notation \textit{NumGroups}[$(n,q)$] = $(\mathit{numGrps},\mathit{lastDepth})$. 
    \item \textit{NumFinalGroups}, a dictionary searchable by $n$ again contains a pair $(\mathit{numGrps}, \mathit{lastDepth})$, where the numbers have similar meaning as in \textit{NumGroups}, but now track the number of groups to the node $n$ reached via any final state of the automaton, and the length of paths in the final group discovered thus far. We use the notation \textit{NumGroups}[$n$] = $(\mathit{numGrps},\mathit{lastDepth})$.
    \item \textit{Solutions}, a dictionary searchable by $n$, which contains an ordered list of (pointers to) search states that are used to enumerate the solutions that reach $n$. The ordering in \textit{Solutions} is needed so that shortest paths are returned prior to longer ones. We use the syntax \textit{Solutions}[$n$].append(sState) to denote that sState is added to the end of the list in \textit{Solutions[$n$]}.
\end{itemize}

\begin{algorithm}[t]
\caption{Evaluation of $Q= \textit{SHORTEST}\ k\ \textit{GROUPS}\ (v,\rgx,?x)$.}
\label{alg:allshortestkgroupsTermination}
\begin{algorithmic}[1]
\Function{k-ShortestGroups}{$G,Q$}
    \State $\cA \gets$ UnambiguousAutomaton($\rgx$) \Comment{\emph{with initial state $q_0$ and final states $F$}}
    \State InitAllStructures() \Comment{\emph{Initialize all data structures}}
        \State startState $\gets$ $(v,q_0,0,\bot)$
        \State \textit{Visited}.push(startState); \textit{Open}.push(startState)
        \State \textit{NumGroups}[$(n,q)$] = $(1,0)$ \Comment{\emph{First group, length is zero}}
    \While{\textit{Open} $\neq \varnothing$}
        \State current $\gets$ \textit{Open}.pop() 
        \Comment{\emph{current = $(n,q,\mathit{depth},\mathit{prevList})$}}
        \If{$q \in F$} 

               \If{$n \notin$ \textit{Solutions}.keys()} \Comment{\emph{First time visiting $n$}}
                    \State \textit{Solutions}[$n$].append(current)
                    \State \textit{NumFinalGroups}[$n$] = $(1,\mathit{depth})$
               \Else
                    \State group $\gets$  \textit{NumFinalGroups}[$n$] \Comment{\emph{group = $(n,\mathit{numGrps},\mathit{lastDepth})$}}
                    \If{$\mathit{depth} > \mathit{lastDepth}$ \textbf{and} $\mathit{numGrps}<k$} \Comment{\emph{New solution group}}
                        \State \textit{NumFinalGroups}[$n$] = $(\mathit{numGrps} + 1,\mathit{depth})$
                        \State \textit{Solutions}[$n$].append(current)
                    \EndIf
                    \If{$\mathit{depth} = \mathit{lastDepth}$}
                    \State \textit{Solutions}[$n$].append(current)
                    \EndIf                    
               \EndIf
            \EndIf
            
 	\For{$(n',q',\mathit{edge}') \in$ Neighbors(current$,G,\cA$)}
            \If{$(n',q',\mathit{depth}+1,*) \in$ \textit{Visited}} \Comment{\emph{Valid paths for an existing group}}
                \State $(n',q',\mathit{depth}+1,\mathit{prevList}') \gets$ \textit{Visited}.get($n',q',\mathit{depth}+1$) 
                    \State $\mathit{prevList}'$.add($\langle$current$,\mathit{edge}'\rangle$)
            \Else \Comment{\emph{First time visiting $(n',q')$ at $\mathit{depth}+1$}}    
\State $(\mathit{oldNumGroups},\mathit{oldDepth}) \gets $ \textit{NumGroups}[$(n',q')$] \Comment{\emph{If $(n',q')$ unvisited returns $(0,-1)$}}
 \If{$oldNumGroups < k$} \Comment{\emph{Creating a new group for $(n',q')$}}
            \State \textit{prevList}.init() 
            \State \textit{prevList}.add($\langle$current$,\mathit{edge}'\rangle$)
            \State newState $\gets$ ($n',q',\mathit{depth}+1,\mathit{prevList}$)
                \State \textit{Visited}.push(newState);
                 \textit{Open}.push(newState)
                \State \textit{NumGroups}[$(n',q')$] = $(\mathit{oldNumGroups}+1, \mathit{depth}+1)$
            \EndIf
            \EndIf
 	\EndFor 
    \EndWhile


\EndFunction




\end{algorithmic}
\end{algorithm}

Algorithm~\ref{alg:allshortestkgroupsTermination} starts from $(v,q_0)$ and scans its neighbors $(n',q')$ in the product graph using the triple $(n',q',\mathit{depth}+1)$ as the search key. Whenever $(n',q')$ was not visited at $\mathit{depth}+1$ (lines 24--31), a new group is created, making sure that no more than $k$ groups have been defined (lines 25--26). We remark that we might have already visited $(n',q')$ with paths of smaller length, so we need to check that no unnecessary groups will be created. If we already visited $(n',q')$ at $\mathit{depth} + 1$, we add these paths since they are paths of an existing group (lines 21--23). We check for solutions after popping from the queue (lines 9--19). The first time we detect that a node $n$ is a solution, we update the dictionary \textit{Solutions}, and update \textit{NumSolutions}[$n$] to record that a single group had been detected thus far, and record the length of paths in this group (lines 10--12). When a node $n$ is discovered as a query answer again, we might be visiting it either with paths of a previously undiscovered length (lines 15--17), or with ones of length that we had already seen (lines 18--19). In the former case, we check that only $k$ groups will be created and record the new group, adding its paths to \textit{Solutions}[$n$]. In the latter case, we add the discovered paths to the final group for $n$. 

If one would like to achieve pipelining, one can check in line 15 whether we are trying to create the group $k+1$ for the first time and in that case enumerate the results. If not, we can simply use \textit{Solutions} when the algorithm terminates, to see which nodes are in the solution set and enumerate their respective groups. Due to the fact that we use BFS style exploration, solutions have always been added in increasing order of path length. As a final remark, notice that we can have $k$ groups to some nodes $n$ reached by a final state of our automaton (stored in \textit{NumFinalGroups}[$n$]), but we might still need to discover $k$ groups to different pairs $(n,q')$ in the product graph, since these might lead to solutions for nodes other than $n$. This is the reason why we keep track of \textit{NumGroups}[$(n,q)$]. Next, we illustrate how the algorithm operates via an example.

\begin{example}
    \label{ex:shortestgroups}
    We again use the graph of Figure~\ref{fig:introNew} and consider the query
   $$Q = \textit{SHORTEST $2$ GROUPS WALK}\ (\texttt{Joe},\ \texttt{follows}^*\cdot \texttt{works},\ ?x)\;.$$   
   
\textcolor{black}{
The \textit{Visited} structure that will be generated by 
Algorithm~\ref{alg:allshortestkgroupsTermination} is in Figure~\ref{fig:shortestgroups}. As before, a search state $(n, \allowbreak q, \allowbreak \mathit{depth}, \allowbreak \mathit{prevList})$ is depicted as the triple $(n,q,\mathit{depth})$, with $\mathit{prevList}$ as outgoing arrows, pointing to previously created search states. The search starts with $(\texttt{Joe},q_0)$, creating its first group of length zero (the number of groups to the $(n,q)$ portion of the search state is above the triple). The immediate neighbors are then explored, and since each is reached for the first time, the algorithm discovers the lengths of their respective first groups. At distance two from the start we visit $(\texttt{Anne},q_0)$ and $(\texttt{Jane},q_0)$ for the first time, but we reach $(\texttt{Paul},q_0)$ and $(\texttt{Joe},q_0)$ for the second time, now with longer paths, so two new groups are created.}

\textcolor{black}{
At distance three, $(\texttt{ENS}\;\texttt{Paris},q_F)$ is discovered as a solution for the first time, while $(\texttt{Anne},q_0)$, $(\texttt{Jane},q_0)$, $(\texttt{John},q_0)$, and $(\texttt{Lily},q_0)$ are all reached with a new group of paths, bringing their group counts to two. At the same time, we do not expand $(\texttt{Joe},q_0)$ to $(\texttt{Paul},q_0)$, because this is blocked in line 26 and would create a third group to it.
}


\textcolor{black}{
At distance four we see $(\texttt{ENS}\;\texttt{Paris},q_F)$ for the second time, but we do not expand to any other nodes. Indeed, we already discovered two groups for all others. Notice that \textit{Solutions}[\texttt{ENS} \texttt{Paris}] contains $(\texttt{ENS}\;\texttt{Paris},q_F,3)$ and $(\texttt{ENS}\;\texttt{Paris},q_F,4)$ in this particular order, so we can first return all three paths of length three (the first group), following with the two paths of length four (the second group).
}
\qed
\end{example}

\noindent\textbf{Enumeration and Runtime.} The enumeration procedure in this case is identical as in Section~\ref{ss:shortestk}, but now more items might be present in \textit{Solutions}$[n]$. Output-linear delay is guaranteed by this method of enumeration and no tracking of the number of returned solutions is needed due to the fact that each solution within the group needs to be returned. Regarding the pre-processing phase, a bound of $O(k\cdot |\cA|\cdot |G|)$ is achieved as when comparing a single shortest to all shortest path runtime. Namely, for each group here we run the all-shortest algorithm for every path length in the group. As in that case, many more solutions are potentially recorded compared to the case of shortest $k$ paths.

\begin{figure}
\begin{center}
\resizebox{.9\columnwidth}{!}{
\begin{tikzpicture}[->,>=stealth',auto, thick, scale = 1.0,initial text= {},    a/.style={ circle,inner sep=2pt }, label distance=-25pt]
        	\node [a, label={north:{\textit{\textcolor{red}{1}}}}] at (0,0) (Joe) {$(\texttt{Joe},q_0,0)$};
        	\node [a, label={north:{\textit{\textcolor{red}{1}}}}] at (4,1.1) (Paul) {$(\texttt{Paul},q_0,1)$};
            \node [a] at (4,-2.2) (John) {$(\texttt{John},q_0,1)$};            
        	\node [a, label={north:{\textit{\textcolor{red}{1}}}}] at (4,0) (Lily) {$(\texttt{Lily},q_0,1)$};    
        	\node [a, label={north:{\textit{\textcolor{red}{1}}}}] at (8,1.1) (Anne) {$(\texttt{Anne},q_0,2)$};         
        	\node [a, label={ north:{\textit{\textcolor{red}{1}}}}] at (8,0) (Jane) {$(\texttt{Jane},q_0,2)$};
        	\node [a] at (12,1.1) (ENS) {$(\texttt{ENS}\;\texttt{Paris},q_F,3)$};

            \node at ($(ENS)+(0,.4)$) {\textit{\textcolor{red}{1}}};
            \node at ($(John)+(0,.4)$) {\textit{\textcolor{red}{1}}};

        \node [a] at (8,-1.1) (PPaul) {$(\texttt{Paul},q_0,2)$};
        \node [a] at ($(PPaul)+(4,1.1)$) (AAnne) {$(\texttt{Anne},q_0,3)$};
        \node [a] at ($(PPaul)+(4,0)$) (JJane) {$(\texttt{Jane},q_0,3)$};
        \node [a] at ($(AAnne)+(4,0)$) (EENS) {$(\texttt{ENS}\;\texttt{Paris},q_F,4)$};

            \node at ($(PPaul)+(0,.4)$) {\textit{\textcolor{red}{2}}};
            \node at ($(AAnne)+(0,.4)$) {\textit{\textcolor{red}{2}}};
            \node at ($(JJane)+(0,.4)$) {\textit{\textcolor{red}{2}}};
            \node at ($(EENS)+(0,.4)$) {\textit{\textcolor{red}{2}}};

            \node [a, label={ north:{\textit{\textcolor{red}{2}}}}] at ($(John)+(4,0)$) (Joe2) {$(\texttt{Joe},q_0,2)$};
            \node [a] at (12,-3.3) (Lily2) {$(\texttt{Lily},q_0,3)$};
            \node [a] at ($(Joe2)+(4,0)$) (John2) {$(\texttt{John},q_0,3)$};

            \node at ($(John2)+(0,.4)$) {\textit{\textcolor{red}{2}}};
            \node at ($(Lily2)+(0,.4)$) {\textit{\textcolor{red}{2}}};

                        

        	        	
		  \path[->] (Paul.west) edge node [above, sloped] {\texttt{e3}} (Joe);	
		  \path[->] (Lily) edge node [above, sloped] {\texttt{e4}} (Joe);	
		  \path[->] (Anne) edge node [above, sloped] {\texttt{e6}} (Paul);	
		  \path[->] (Jane) edge node [above, sloped] {\texttt{e7}} (Paul);	
		  \path[->] (Jane) edge node [above, sloped] {\texttt{e8}} (Lily);	
		  \path[->] (ENS) edge node [above, sloped] {\texttt{e12}} (Anne.east);	    
		  \path[->] (ENS) edge node [above, sloped] {\texttt{e11}} (Jane);	
		  \path[->] (John.west) edge node [above, sloped] {\texttt{e2}} (Joe);
          

 		  \path[->] (Lily2) edge node [above, sloped] {\texttt{e4}} (Joe2);	
 		  \path[->] (John2) edge node [above, sloped] {\texttt{e2}} (Joe2);
         \path[->] (Joe2) edge node [above, sloped] {\texttt{e1}} (John);



		  \path[->] 
            (AAnne) edge node [above, sloped] {\texttt{e5}} (PPaul)
            (PPaul) edge node [above, sloped] {\texttt{e12}} (Lily)
            (JJane) edge node [above, sloped] {\texttt{e6}} (PPaul)
            (EENS) edge node [above, sloped] {\texttt{e10}} (JJane)
            (EENS) edge node [above, sloped] {\texttt{e11}} (AAnne)
        ;

\end{tikzpicture}
}
\end{center}
    \vspace{-15pt}
    \caption{\textit{Visited} when running Algorithm~\ref{alg:allshortestkgroupsTermination} in Example~\ref{ex:shortestgroups}. The number above each search state is the content of \textit{NumGroups} for the $(n,q)$ portion of the search state.}
    \label{fig:shortestgroups}
\end{figure}

\section{TRAIL, SIMPLE, and ACYCLIC}
\label{sec:simple}
We now devise algorithms for finding trails, simple paths, or acyclic paths. It is well known that even checking whether there is a single path between two nodes that conforms to a regular expression and is a simple path, trail, or acyclic path is NP-complete~\cite{Baeza13,CruzMW87,MartensNT-stacs20,BaganBG20,MendelzonW89}, even for undirected graphs~\cite{MartensP22}, so we do not know any worst-case polynomial time algorithms. These hardness results already hold for fixed regular expressions, such as $(aa)^*$ or $a^*ba^*$~\cite{MartensNT-stacs20,BaganBG20,MendelzonW89}. 
On the other hand, regular expressions for which these problems are NP-hard are rare in practice~\cite{BonifatiMT-www19,MartensT-icdt18,MartensT-sigmodrecord19,HammererM25} and, moreover, regular expressions for which these problems are tractable can be evaluated by cleverly enumerating paths in the product graph~\cite{MartensT-icdt18,MartensT-tods19}.
Our algorithms will follow this intuition and prune the search space whenever possible. In the worst case, all such algorithms will be exponential but we will show that, on real-world graphs, the number of paths is manageable. (That said, queries that ask to return a potentially large number of paths are at the responsibility of the user. If the user wants to see a large output, the system should do its best to give it to them.) Our pruning technique technique will ensure that all dead ends are discarded and we keep the computational overhead under control. We begin by showing how to find \emph{all} trails and simple/acyclic paths.

\subsection{Returning All Paths}
\label{ss:algoSimple}
We start with queries of the form
$$Q = (\ALL\ \SHORTEST)?\  \res\  (v,\rgx,?x)$$
where the restrictor $\res$ is \TRAIL, \SIMPLE, or \ACYCLIC. Algorithm~\ref{alg:all} shows how to evaluate such queries. Intuitively, the algorithm explores the product graph by enumerating all paths starting in $(v,q_0)$ \textcolor{black}{but pruning as soon as the respective paths are no longer trails, simple, or acyclic.}   This time, our \emph{search state} is a  tuple $(n, q, \mathit{depth}, e, \mathit{prev})$, where $n$ is a node in the input graph, $q$ is a state automaton of the automaton $\cA$ for $\rgx$, the length of a shortest path from $(v,q_0)$ to $(n,q)$ in $G_\times$ is $\mathit{depth}$, the edge $e$ is the edge we used to arrive in node $n$, and $\mathit{prev}$ is a pointer to another search state stored in \textit{Visited}. The latter is a set containing search states that we have already visited. Similarly to Algorithm~\ref{alg:allshortest}, we use a dictionary \textit{ReachedFinal} containing pairs $(n,\mathit{depth})$, with key $n$. Here, $n$ is a node reached in some query answer and $\mathit{depth}$ is the length of a shortest path from $v$ to $n$ in $G$.

\begin{algorithm}[tb]
\caption{Evaluation for $Q = (\ALL\ \SHORTEST)?\ \res\ (v,\rgx,?x)$.}
\label{alg:all}
\begin{algorithmic}[1]
\Function{AllRestrictedPaths}{$G,Q$}
    \State $\mathcal{A} \gets$ UnambiguousAutomaton($\rgx$) \Comment{\emph{with initial state $q_0$ and final states $F$}}
    \State \textit{Open}.init(); \textit{Visited}.init(); \textit{ReachedFinal}.init()
	\State startState $\gets$ ($v,q_0,0,\mathit{null},\bot$)
    \State \textit{Visited}.push(startState); \textit{Open}.push(startState)
    \While{\textit{Open} $\neq \varnothing$}
        \State current $\gets$ \textit{Open}.pop() \Comment{\emph{current = ($n,q,\mathit{depth},e,\mathit{prev}$)}}
                \If{$q \in F$} 
                    \If{$!$(\ALL\ \SHORTEST)}
        		      \State \textit{Solutions}.add(current)
                    \ElsIf{$n \notin$ \textit{ReachedFinal}}
                        \State \textit{ReachedFinal}.add($\langle n,\mathit{depth} \rangle$)
        		      \State \textit{Solutions}.add(current)
                    \Else
                        \State $\mathit{optimalDepth} \gets$ \textit{ReachedFinal}.get($n$).$\mathit{depth}$
                        \If{$\mathit{depth} = \mathit{optimalDepth}$} 
            		      \State \textit{Solutions}.add(current)
                        \EndIf
                    \EndIf
    	    	\EndIf
 	\For{next $\gets (n', q', edge') \in$ Neighbors(current$,G,\mathcal{A}$)}
 		\If{isValid(current, next, \res)} \Comment{\emph{$\res$ is the restrictor}}
                \State new $\gets$ ($n', q', \mathit{depth}+1, \mathit{edge}',$ current)
                \State \textit{Visited}.push(new);
                       \textit{Open}.push(new)
        \EndIf
 	\EndFor 
    \EndWhile   
\EndFunction
\end{algorithmic}
\end{algorithm}

The execution is similar to Algorithm~\ref{alg:any}, but \textit{Visited} is not used to discard solutions. Instead, when checking whether the current node can be extended to a neighbor (lines 18--21), we call isValid, which checks whether restrictor $\res$ allows the path to the current node $n$ in \textit{Visited} to be extended to the next node $n'$. Notice that we need to check whether the path in the \emph{original graph} $G$ satisfies \res, and not the path in the product graph. If isValid confirms that this extension to $e'$ still satisfies $\res$, we add the new search state to \textit{Visited} and \textit{Open} (line 21). When popping from \textit{Open}, we also check if a potential solution is reached (line 8).

If \ALL\ \SHORTEST\ is \emph{not} present in the query, we simply add the newly found solution. If it is, we need to make sure to add only shortest paths. The ReachedFinal dictionary \textit{ReachedFinal} is used to track the already discovered nodes. If the node is seen for the first time, the dictionary is updated and a new solution is added (lines 11--13). Upon discovering the same node again (lines 14--17), a new solution is added only if it is shortest. Once all paths have been explored ($\mathit{Open}=\emptyset$), we can enumerate the solutions, just as in the previous algorithms.

\begin{example}\label{ex:simple}
\textcolor{black}{
   Consider again the graph $G$ in Figure~\ref{fig:introNew} and the query
    $$Q = \SIMPLE\ (\texttt{John},\ \texttt{follows}^+\cdot \texttt{lives},\ ?x).$$
Namely, we use the same regular expression as in Example~\ref{ex:anywalk}, but we allow only simple paths. The automaton $\cA$ is therefore identical to the one in Example~\ref{ex:anywalk}. The algorithm starts by visiting $(\texttt{John},q_0)$, followed by $(\texttt{Joe},q_1)$. After this we will visit $(\texttt{John},q_1)$, $(\texttt{Paul},q_1)$ and $(\texttt{Lily},q_1)$. In the next step we will try to expand $(\texttt{John},q_1)$, but will detect that this leads to a path which is not simple (we could have detected this in the previous step, but this way the pseudo-code is more compact). We will continue exploring neighbors, building the \textit{Visited} structure depicted in Figure~\ref{fig:visitedsimple}. In Figure~\ref{fig:visitedsimple} we use the same notion for $\mathit{prev}$ pointers as in previous examples. For brevity, we do not show $\mathit{depth}$, but this is simply the length of the path needed to reach $(\texttt{John},q_0)$ in Figure~\ref{fig:visitedsimple}. We note that several nodes, e.g, $(\texttt{Jane},q_1)$, appear multiple times since they will be present differently in the search state, e.g., as $(\texttt{Jane},q_1,3,\texttt{e6},\mathit{prev})$,  $(\texttt{Jane},q_1,3,\texttt{e7},\mathit{prev}')$, and $(\texttt{Jane},q_1,4,\texttt{e6},\mathit{prev}'')$.} 
\qed
\end{example}

\begin{figure}
    \centering
    \resizebox{0.95\columnwidth}{!}{
\begin{tikzpicture}[->,>=stealth',auto, thick, scale = 1.0,initial text= {},    a/.style={ circle,inner sep=2pt    }]
        	\node [a] at (0,0) (John) {$(\texttt{John},q_0)$};
        	\node [a] at (4,0) (Joe) {$(\texttt{Joe},q_1)$};
            \node [a] at (8,1.3) (John2) {$(\texttt{John},q_1)$};
        	\node [a] at (8,0) (Paul) {$(\texttt{Paul},q_1)$};    
        	\node [a] at (8,-1.3) (Lily) {$(\texttt{Lily},q_1)$};         
        	\node [a] at (12,1.3) (Anne) {$(\texttt{Anne},q_1)$};
        	\node [a] at (12,0) (Jane1) {$(\texttt{Jane},q_1)$};
            \node [a] at (12,-1.3) (Jane2) {$(\texttt{Jane},q_1)$};
            \node [a] at (16,1.3) (Rome) {$(\texttt{Rome},q_F)$};
            \node [a] at ($(Lily)+(4,-1.3)$) (Paul2) {$(\texttt{Paul},q_1)$};
            \node [a] at ($(Paul2)+(4,1.3)$) (Anne2) {$(\texttt{Anne},q_1)$};
            \node [a] at ($(Paul2)+(4,0)$) (Jane3) {$(\texttt{Jane},q_1)$};
            \node [a] at ($(Anne2)+(4,0)$) (Rome2) {$(\texttt{Rome},q_F)$};

		  \path[->] (Joe) edge node [above, sloped] {\texttt{e1}} (John);	
		  \path[->] (John2.west) edge node [above, sloped] {\texttt{e2}} (Joe);	
		  \path[->] (Paul) edge node [above, sloped] {\texttt{e3}} (Joe);	
		  \path[->] (Lily.west) edge node [above, sloped] {\texttt{e4}} (Joe);	
		  \path[->] (Jane1) edge node [above, sloped] {\texttt{e6}} (Paul);	
		  \path[->] (Anne.west) edge node [above, sloped] {\texttt{e5}} (Paul);	    
		  \path[->] (Jane2) edge node [above, sloped] {\texttt{e7}} (Lily);	
		  \path[->] (Rome) edge node [above, sloped] {\texttt{e9}} (Anne)
            (Paul2.west) edge node [above, sloped] {\texttt{e12}} (Lily)
            (Anne2.west) edge node [above, sloped] {\texttt{e5}} (Paul2)
            (Jane3) edge node [above] {\texttt{e6}} (Paul2)
            (Rome2) edge node [above] {\texttt{e9}} (Anne2)
          ;
\end{tikzpicture}
}
    \vspace{-10pt}
    \caption{\textit{Visited} after running Algorithm~\ref{alg:all} in Example~\ref{ex:simple}.}
    \label{fig:visitedsimple}
\end{figure}

\noindent\textbf{Enumeration and Runtime.} 
When dealing with \TRAIL, \SIMPLE, and \ACYCLIC, we need to distinguish between the complexity to construct \textit{Solutions} and enumerating the output once \textit{Solutions} has been constructed. Constructing \textit{Solutions} takes time $O\big((|\mathcal{A}|\cdot |G|)^{|G|}\big)$, which is exponential. Unless P $=$ NP, getting rid of the exponential is impossible, since already deciding whether a trail, simple path, or acyclic path exists that matches a (fixed) regular expression is NP-complete~\cite{CruzMW87,BaganBG20,MartensNP23}.
The algorithm for constructing \textit{Visited} and \textit{Solutions} terminates since eventually all paths that are valid according to the restrictor $\res$ will be explored, and no new search states will be added to \textit{Open}. 
Once \textit{Visited} is constructed, the enumeration algorithm is essentially the same as in Section~\ref{ss:anyWalk}. 
%
Output-linear complexity of enumeration (after \textit{Solutions} has been constructed) is achieved analogously to Algorithm~\ref{alg:any} since each solution defines a single path.

Due to the fundamental NP-hardness of the problem, the pipelined version of the algorithm does not achieve linear delay or output-linear complexity. Indeed, the worst-case delay or computation time for the next output is exponential, which is unavoidable unless P $=$ NP.

\subsection{Adding \ANY and \ANY\ \SHORTEST} Treating queries of the form
$$Q = \ANY\ (\SHORTEST)?\  \res\  (v,\rgx,?x)$$
can be done with minimal changes to Algorithm~\ref{alg:all}. Namely, we would use \textit{ReachedFinal} as a \emph{set} that stores the node first time a solution is found in order to not repeat any results. In addition, we would replace lines 8--17 with:
\begin{center}
{
\begin{algorithmic}
\If{$q \in F$ and $n\notin$ \textit{ReachedFinal}}
\State \textit{ReachedFinal}.add($n$)
\State \textit{Solutions}.add(current)
\EndIf
\end{algorithmic}
}
\end{center}

\noindent\textbf{Enumeration and Runtime.} The analysis is the same as for Algorithm~\ref{alg:all}. Its worst-case runtime is $O\big((|\mathcal{A}|\cdot |G|)^{|G|}\big)$, which is also the best known runtime due to NP-completeness of the problem~\cite{CruzMW87}. Enumeration is analogous to Algorithm~\ref{alg:all}.

\subsection{\SHORTEST $k$ and \ANY $k$}\label{ss:shortestKTrails}

Here we show how to deal with queries of the form
$$\SHORTEST\ k\ \res\ (v,\rgx,?x)\;,$$
where $\res$ is one of \TRAIL, \SIMPLE, or \ACYCLIC. We note that solving these cases also solves the \ANY $k$ case with a restriction to \TRAIL, \SIMPLE, or \ACYCLIC. 

The idea is presented in Algorithm~\ref{alg:shortestKretrictor} and is similar to finding all paths according to one of the three restrictors (Algorithm~\ref{alg:all}).
The difference is that we only track up to $k$ solutions for each node $n$ that is reachable from $v$ by a path that matches \rgx. As before, \textit{Visited} will be a set allowing us to reconstruct solution paths. To find shortest paths, we require \textit{Open} to be a queue, meaning that we do a BFS-style exploration. The structure of search state is again $(n,q,e,\mathit{prev})$, with $n$ a node in the graph, $q$ a state of the automaton, $e$ an edge used to reach $n$ and $\mathit{prev}$ a pointer to the predecessor search state in \textit{Visited} used to reach the current state. Additionally, we will use the following structures to keep track of $k$ shortest paths for each reachable node: 
\begin{itemize}
    \item \textit{Reachable} is a set that keeps track of each node $n$ reachable from $v$ by our query.
    \item \textit{ReachedFinal} is a dictionary with $n$ being the search key and storing the number of paths reaching $n$ with an accepting state. For example, \textit{ReachedFinal}[$n$] = 5 means we found five paths that are an answer to our query reaching the node $n$.
    \item \textit{Solutions} is a dictionary with $n$ being the search key. Each entry \textit{Solutions[$n$]} stores a list of search states which visit the node $n$ via an accepting state of our automaton.
\end{itemize}

\noindent\textbf{Enumeration and Runtime.} Basically, we do the same exploration as in Algorithm~\ref{alg:all}, but when we find a solution, we check that no more than $k$ shortest solutions are stored (lines 8--15). The program relies on the unambiguity on $\cA$, which can be achieved by determinizing it. Since we need to enumerate all the paths, the program's worst-case complexity is again $O\big((|\mathcal{A}|\cdot |G|)^{|G|}\big)$. Notice that no factor $k$ is present in this bound since it already includes exploring all possible paths and we only need to count the number of solutions to each reached node. For enumeration, once \textit{Solutions} is computed, we can traverse the list \textit{Solutions}$[n]$, which stores a search state with a single path which can again be enumerated with output-linear delay.

Similarly to Algorithm~\ref{alg:all}, output-linear delay for pipelined execution is impossible unless P $\neq$ NP.

\begin{algorithm}
\caption{Evaluating $Q= \SHORTEST\ k\ \res\ (v,\rgx,?x)$ over $G$.}
\label{alg:shortestKretrictor}
\begin{algorithmic}[1]
\Function{Shortest $k$ Restricted Paths}{$G,Q$}
    \State $\mathcal{A} \gets$ UnambiguousAutomaton($\rgx$) \Comment{\emph{with initial state $q_0$ and final states $F$}}
    \State Initialize() \Comment{\emph{Initialize all data structures}}
	\State start $\gets$ ($v,q_0,null,\bot$)
    \State \textit{Visited}.push(start); \textit{Open}.push(start)
    \While{\textit{Open} $\neq \varnothing$}
        \State current $\gets$ \textit{Open}.pop() \Comment{\emph{current = ($n,q,e,\mathit{prev}$)}}

        \If{$q\in F$} 
        \If{$n\notin$ \textit{Reachable}}
        \State \textit{Reachable}.add($n$)
        \State \textit{ReachedFinal}[$n$] = 1
        \State \textit{Solutions}[$n$].append(current)
        \ElsIf{\textit{ReachedFinal}[$n$] $ < k$} \Comment{Only return $k$ paths to $n$}
        \State \textit{ReachedFinal}[$n$] += 1
        \State \textit{Solutions}[$n$].append(current)
        \EndIf          
        \EndIf
        
 	\For{next = ($n',q',\mathit{edge}'$) $\in$ Neighbors(current$,G,\mathcal{A}$)}
 		\If{\text{isValid}(current, next, \res)} \Comment{\emph{This assures termination!}}   
            \State new $\gets$ ($n', q', \mathit{edge}',$ current)  
        \State \textit{Visited}.push(new); \textit{Open}.push(new)  
        \EndIf
 	\EndFor 
    \EndWhile
 \EndFunction

\end{algorithmic}
\end{algorithm}


\subsection{Shortest $k$ Groups}

The final class of queries we cover is of the form
$$\text{\SHORTEST $k$ \GROUPS $\res$ $(v,\rgx,?x)$},$$ where $\res$ is one of \TRAIL, \SIMPLE, \ACYCLIC. The corresponding algorithm, shown in Algorithm~\ref{alg:kgroupsrestrictor} is virtually identical to Algorithm~\ref{alg:shortestKretrictor}, but keeps track of groups instead of paths. Its main data structures are:
\begin{itemize}
    \item \textit{Reachable} is a set that keeps track of each node $n$ reachable from $v$ by our query.
    \item \textit{Groups} is a dictionary using nodes $n$ of our graph as the search key, and storing two components: (i) \textit{Groups}[$n$].numGroups is the number of different path lengths that we discovered reaching $n$ in a final state of the automaton; and (ii) \textit{Groups}[$n$].\text{lastDepth} is the length of the longest/last path we discovered reaching $n$ in a final state of the automaton.
    \item \textit{Solutions} is a dictionary with search key $n$. Each entry \textit{Solutions}[$n$] is a list of search states that visit the node $n$ in an accepting state of our automaton.
\end{itemize}

To ensure that only $k$ groups are returned for each $n$, we limit \textit{Groups}[$n$].numGroups to at most $k$ (line 14). When a solution $n$ is discovered for the first time (lines 9--13), the first solution group for $n$ is created. If we already discovered $n$ and no more than $k$ groups have been created (line 14), we consider the two viable options: (i) we are creating a new group, in which case the recorded number of solution groups needs to be strictly less than $k$ (lines 15--18); or (ii) we are extending the last group we created (lines 19--20). For correctness, we need the automaton to be unambiguous and the complexity and enumeration is the same as for Algorithm~\ref{alg:shortestKretrictor}.

\begin{algorithm}
\caption{Evaluating a query $Q= \SHORTEST\ k\ \GROUPS\  \res\ (v,\rgx,?x)$.}
\label{alg:kgroupsrestrictor}
\begin{algorithmic}[1]
\Function{Shortest $k$ Groups Restricted Paths}{$G,Q$}
    \State $\mathcal{A} \gets$ UnambiguousAutomaton($\rgx$) \Comment{\emph{with initial state $q_0$ and final states $F$}}
    \State Initialize() \Comment{\emph{Initialize all data structures}}
	\State start $\gets$ ($v,q_0,null,0,\bot$)
    \State \textit{Visited}.push(start); \textit{Open}.push(start)
    \While{\textit{Open} $\neq \varnothing$}
        \State current $\gets$ \textit{Open}.pop() \Comment{\emph{current = ($n,q,e,\mathit{depth},\mathit{prev}$)}}

        \If{$q\in F$} 
            \If{$n\notin$ \textit{Reachable}}
                \State \textit{Reachable}.add($n$)
                \State \textit{Groups}[$n$].numGroups = 1
                \State \textit{Groups}[$n$].lastDepth = $\mathit{depth}$
                \State \textit{Solutions}[$n$].append(current)
                \ElsIf{\textit{Groups}[$n$] $\leq k$}
                \If{(\textit{Groups}[$n$].lastDepth $< \mathit{depth}$ \textbf{and} Groups[$n$].numGroups $<k$)} \Comment{\emph{New group}}
                \State \textit{Groups}[$n$].numGroups += 1
                \State \textit{Groups}[$n$].lastDepth $= \mathit{depth}$
                \State \textit{Solutions}[$n$].append(current)             

                \ElsIf{\textit{Groups}[$n$].lastDepth $= \mathit{depth}$} \Comment{\emph{Same group, new solution}}
                \State \textit{Solutions}[$n$].append(current)     
                
                \EndIf    
                
            \EndIf       
        \EndIf

 	\For{next = ($n',q',\mathit{edge}'$) $\in$ \textit{Neighbors}(current$,G,\mathcal{A}$)}
 		\If{\text{isValid}(current, next, \res)} \Comment{\emph{This assures termination}}   
            \State new $\gets$ ($n', q', \mathit{edge}', \mathit{depth} +1, $ current)

        \State \textit{Visited}.push(new); \textit{Open}.push(new)        
        \EndIf
 	\EndFor 
    \EndWhile
 \EndFunction
\end{algorithmic}
\end{algorithm}

\section{Syntax for Returning Paths and Implementation Details}
\label{sec:implementation}
\label{ss:implementation}

\mdb is implemented as the path processing engine of \millennium, an open-source graph database engine~\cite{mdb}.  \millennium provides the infrastructure necessary to process generic queries such as RPQs and takes care of parsing, generation of execution plans, and data storage, while \mdb executes path queries as described in this paper. 
\millennium supports both the property graph data model coupled with an expressive fragment of Cypher~\cite{FrancisGGLLMPRS18} and RDF with a \textcolor{black}{fairly complete implementation of SPARQL 1.1 query (see the system documentation for a list of currently unsupported features)}. The system also provides updates and a light transactional management mechanism in terms of multi-version concurrency control with a single writer thread. While the Cypher-like syntax supports returning paths as described in this paper, we also extended the SPARQL syntax to allow it.

\medskip
\noindent \textbf{Syntax for Returning Paths in Property Graphs.} We extend \millennium's Cypher-like syntax with a new construct that allows returning paths in RPQ query answers by adding patterns of the following form into the \texttt{MATCH} clause:
\begin{center}
\verb|(<src>)=[<PathMode> <PathVariable> <RegularPathQuery>]=>(<tgt>)|
\end{center}
Here \texttt{src} and \texttt{tgt} are either graph nodes or variables, \texttt{PathMode} can be any GQL path mode, \texttt{PathVariable} is the variable used to bind and return the path, and \texttt{RegularPathQuery} is an arbitrary regular expression constructed from edge labels ($\rgx$ in Section~\ref{sec:prelim}). 
 \millennium uses a the \texttt{=[]=>} arrows for path queries, instead of the usual \texttt{-[]->} used in Cypher and GQL. For example, \millennium's syntax for a query that returns all shortest paths starting at node \texttt{Joe} in Figure~\ref{fig:introNew} that match \texttt{:follows+/:works} is
\begin{center}
\begin{verbatim}
    MATCH (?src  {name:"Joe"})=[ALL SHORTEST ?p :follows+/:works*]=>(?tgt)
    RETURN ?tgt.name, ?p
\end{verbatim}
\end{center}
Here, we assume that nodes in Figure~\ref{fig:introNew} have an attribute \texttt{name} that identifies their value as commonly used in property graphs.

\medskip
\noindent \textbf{Extending SPARQL with Path Returning Capabilities.} To support returning paths in SPARQL, we introduce a new triple pattern with the following abstract definition:
\begin{center}
\verb|<src> <PathMode> (<PropertyPath> AS <PathVariable>) <tgt>|
\end{center}
Similarly as in the case of property graphs,  \texttt{src} and \texttt{tgt} are either nodes (i.e., IRIs or literals) or variables, \texttt{PathMode} is a GQL path mode, \texttt{PathVariable} is the variable used to bind and return the path, and \texttt{PropertyPath} is a property path pattern. Our implementation currently supports property path patterns that do not use negated property sets. Furthermore, we do not yet implement the \TRAIL mode, as it is not clear whether a triple in RDF represents an edge, or whether a more subtle definition should be made, as in the case of property graphs. Path variables are used only to return paths. Such triple patterns can then be added to the SPARQL \texttt{WHERE} clause. An example query that extracts the names of people connected to \texttt{Joe} in an RDF dataset similar to the one in Figure~\ref{fig:introNew} would be:
\begin{center}
\begin{verbatim}
SELECT ?name ?p
WHERE {
  ex:Joe ANY SHORTEST (:follows+/:works AS ?p) ?tgt .
  ?tgt :hasName ?name .
}
\end{verbatim}
\end{center}


\medskip
\noindent \textbf{Data Access.}
By default, \millennium stores graph data on disk using B+trees and loads the necessary pages into a main memory buffer during query execution. The B+trees we use are clustered on initial loading of data and their leaves store the required relations using node/edge IDs. All literal values (strings, numbers, etc.) are also represented via numeric IDs and converted to their original representation as needed. All B+trees in \millennium are accessed using the standard linear iterator interface used in relational databases~\cite{ramakrishnan00}. Depending on the data model used, a different set of relations is stored as we explain next. 

\paragraph*{Property Graphs.} To represent property graphs, \millennium indexes several relations, however, our algorithms only use the following two:
\begin{center}
$\textsc{Edges(NodeFrom, Label, NodeTo, EdgeId)}$\\
$\textsc{Nodes(NodeId)}$
\end{center}
The first relation allows us to find neighbors of a certain node reachable by an edge with a specific label, as used, for instance, in Algorithm~\ref{alg:any} (line 14). The second table  keeps track of nodes that are materialized in the database. In essence, the table \textsc{Nodes} is only used when checking whether the start node of an RPQ actually exists in the database (see e.g., Algorithm~\ref{alg:any}, line 9). 

\paragraph*{Two-Way Navigation for Property Graphs.} All algorithms in this paper can be easily extended with the ability to traverse edges backwards, as required, for instance, by C2RPQs~\cite{CalvaneseGLV02} and SPARQL property paths~\cite{HarrisS13}. To support this, we also index the inverse \textsc{Edges} relation $\textsc{Edges$^-$(NodeTo, Label, NodeFrom, EdgeId)}.$ Using this index, all of the described algorithms can be extended by using the \textsc{EDGES}$^-$ table whenever looking for a neighbor accessed via a backward edge.

\paragraph*{RDF Data.} When storing RDF data we use four different permutations of the $\textsc{Triples(s,p,o)}$ relation. Most important for us are the $\mathit{SPO}$ and $\mathit{OPS}$ permutations, which mimic the \textsc{EDGES} and \textsc{EDGES}$^-$ relations from property graphs.

\medskip
\noindent \textbf{Pipelined Execution.} In \mdb, all the algorithms are implemented in a pipelined fashion using linear iterators. This means that the solution is returned to the user as soon as it is encountered. This requires, for instance, that the main while loop of Algorithm~\ref{alg:any} be halted as soon as a new solution is detected in lines 9--10, and similarly for Algorithm~\ref{alg:all}. In the case of Algorithm~\ref{alg:allshortest} the situation is a bit more complex, as solutions can be detected while scanning the neighbors (lines 14--23), instead of upon popping an element from the queue (lines 11-13). All of these issues can be resolved by noting that linear iterators for neighbors of a node can be paused, and their execution resumed. For completeness, we include expanded pseudo-code for pipelined execution of our algorithms in our online appendix~\cite{repo}. The main benefit of the pipelined execution is that paths can be encountered on demand and the entire solution set does not need to be constructed in advance.


\medskip
\noindent \textbf{Query Variables.} Throughout the paper, we assumed that the RPQ part of our query takes the form $(v,\rgx,?x),$
where $v$ is a node identifier, and $?x$ is a variable. In other words, we were searching for nodes reachable from a fixed node $v$. It is straightforward to extend our algorithms to patterns of the form $(v,\rgx,v')$, where both endpoints of the path are known: we can run any of the algorithms as described in the paper, and check whether $v'$ is a query answer.  The case when both ends are variables; namely,  $(?x,\rgx,?y)$, is more challenging to handle, and the techniques we developed only allow supporting it by iterating over all relevant start nodes. Some immediate optimizations are possible here; for instance, by only considering start nodes of edges labeled by transitions that leave the initial state of the automaton for \rgx. 
We leave the study of this problem for future work.

\section{Experimental Evaluation}
\label{sec:exp}
We empirically evaluate \mdb and show that the approach scales on a broad range of real-world and synthetic data sets and queries. We describe our experimental setup and discuss of the obtained results. For our code, data and queries see~\cite{ANONrepo}.

\subsection{Experimental Setup}
We perform three sets of experiments:
\begin{itemize}
    \item \textsf{Pokec}, which tests the effect of path length on performance;
    \item \textsf{Wikidata}, which tests the performance over a large real-world graph and user supplied queries; and
    \item \textsf{Diamond}, which tests the effect of having a large number of paths in the graph.
\end{itemize}
Next we describe each set of experiments in more detail.

\medskip
\noindent \textbf{(1) \textsf{Pokec.}} This set of experiments uses the Pokec social network graph from the SNAP graph collection~\cite{snapnets}. Pokec is a Slovakian social network which records (directed) user connections, similar to the graph of Figure~\ref{fig:introNew}, but with a single type of edge label (we call this label \textit{follows}). The graph contains around 1.6 million nodes and 30 million edges. In our experiment, we select the node that is median in terms of a simple centrality measure,\footnote{More precisely, we computed the number of edges in which each node participates and selected one whose count is the median for the dataset.} and traverse \textit{follows}-edges from this node. We run our algorithms until 100,000 paths that witness these connections are returned. We explore paths of length 1 through $k$, where $k$ ranges from 1 to 12. Longer paths are not very interesting in this graph, since the graph's diameter is 11. We pair these queries with the path modes described in Section~\ref{sec:prelim}. 
The  idea behind this experiment is to test what happens with query performance as we seek longer paths in a real-world graph of intermediate size.

\medskip
\noindent \textbf{(2) \textsf{Wikidata.}} Here, we want to check performance over a large real-world graph. For this we use Wikidata~\cite{VrandecicK14} and queries from its public SPARQL query log~\cite{MalyshevKGGB18}. We use WDBench~\cite{wdbench}, a recently proposed Wikidata SPARQL benchmark. WDBench provides a  curated version of the data set based on the truthy dump of Wikidata~\cite{WikiDataData}, which is an edge-labeled graph with 364 million nodes and 1.257 billion edges, using more than 8,000 different edge labels. The data set is publicly available~\cite{wddata}. WDBench provides multiple sets of queries extracted from the Wikidata's public endpoint query log.  We use the \textit{Paths} query set, which contains 659  2RPQs patterns.\footnote{We remove the single pattern from the original query set that uses negated property sets, a feature we currently do not support.} From these, 592 have a fixed starting point or ending point (or both), while 67 have both endpoints free. We note that these queries require general regular expressions which cannot be expressed in some of the tested systems. The 659 patterns are then used in our tests under the restrictor and selector options described in Section~\ref{sec:prelim}.

\medskip
\noindent \textbf{(3) \textsf{Diamond.}} Here we test what happens when there is a large number of paths present in our graph. The database we use, taken from~\cite{pmr}, is presented in Figure~\ref{fig:diamond}. The queries we consider look for paths from \texttt{start} to \texttt{end} using \textit{a}-labeled edges. Notice that all such paths are, at the same time, shortest, trails and simple paths, and have length $2n$. Furthermore, there are $2^n$ such paths, while the graph only has $3n+1$ nodes and $4n$ edges. We test our query with the path modes from Section~\ref{sec:prelim}, while scaling $n$ (and thus path length) from $1$ to $40$. While returning all these paths is unfeasible for any algorithm, we test whether a portion of them (100,000 in our experiments) can be retrieved efficiently. 

\begin{figure}
    \centering
\resizebox{.7\columnwidth}{!}{
		\begin{tikzpicture}[->,>=stealth',auto, thick, scale = 1.0,initial text= {},    node/.style={ square,inner sep=2pt    }]
  \tikzstyle{every state}=[draw,thick,rectangle,rounded corners,fill=white!85!black,minimum size=5mm, text=black, font=\ttfamily, inner sep=2pt]
            \tikzstyle{every node}=[font=\ttfamily]
		\def\y{.7}	
		\def\x{1.2}	
		
			\node [state](A) at (-0.2,0) [draw] {start}; 
			\node [state](q1) at (1*\x,1*\y)[draw] {$u_1$}; 
			\node [state](q2) at (1*\x,-1*\y) [draw] {$v_1$}; 	
			\node [state] (q3) at (2*\x,0*\y) [draw] {{$w_1$}};	
			\node [state](q4) at (3*\x,1*\y) [draw] {$u_2$}; 
			\node [state](q5) at (3*\x,-1*\y) [draw] {$v_2$}; 
			\node [state](q6) at (4*\x,0*\y) [draw] {{$w_2$}}; 
			
			\node [](q7) at (5*\x,0*\y) [] {$\cdots$}; 
			
			\node [state](q7) at (6*\x,0*\y) [draw] {{$w_{n-1}$}}; 
			\node [state](q8) at (7*\x,1*\y)[draw] {$u_n$}; 
			\node [state](q9) at (7*\x,-1*\y) [draw] {$v_n$}; 
			\node [state](B) at (8*\x,0*\y) [draw] {end};

			\path (A) edge   node[above] {a}  (q1);
			\path (A) edge   node[below] {a} (q2);
			\path (q1) edge   node[above]  {a} (q3);
			\path (q2) edge  node[below]  {a} (q3);
			
			\path (q3) edge   node[above] {a}  (q4);
			\path (q3) edge   node[below] {a} (q5);
			\path (q4) edge   node[above]  {a} (q6);
			\path (q5) edge  node[below]  {a} (q6);
			
			\path (q7) edge   node[above] {a}  (q8);
			\path (q7) edge   node[below] {a} (q9);
			\path (q8) edge   node[above]  {a} (B);
			\path (q9) edge   node[below]  {a} (B);
		\end{tikzpicture}
    }
    \caption{Graph database with exponentially many paths from \texttt{start} to \texttt{end}.}
    \label{fig:diamond}
\end{figure}

\medskip
\noindent \textbf{Tested Systems.} \emph{In all the experiments, we use the property graph version of \mdb.} Recall that \mdb supports both BFS traversal 
and DFS traversal. 
We denote the two versions as \mdbbfs and \mdbdfs, respectively. When there is only one algorithm (e.g., for all shortest walks), we simply write \mdb. All the versions assume the data to be stored on disk and being buffered into main memory as required. 

In order to compare with state of the art in the area, we selected six publicly available graph database systems that allow for benchmarking with no legal restrictions. 
These are:
\begin{itemize}
    \item \sysneo: \textcolor{black}{Neo4J Community Edition version 5.26.26}~\cite{Webber12};
  \item \sysnebula: NebulaGraph version 3.5.0~\cite{nebula};
  \item \syskuzu: Kuzu version 0.0.6~\cite{kuzu};
  \item \sysjena: Jena TDB version 4.1.0~\cite{JenaTDB};
  \item \sysblaze: Blazegraph version
  2.1.6~\cite{ThompsonPC14}; and
  \item \sysvirtuoso: Virtuoso version
  7.2.6~\cite{Erling12}.
\end{itemize}
From these systems, \sysneo and \sysnebula use the \ALL \TRAIL semantics by default. \sysneo and \syskuzu support \ANY \SHORTEST \WALK and \ALL \SHORTEST \WALK modes. \syskuzu also supports \ALL \WALK, but limits paths to length at most 30. SPARQL systems (\sysjena, \sysblaze, \sysvirtuoso) support arbitrary RPQs but do not return paths. Following the SPARQL semantics~\cite{HarrisS13}, they detect pairs of nodes connected by an arbitrary walk. A brief summary of supported features can be found in Table~\ref{tab:intro} in the Introduction. Other systems we considered are \textsc{DuckDB}~\cite{duckDB}, Oracle Graph Database~\cite{oracle} and Tiger Graph~\cite{TigerGraph}, which support (parts of) SQL/PGQ. Unfortunately, \cite{oracle} and \cite{TigerGraph} are commercial systems with limited free versions, while the SQL/PGQ module for \textsc{DuckDB}~\cite{duckDB} is still in development.

\textcolor{black}{We note that \emph{none of the existing systems} support path variables to the extent that our system does. This is essentially because SPARQL systems don't have path variables and systems based on GQL or SQL/PGQ don't deal with all regular expressions or cap the maximal path length as a hardcoded value. Since the current GQL standard consider nesting of Kleene star as a language opportunity, existing systems simply don't implement it. \mdb shows that, when it comes to path variables, nesting of Kleene star is, in principle, not an issue.}



\medskip
\noindent \textbf{Experimental Setup.} The experiments were run on a commodity server with an Intel\textregistered Xeon\textregistered \allowbreak Silver 4110 CPU, and 128GB of
DDR4/2666MHz RAM, running Linux Debian 10 with the kernel version 5.10. The hard
disk used to store the data was a SEAGATE model  ST14000NM001G with 14TB capacity. Note that this is a classical HDD, and not an SSD. Custom indexes for speeding up the queries were created for \sysneo, \sysnebula and \syskuzu, and the four systems were run with the default settings and no limit on RAM usage. \sysjena, \sysblaze, \sysvirtuoso and \mdb were assigned 64GB of RAM for buffering. Since we run large batches of queries, these are executed in succession, in order to simulate a realistic load to a database system. 
\begin{center}
\emph{All queries were run with a limit of 100,000 results and a timeout of 1 minute}.
\end{center}

\subsection{Pokec: Scaling Path Length}
\label{ss:pokec}
\begin{figure}
    \centering
    \resizebox{0.7\columnwidth}{!}{
    \includegraphics{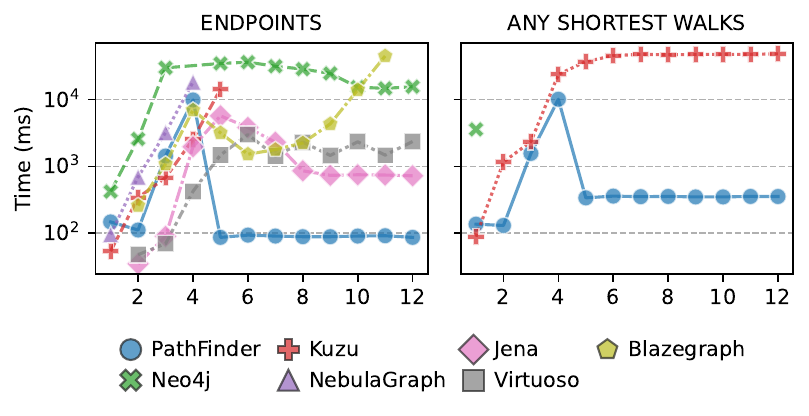}
    }
    \vspace{-10pt}
    \captionsetup{justification=centering}
    \caption{Runtimes over the Pokec dataset -- endpoints and a single (shortest) path experiment.}
    \label{fig:pokecsec5}
\end{figure}

\begin{figure}
    \centering
    \resizebox{0.7\columnwidth}{!}{
    \includegraphics{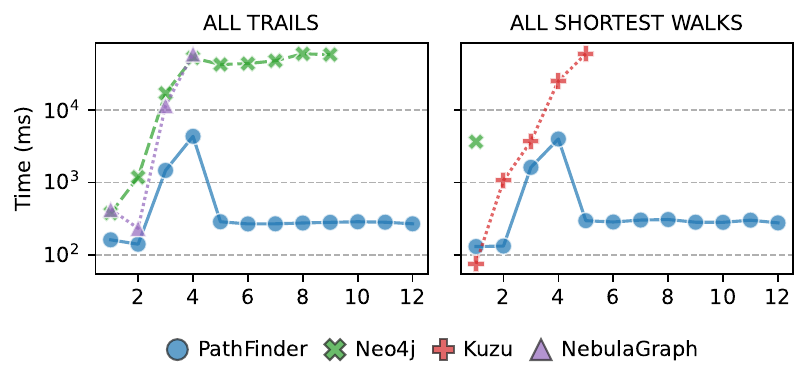}
    }
    \vspace{-10pt}
    \captionsetup{justification=centering}
    \caption{Performance of graph engines in the Pokec dataset.}
    \label{fig:pokecintro}
\end{figure}

Here we take a highly connected graph of medium size and test what happens if we ask for paths of increasing length. 
All tested systems easily loaded this data set.
Given that SPARQL systems cannot return paths, we compare with them when the system is only asked to retrieve the reachable nodes (the \textsf{ENDPOINTS} experiment). Given that other systems only support trails and walks, we retrieve paths according to these two modes. Our results are presented in Figures~\ref{fig:pokecsec5} and~\ref{fig:pokecintro}. 

\medskip
\noindent \textbf{\textsf{ENDPOINTS}.} Figure~\ref{fig:pokecsec5} (left) shows our results for retrieving reachable nodes, \textcolor{black}{that is, the set of nodes $v'$ for that bind to $?x$ in RPQs of the form $(v,\rgx,?x)$.} We see that SPARQL systems handle this use case relatively well (only \sysblaze timed out for the largest path length). Likewise, \sysneo shows no timeouts. In contrast, \sysnebula and \syskuzu start timing out for paths of length 5 and 6, respectively. 

\mdb and several other systems exhibit a spike in runtime for lengths 3 and 4. For \mdb, this happens due to a large portion of pages being fetched from disk into the buffer. Once the data needed to compute 100,000 paths has been fetched, the performance stabilizes. \mdb shows superior performance \textcolor{black}{(around 10x--100x faster than the other systems)} once the runtime stabilizes when all data required to run the queries is loaded from disk. 
We note that \mdb does \emph{not} use caching but rather recomputes the query for each length larger than 5. 

\medskip
\noindent \textbf{\textsf{WALKS}.}
The results for \ANY \SHORTEST \WALK are given in Figure~\ref{fig:pokecsec5} (right). As we can see, \syskuzu has a highly performant algorithm that handles this use case well, while \sysneo times out quickly. \mdb is around 100x faster than all other systems, with stable performance for longer lengths. Again, the performance spike for \mdb around lengths 3 and 4 happens because much data is still being loaded onto disk during these first experiments. This data stays in the buffer for longer paths, which causes the run-time to stabilize.
The case of \ALL \SHORTEST \WALK is presented in Figure~\ref{fig:pokecintro} (right). The picture here is similar. \mdb scales very well, while no other system is able to handle paths of length 6 or more within the 1-minute timeout. The data loading spike is again present for length 3 and 4, but it still results in fast performance. Whereas the present experiment 
aims at returning the paths of length at most $k$ (for $k = 1 , \ldots,12$), we also conducted an experiment where we look for paths of length precisely $k$. Since the latter experiment generated virtually identical plots, we omit it. 
Overall, we can conclude that \mdb offers stable performance, and unlike the other systems, does not get into issues as the path lengths increaase.
\sysnebula is not used in this experiment since it does not support the \WALK semantics. 

\medskip
\noindent \textbf{\textsf{TRAILS}.} The results for \ALL \TRAIL are shown in Figure~\ref{fig:pokecintro} (left). The performance of \sysneo  is significantly better here than for \SHORTEST \WALK, with timeouts occurring later. \sysnebula started timing out for paths of length 5. For \mdb, we see the performance of \mdb of the BFS version of Algorithm~\ref{alg:all}. For the DFS version (not shown in the figure) the picture is similar.
As in other experiments, we see that \mdb handles the query load with no major issues. \textcolor{black}{It is about 100x faster than all other systems.}

\subsection{Wikidata: The Effect of Big Graphs}
\label{ss:pokes}
\begin{figure}
    \centering
    \resizebox{0.7\columnwidth}{!}{
    \includegraphics{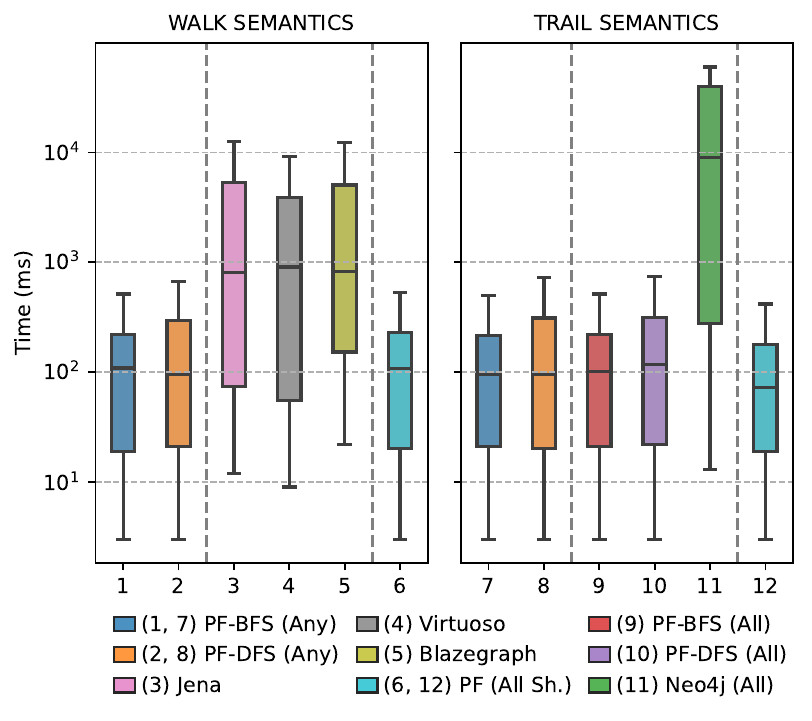}
    }
    \vspace{-10pt}
    \captionsetup{justification=centering}
    \caption{Runtimes for the Wikidata experiment.}
    \label{fig:wikidata}
\end{figure}

In this experiment, we test whether returning paths is feasible in big real-world graphs. We encountered significant issues when loading the data set into some engines. For \sysnebula we ran into the well documented storage issue~\cite{NebulaIssue} that we could not resolve, 
 while \syskuzu ran out of memory while loading the data set. We even tried splitting the data set into smaller chunks, with one file for each distinct edge label. In this case we only managed to load the ten biggest edge sets into \syskuzu, but this amounts to less than a third of the total number of edges, so we excluded \syskuzu from this experiment. 
The systems that could load the data were \mdb, \sysneo, \sysblaze, \sysjena, and \sysvirtuoso. Given that \sysneo only supports the \WALK and \TRAIL restrictors by default, we ran our queries under these two path modes. Out of 659 queries, 20 could not be expressed in \sysneo since they were complex RPQs. 
The experiment results are in Figure~\ref{fig:wikidata}. 

\medskip
\noindent \textbf{\textsf{WALKS}.} In Figure~\ref{fig:wikidata} (left) we show the results for the \WALK restrictor. Since we run 659 different queries, we present box plots of our results. The first two columns represent the BFS and DFS version of Algorithm~\ref{alg:any} in \mdb, which corresponds to \ANY (\SHORTEST) \WALK mode. \textcolor{black}{It is around 10x faster than the SPARQL engines (the next three columns). We find this difference remarkable, because the SPARQL engines only return endpoint pairs, whereas \mdb returns more information: it additionally returns a witnessing path for each such endpoint pair.} 
The number of timeouts we observed tell a similar story. Here both \mdbbfs and \mdbdfs had 12 timeouts. In contrast, \sysjena timed out 95 times, and \sysblaze and \sysvirtuoso 86 and 24 times, respectively. We remark that while \sysneo supports \ANY \SHORTEST \WALK mode, it timed out in 657 out of 659 queries, so we did not include it in the graphs. 

The rightmost column in Figure~\ref{fig:wikidata} (left) shows the performance of \mdb for the \ALL \SHORTEST mode of Algorithm~\ref{alg:allshortest}. Again, surprisingly, despite having to fulfil a more complex task (returning all matching endpoint pairs, together with \emph{all} shortest witnessing paths), finding 100,000 paths under this mode shows almost identical performance to finding a single shortest path for each reached node. The number of timeouts for \mdb was 10. \textcolor{black}{This number is lower than for the \ANY \SHORTEST \WALK mode because we only search for 100,000 paths. Under \ALL \SHORTEST, we do not need to find as many node pairs to find this amount of paths.} Again, while \sysneo does support this path mode, it could only complete 2 out of 659 queries. \textcolor{black}{Overall, in this experiment, 
\begin{itemize}
\item returning witnessing walks is feasible, even on big graphs and
\item \mdb presents a stable strategy for finding such walks, even outperforming systems that do not return witnessing walks.
\end{itemize}
}

\medskip
\noindent \textbf{\textsf{TRAILS}.} The results for the \TRAIL semantics are shown in Figure~\ref{fig:wikidata} (right). The first two columns correspond to \ANY \SHORTEST \TRAIL and \ANY \TRAIL in \mdb. This performance is almost identical to the \ANY \WALK case, with only 26 and 27 timeouts for the BFS and DFS versions, respectively. 
The next three columns correspond to the \ALL \TRAIL mode, which is only supported by \mdbbfs, \mdbdfs, and \sysneo. \textcolor{black}{Here we compare only with \sysneo since it is the only engine that could load the dataset and supports the \TRAIL semantics.}  
\mdb performs around 10x faster than \sysneo. This is reflected in the number of timeouts with 134 for \sysneo, and only 11 for \mdbbfs and 13 for \mdbdfs. 
The rightmost column corresponds to \ALL \SHORTEST \TRAIL mode in \mdb, which again shows similar performance to other \TRAIL-based modes, with only 21 timeouts. 

Overall, we see that \mdb shows remarkably stable performance when returning trails. Interestingly, while the theoretical literature classifies the \TRAIL mode as intractable~\cite{MartensNT-stacs20,MartensT-tods19}, and algorithms proposed in Section~\ref{sec:simple} take a brute-force approach to solving them, over real-world data they do not seem to fare significantly worse than algorithms for the \WALK restrictor. This is most likely due to the fact that they can either detect 100,000 results rather fast, or because the theoretical hardness requires a certain kind of interaction between the query and the data that does not occur in this experiment. 

We remark that we also ran the experiments for \SIMPLE and \ACYCLIC restrictors in \mdb, with identical results as in the \TRAIL case, showing that Algorithm~\ref{alg:all} is indeed a good option for real-world use cases.

\subsection{Diamond: Scaling the Number of Paths}
\label{ss:diamond}
In this experiment, we test the performance of the query looking for paths between the node \textsf{start} and the node \textsf{end} in the graph of Figure~\ref{fig:diamond}. We scale the size of the database by setting $n=1,\ldots ,40$. This allows us to test how the algorithms perform when the number of paths is large, i.e.,  $2^n$. For each value of $n$ we will look for the first 100,000 results. To compare with other engines, we focus on the \WALK restrictor and the \TRAIL restrictor. All the other paths modes in \mdb, which is the only system supporting them, have identical performance as in the \TRAIL case, since they are all derivatives of Algorithm~\ref{alg:all}. Since SPARQL systems cannot return paths, we exclude them from this experiment. 

\begin{figure}
     \centering
         \includegraphics[width=0.7\columnwidth]{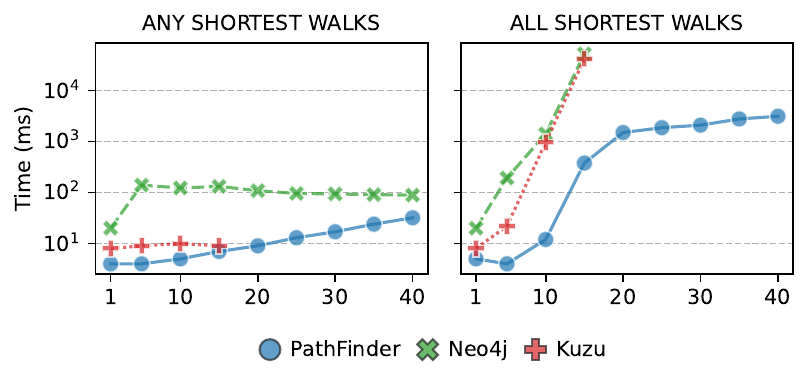} 
     	\vspace{-10pt}
        \caption{\WALK queries on the graph of Figure~\ref{fig:diamond}.}
        \label{fig:walkdiamond}
\end{figure}

\medskip
\noindent \textbf{\textsf{SHORTEST WALKS}.}
The runtimes for the \ANY \SHORTEST \WALK and \ALL \SHORTEST \WALK modes is presented in Figure~\ref{fig:walkdiamond}. Here we compare \mdb, \sysneo, and \syskuzu, since \sysnebula does not support the mode \textcolor{black}{and the SPARQL systems do not return paths}.  Due to the small size of the graph, we run each query twice and report the second result. This is due to minuscule runtimes which get heavily affected by initial data loading. As we can observe, for \ANY \SHORTEST \WALK (Figure~\ref{fig:walkdiamond} (left)) all the engines perform well. We even pushed this experiment to $n=1000$ with no issues for \sysneo and \mdb. \syskuzu only works up to $n=15$ since the longest paths it supports are of length 30. 
In the case of \ALL \SHORTEST \WALK (Figure~\ref{fig:walkdiamond} (right)), \sysneo times out for $n=16$. \syskuzu stops at $n=15$ with a successful execution because of its hard limit of path length 30. \mdb seems to have a linear time curve in this experiment, which is in line with its theoretical principles~\cite{pmr}. The the other engines' run times grow exponentially, showing the full power of Algorithm~\ref{alg:allshortest} when returning 100,000 paths. We scaled to $n=1000$ and the results for \mdb were quite similar. 

\begin{figure}
     \centering
         \includegraphics[width=0.7\columnwidth]{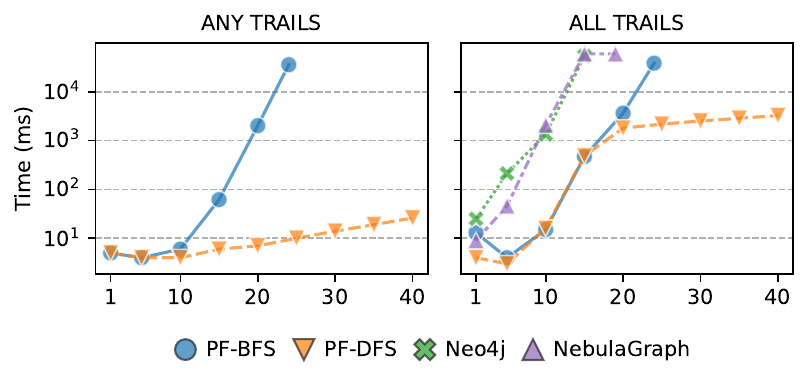}
     	\vspace{-10pt}
        \caption{\TRAIL queries on the graph of Figure~\ref{fig:diamond}.}
        \label{fig:traildiamond}
\end{figure}

\medskip
\noindent \textbf{\textsf{TRAILS}.} In addition to comparing with other systems, this experiment allows to determine which traversal strategy (BFS or DFS) is better suited for Algorithm~\ref{alg:all} in extreme cases such as the graph of Figure~\ref{fig:diamond}. We present the results for \ANY \TRAIL and \TRAIL in Figure~\ref{fig:traildiamond}. Considering first the \ANY \TRAIL case, which is only supported by \mdb, 
the BFS-based algorithm will time out for $n=26$. This is to be expected, since it constructs all paths of length $1,2,\ldots ,25$, before considering the first path of length 26. In contrast, DFS will find the required paths rather quickly. 

Concerning the \TRAIL mode, which retrieves \emph{all} trails, the situation is similar. Here we also compare with other engines that find trails. As we can see, only \mdbdfs could handle the entire query load. All others exhibit an exponential performance curve.  This illustrates that, for a huge number of \emph{trails}, DFS is the strategy of choice. 


\subsection{Top-$k$ Paths and Top-$k$ Groups}
\label{ss:topK}
To test the top-$k$ and top-$k$ groups modes we run all of the query patterns described above, but now returning top-$k$ paths and top-$k$ groups with different values for $k$. In order not to overcrowd the section, we present only a selection of the obtained results.

\begin{figure}
     \centering
         \includegraphics[width=0.7\columnwidth]{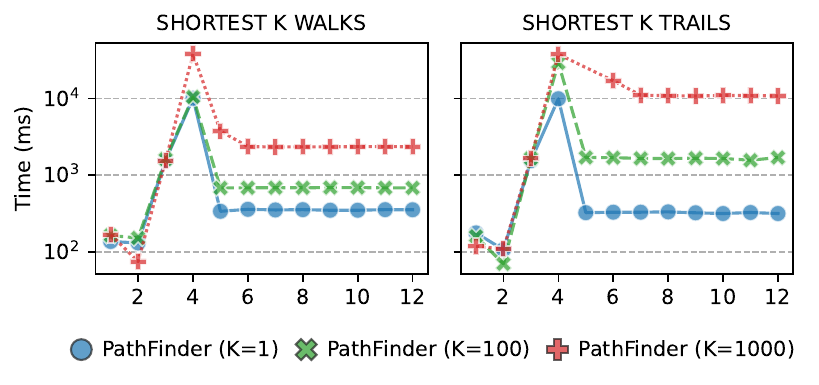} 
     	\vspace{-10pt}
        \caption{Top $k$ walks and trails over the Pokec dataset.}
        \label{fig:topkPokec}
\end{figure}

\medskip
\noindent \textbf{The Pokec Dataset.} We begin by testing what is the cost of retrieving shortest $k$ walks and trails over the Pokec dataset with $k$ equal to 1, 100, and 1000, respectively. The results are presented in Figure~\ref{fig:topkPokec}. The value of $k=1$ is equivalent to returning a single shortest walk or trail and is there for comparison with base algorithms. As expected, requiring more results requires more time, with the \WALK semantics slightly outperforming the \TRAIL semantics. \textcolor{black}{If comparing to \emph{all} shortest walks in Figure~\ref{fig:pokecintro}, we can see that the algorithm is actually slower when many paths are required (100 or more). This is consistent with comparing the code of Algorithm~\ref{alg:allshortest} (all shortest) vs Algorithm~\ref{alg:allshortestkpaths} (shortest $k$). Namely, the latter does significantly more book keeping and is required to expand over longer and longer paths as discussed in Example~\ref{ex:shortestk}. Again, we stress here that \SHORTEST $k$ can contain many more paths than \ALL \SHORTEST. This is because \SHORTEST $k$ contains paths of varying lengths, requiring the search algorithm to continue its operation until $k$ paths are discovered. On the other hand, \SHORTEST only considers shortest paths (of which there is generally only a few). This illustrates that the correct implementation of \SHORTEST $k$ could lead to unintuitive slowdowns, given that it will require further graph exploration.}

\begin{figure}
     \centering
         \includegraphics[width=0.7\columnwidth]{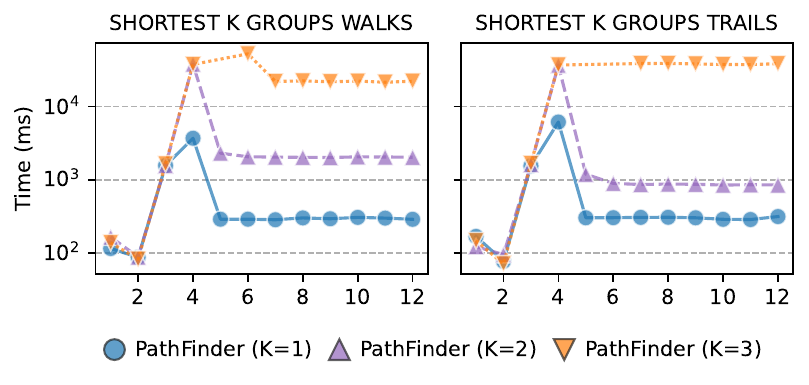} 
     	\vspace{-10pt}
        \caption{Top $k$ groups for walks and trails over the Pokec dataset.}
        \label{fig:topkGroupsPokec}
\end{figure}

The results for shortest $k$ groups of walks and trails is given in Figure~\ref{fig:topkGroupsPokec}. Here we test for values of $k=1,2,3$. The line for $k=1$ is again our baseline experiment which returns \emph{all} shortest walks/trails. As expected, more groups requires more runtime. 

\begin{figure}
     \centering
         \includegraphics[width=0.7\columnwidth]{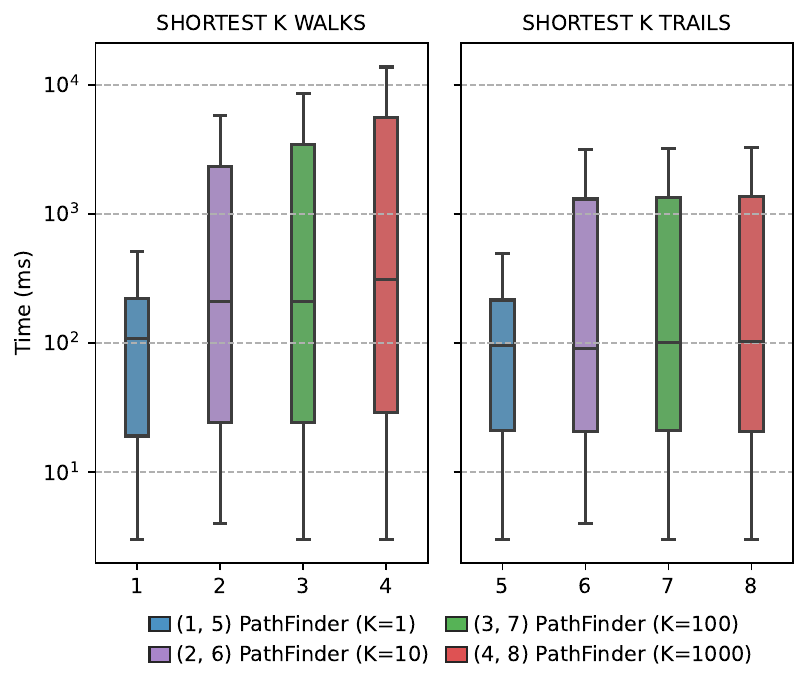} 
     	\vspace{-10pt}
        \caption{Top $k$ walks and trails over the Wikidata dataset.}
        \label{fig:shortestKWiki}
\end{figure}

\medskip
\noindent \textbf{The Wikidata Experiment.} Over Wikidata, we obtain similar results as for Pokec. For shortest $k$ walks and trails, the results are in Figure~\ref{fig:shortestKWiki}. \textcolor{black}{The conclusion here is similar as in the case of the Pokec dataset, however the execution time is much more similar to the \ALL \SHORTEST mode, particularly in the mean, as Wikidata usually contains many more paths than Pokec.} 
Interestingly, it becomes apparent here that the \WALK semantics can be (marginally) slower than the \TRAIL semantics when shortest $k$ paths are required. This is most likely due to the fact that for the \WALK semantics (Section~\ref{ss:shortestk}) one might be required to keep expanding paths between a pair of nodes to a length that is longer than the number of nodes in the graph, which is prohibited for \TRAILS (Section~\ref{ss:shortestKTrails}). 
The results for the group modes are very similar, so we omit them for brevity.

\subsection{Conclusions of the Experiments}
\label{ss:concl}
Based on our experiments, we believe that one can conclude that \mdb offers a sound strategy for dealing with path-returning queries in graphs. It is highly performant on all the query loads we considered, and runs faster than every other system in every scenario we tested, typically with a 10x or 100x increase in speed. This is particularly true for the \WALK semantics, which runs very fast and with few timeouts, even on huge datasets such as Wikidata. When it comes to \TRAIL, one has to be careful selecting breadth-first search (BFS) or depth-first search (DFS). The former is a good candidate for highly connected graphs with few hops, and the latter is better able to handle a huge number of paths. Finally, we remark that \mdb was tested only as a disk-based system that loads data into a main memory buffer as required by the queries. 
Our code~\cite{ANONrepo} also includes an in-memory version of \mdb, which uses the Compressed Sparse Row representation of graphs~\cite{csr} in order to store the data in memory, and which runs about twice as fast as the results we presented (results not included for brevity). 


\section{Related Work}
\label{sec:related}
The topic of returning paths that match regular path queries is fairly new. 
To the best of our knowledge, the complexity of the problem was first formally studied in \cite{MartensT-icdt18}. 
Most existing work focuses on finding node pairs connected by a path that conforms to the regular path query~\cite{CalvaneseGLV00,CalvaneseGLV02,YakovetsGG16,Gubichev15,FiondaPG15,BaganBG20,MartensNT-stacs20,CaselS-icdt21,FigueiraGKMNT20}. The most notable exceptions are Eppstein's algorithm for returning the $k$ shortest walks~\cite{Eppstein98} and Yen's algorithm~\cite{yen}, which returns $k$ shortest simple paths. None of these match these paths against a regular path query. Extensions of Yen's algorithm that simultaneously check if the paths match a regular path query have been developed in \cite{MartensT-tods19,MartensNP23}, both for simple paths and for trails. The corresponding problems for undirected paths were studied in \cite{MartensP22}. Returning shortest paths that match a given RPQ was studied in~\cite{MartensT-tods19}, but not specifically using Eppstein's data structure.

A closely related problem to ours is deciding if there exists a simple path or trail from a given source to target node for a regular path query. If the regular path query is part of the input, the problem is trivially NP-hard because it generalizes the Hamiltonian Path problem, but a closer analysis reveals fixed-parameter tractability in many real-world cases~\cite{MartensT-tods19}. For fixed regular path queries $r$, several studies aimed at understanding which regular expressions $r$ make the problem hard or polynomial-time solvable. In these problems, one is only given as input the graph $G$, a source node $s$, a target node $t$, and the question is if there exists a simple path or trail from $s$ to $t$ that matches $r$. In this context, Bagan et al.~\cite{BaganBG20} provided a complete trichotomy for simple paths, whereas Martens et al.~\cite{MartensNP23} provided one for trails. For undirected paths, pinning down for which regular path queries the problem is easy versus hard is much more challenging~\cite{MartensP22}. Resolving this problem required solving the 30-year open problem of deciding if an undirected graph has a simple path of length zero modulo $k$~\cite{ArkinPY91}, which was recently shown to be in polynomial time~\cite{LiuY25}. A complete classification in the style of \cite{BaganBG20,MartensNP23} is still open.

On the systems side, Gubichev et al.~\cite{GubichevN11} focused on returning paths in graphs, but not according to an RPQ pattern, and \cite{SavenkovMUP17} uses a BFS-style exploration to find the first $k$ paths, which means that some non-shortest paths will be returned. A similar approach for top-$k$ results is presented in~\cite{katjaJedi}, but not preferring shortest paths. Furthermore, we should mention that Neo4j's system implementing Cypher has been returning paths since its early release in 2010--2011. These path returning capabilities focused on returning trails and were implemented for a restricted class of regular path queries, for example, disallowing concatenation under Kleene star.

Closest to our work is~\cite{pmr}, where a compact representation of RPQ-conforming paths (called a path multiset representation or PMR) is presented. In a nutshell, a PMR consists of a graph $R$, a homomorphism $h$ from $R$ to the graph database $G$, and sets of start and end nodes. PMRs can represent the resulting paths for all GQL path modes. Whereas \cite{pmr} only studied how to compute the PMRs for variants of the \SHORTEST mode, the present work supports \emph{all} GQL path modes and presents implementable algorithms. To make this more concrete, the \textit{Visited} structure of our algorithms in fact encodes a PMR of~\cite{pmr} for any GQL path mode. 

Additional important approaches that share similarity with our algorithmic toolbox are~\cite{FiondaPG15}, where a graph crawl based on the product graph construction is also exploited, and~\cite{HofmanM15,Pirro20}, where frameworks for proposing witnesses to an RPQ answer (or the lack thereof) is explored. Finally, we mention that~\cite{DavidFM24} studied an equivalent of Algorithm~\ref{alg:allshortest} which removes the restriction of unambiguity on the automaton used to find all shortest walks between a fixed pair of nodes. 

\section{Conclusions}
\label{sec:concl}
We present \mdb, a unifying framework for returning paths in answers to regular path queries (RPQs). To the best of our knowlede, \mdb is the first system that allows returning paths under \emph{every} mode prescribed by the GQL and SQL/PGQ query standards~\cite{GQL}. Our experimental evaluation shows the approach to be highly competitive on realistic workloads, \textcolor{black}{outperforming other engines by one or two orders of magnitude}.  While our work was developed in the context of property graphs, it is straightforward to implement it on top of an existing SPARQL engine (see~\cite{ANONrepo} for an example). 
%

We showed that returning paths that match RPQs can be practically viable, which opens the question to which extent we want to explore similar functionality for SPARQL~\cite{HarrisS13}.  While our initial results do show that supporting path retrieval in SPARQL is feasible, it remains to be studied how this functionality mixes with the property path syntax and semantics (which differs significantly from RPQs) and how such an extension could be incorporated in the SPARQL standard. In the future, we plan to look at more expressive GQL path queries that take into consideration the data residing in nodes and edges and try to develop a toolbox of algorithms to handle these cases. 

 \begin{acks}
 Far\'ias, Rojas and Vrgo\v{c} were supported by ANID -- Millennium Science Initiative Program -- Code ICN17\_002. Vrgo\v{c} was also supported by the ANID Fondecyt Regular project 1240346. Martens was supported by ANR project EQUUS ANR-19-CE48-0019; funded by the Deutsche
    Forschungsgemeinschaft (DFG, German Research Foundation), project number
    431183758.
    \end{acks}

\bibliographystyle{ios1}           
\bibliography{nourlbiblio}        

@inproceedings{LosemannM14,
  author       = {Katja Losemann and
                  Wim Martens},
  title        = {{MSO} queries on trees: enumerating answers under updates},
  booktitle    = {Symposium on Logic in Computer Science (LICS)},
  pages        = {67:1--67:10},
  publisher    = {{ACM}},
  year         = {2014},
  doi          = {10.1145/2603088.2603137}
}

@techreport{laundering,
    author = {Darryl Salas},
    title = {How to Combat Money Laundering Using Graph Technology},
    institution = {Neo4j},
    year = 2021,
    note = {\url{https://neo4j.com/whitepapers/anti-money-laundering-framework-solution-guide/}}
}

@inproceedings{PODSgems,
    author = {Leonid Libkin and Wim Martens and Filip Murlak and Liat Peterfreund and Domagoj Vrgo\v{c}},
    title = {Querying Graph Data: Where We Are and Where To Go},
    booktitle = {Symposium on Principles of Database Systems (PODS)},
    year = {2025} 
}

@inproceedings{cypher,
  author       = {Nadime Francis and
                  Alastair Green and
                  Paolo Guagliardo and
                  Leonid Libkin and
                  Tobias Lindaaker and
                  Victor Marsault and
                  Stefan Plantikow and
                  Mats Rydberg and
                  Petra Selmer and
                  Andr{\'{e}}s Taylor},
  title        = {Cypher: An Evolving Query Language for Property Graphs},
  booktitle    = {International Conference on Management of
                  Data (SIGMOD)},
  pages        = {1433--1445},
  publisher    = {{ACM}},
  year         = {2018},
  url          = {https://doi.org/10.1145/3183713.3190657},
  doi          = {10.1145/3183713.3190657},
  bibsource    = {dblp computer science bibliography, https://dblp.org}
}

@book{HopcroftUllman,
    author = {John Hopcroft and Jeffrey Ullman},
    title = {Introduction to Automata Theory, Languages, and Computation},
    publisher = {Addison-Wesley},
    year = {1979}
}

@inproceedings{MartensNT-stacs20,
  author       = {Wim Martens and
                  Matthias Niewerth and
                  Tina Trautner},
  title        = {A Trichotomy for Regular Trail Queries},
  booktitle    = {International Symposium on Theoretical Aspects of Computer Science (STACS)},
  doi          = {10.4230/LIPICS.STACS.2020.7},
  pages        = {7:1--7:16},
  year         = {2020}
}

@inproceedings{CaselS-icdt21,
  author       = {Katrin Casel and
                  Markus L. Schmid},
  title        = {Fine-Grained Complexity of Regular Path Queries},
  booktitle    = {International Conference on Database Theory (ICDT)},
  doi          = {10.4230/LIPICS.ICDT.2021.19},
pages        = {19:1--19:20},
  year         = {2021}
}

@article{ArkinPY91,
  author       = {Esther M. Arkin and
                  Christos H. Papadimitriou and
                  Mihalis Yannakakis},
  title        = {Modularity of Cycles and Paths in Graphs},
  journal      = {J. {ACM}},
  volume       = {38},
  number       = {2},
  pages        = {255--274},
  year         = {1991},
  url          = {https://doi.org/10.1145/103516.103517},
  doi          = {10.1145/103516.103517},
  bibsource    = {dblp computer science bibliography, https://dblp.org}
}

@article{BaganBG20,
  author       = {Guillaume Bagan and
                  Angela Bonifati and
                  Beno{\^{\i}}t Groz},
  title        = {A trichotomy for regular simple path queries on graphs},
  journal      = {J. Comput. Syst. Sci.},
  volume       = {108},
  pages        = {29--48},
  year         = {2020},
  url          = {https://doi.org/10.1016/j.jcss.2019.08.006},
  doi          = {10.1016/J.JCSS.2019.08.006},
  bibsource    = {dblp computer science bibliography, https://dblp.org}
}

@article{MartensNP23,
  author       = {Wim Martens and
                  Matthias Niewerth and
                  Tina Popp},
  title        = {A Trichotomy for Regular Trail Queries},
  journal      = {Log. Methods Comput. Sci.},
  volume       = {19},
  number       = {4},
  year         = {2023},
  url          = {https://doi.org/10.46298/lmcs-19(4:20)2023},
  doi          = {10.46298/LMCS-19(4:20)2023},
  bibsource    = {dblp computer science bibliography, https://dblp.org}
}

@inproceedings{LiuY25,
  author       = {Chun{-}Hung Liu and
                  Youngho Yoo},
  editor       = {Michal Kouck{\'{y}} and
                  Nikhil Bansal},
  title        = {Disjoint Paths Problem with Group-Expressable Constraints},
  booktitle    = {Proceedings of the 57th Annual {ACM} Symposium on Theory of Computing,
                  {STOC} 2025, Prague, Czechia, June 23-27, 2025},
  pages        = {1933--1943},
  publisher    = {{ACM}},
  year         = {2025},
  url          = {https://doi.org/10.1145/3717823.3718109},
  doi          = {10.1145/3717823.3718109},
  bibsource    = {dblp computer science bibliography, https://dblp.org}
}

@inproceedings{MartensT-icdt18,
  author       = {Wim Martens and
                  Tina Trautner},
  title        = {Evaluation and Enumeration Problems for Regular Path Queries},
  doi          = {10.4230/LIPICS.ICDT.2018.19},
booktitle    = {International Conference on Database Theory (ICDT)},
  pages        = {19:1--19:21},
  year         = {2018}
}

@article{MartensT-sigmodrecord19,
  author       = {Wim Martens and
                  Tina Trautner},
  title        = {Bridging Theory and Practice with Query Log Analysis},
  journal      = {{SIGMOD} Rec.},
  volume       = {48},
  number       = {1},
  pages        = {6--13},
  year         = {2019},
  doi          = {10.1145/3371316.3371319},
  bibsource    = {dblp computer science bibliography, https://dblp.org}
}

@article{MartensT-tods19,
  author       = {Wim Martens and
                  Tina Trautner},
  title        = {Dichotomies for Evaluating Simple Regular Path Queries},
  journal      = {{ACM} Trans. Database Syst.},
  volume       = {44},
  number       = {4},
  pages        = {16:1--16:46},
  year         = {2019},
  doi          = {10.1145/3331446},
  bibsource    = {dblp computer science bibliography, https://dblp.org}
}

@inproceedings{BonifatiMT-www19,
  author       = {Angela Bonifati and
                  Wim Martens and
                  Thomas Timm},
  title        = {Navigating the Maze of Wikidata Query Logs},
  booktitle    = {The World Wide Web Conference (WWW)},
  pages        = {127--138},
  publisher    = {{ACM}},
  year         = {2019},
  doi          = {10.1145/3308558.3313472},
  bibsource    = {dblp computer science bibliography, https://dblp.org}
}

@inproceedings{AnglesABBFGLPPS18,
  author    = {Renzo Angles and
               Marcelo Arenas and
               Pablo Barcel{\'{o}} and
               Peter A. Boncz and
               George H. L. Fletcher and
               Claudio Guti{\'{e}}rrez and
               Tobias Lindaaker and
               Marcus Paradies and
               Stefan Plantikow and
               Juan F. Sequeda and
               Oskar van Rest and
               Hannes Voigt},
  title     = {{G-CORE:} {A} Core for Future Graph Query Languages},
  booktitle = {International Conference on Management of Data (SIGMOD)},
  doi          = {10.1145/3183713.3190654},
  year      = {2018}
}

@misc{millenniumDB,
  author    = {MillenniumDB Team},
  title     = {{MillenniumDB Source Code}},
  url       = {https://github.com/MillenniumDB/MillenniumDB},
  year      = {2021}
}

@article{mdb,
  author       = {Domagoj Vrgoc and
                  Carlos Rojas and
                  Renzo Angles and
                  Marcelo Arenas and
                  Diego Arroyuelo and
                  Carlos Buil{-}Aranda and
                  Aidan Hogan and
                  Gonzalo Navarro and
                  Cristian Riveros and
                  Juan Romero},
  title        = {MillenniumDB: An Open-Source Graph Database System},
  journal      = {Data Intell.},
  volume       = {5},
  number       = {3},
  pages        = {560--610},
  year         = {2023},
  url          = {https://doi.org/10.1162/dint\_a\_00229}
}

@inproceedings{FariasMRV24,
  author       = {Benjam{\'{\i}}n Farias and
                  Wim Martens and
                  Carlos Rojas and
                  Domagoj Vrgoc},
  editor       = {Gianluca Demartini and
                  Katja Hose and
                  Maribel Acosta and
                  Matteo Palmonari and
                  Gong Cheng and
                  Hala Skaf{-}Molli and
                  Nicolas Ferranti and
                  Daniel Hern{\'{a}}ndez and
                  Aidan Hogan},
  title        = {PathFinder: Returning Paths in Graph Queries},
  booktitle    = {The Semantic Web - {ISWC} 2024 - 23rd International Semantic Web Conference,
                  Baltimore, MD, USA, November 11-15, 2024, Proceedings, Part {II}},
  series       = {Lecture Notes in Computer Science},
  volume       = {15232},
  pages        = {135--154},
  publisher    = {Springer},
  year         = {2024},
  url          = {https://doi.org/10.1007/978-3-031-77850-6\_8},
  doi          = {10.1007/978-3-031-77850-6\_8},
  bibsource    = {dblp computer science bibliography, https://dblp.org}
}

@article{AnglesABHRV17,
  author    = {Renzo Angles and
               Marcelo Arenas and
               Pablo Barcel{\'{o}} and
               Aidan Hogan and
               Juan L. Reutter and
               Domagoj Vrgo\v{c}},
  title     = {{Foundations of Modern Query Languages for Graph Databases}},
  journal   = {{ACM} Comput. Surv.},
  volume    = {50},
  number    = {5},
  year      = {2017},
  doi       = {10.1145/3104031},
  bibsource = {dblp computer science bibliography, https://dblp.org}
}

@inproceedings{HofmanM15,
  author       = {Piotr Hofman and
                  Wim Martens},
  editor       = {Marcelo Arenas and
                  Mart{\'{\i}}n Ugarte},
  title        = {Separability by Short Subsequences and Subwords},
  booktitle    = {18th International Conference on Database Theory, {ICDT} 2015, Brussels,
                  Belgium, March 23-27, 2015},
  series       = {LIPIcs},
  pages        = {230--246},
  publisher    = {Schloss Dagstuhl - Leibniz-Zentrum f{\"{u}}r Informatik},
  year         = {2015},
  url          = {https://doi.org/10.4230/LIPIcs.ICDT.2015.230},
  doi          = {10.4230/LIPICS.ICDT.2015.230},
  bibsource    = {dblp computer science bibliography, https://dblp.org}
}

@inproceedings{Baeza13,
  author    = {Pablo Barcel{\'{o}} Baeza},
  title     = {Querying graph databases},
  booktitle = {Symposium on Principles of Database Systems (PODS)},
  pages     = {175--188},
  year      = {2013},
  !crossref  = {DBLP:conf/pods/2013},
  doi       = {10.1145/2463664.2465216},
  timestamp = {Tue, 06 Nov 2018 16:58:02 +0100},
  biburl    = {https://dblp.org/rec/conf/pods/Baeza13.bib},
  bibsource = {dblp computer science bibliography, https://dblp.org}
}

@inproceedings{CruzMW87,
  author    = {Isabel F. Cruz and
               Alberto O. Mendelzon and
               Peter T. Wood},
  title     = {A Graphical Query Language Supporting Recursion},
  booktitle = {International Conference on Management of Data (SIGMOD)},
doi          = {10.1145/38713.38749},
pages     = {323--330},
  year      = {1987}
}

@inproceedings{KimelfeldMN25,
  author       = {Benny Kimelfeld and
                  Wim Martens and
                  Matthias Niewerth},
  title        = {A Formal Language Perspective on Factorized Representations},
  booktitle    = {International Conference on Database Theory {(ICDT)}},
  series       = {LIPIcs},
  pages        = {20:1--20:20},
  publisher    = {Schloss Dagstuhl - Leibniz-Zentrum f{\"{u}}r Informatik},
  year         = {2025},
  hideurl          = {https://doi.org/10.4230/LIPIcs.ICDT.2025.20},
  doi          = {10.4230/LIPICS.ICDT.2025.20}
}

@article{Leung05,
  author       = {Hing Leung},
  title        = {Descriptional complexity of {NFA} of different ambiguity},
  journal      = {Int. J. Found. Comput. Sci.},
  volume       = {16},
  number       = {5},
  pages        = {975--984},
  year         = {2005},
  hideurl          = {https://doi.org/10.1142/S0129054105003418},
  doi          = {10.1142/S0129054105003418}
}

@inproceedings{HammererM25,
  author       = {Janik Hammerer and
                  Wim Martens},
  title        = {A Compendium of Regular Expression Shapes in {SPARQL} Queries},
  booktitle    = {Graph Data Management Experiences {\&} Systems {(GRADES)} and Network Data Analytics (NDA)},
  pages        = {4:1--4:10},
  publisher    = {{ACM}},
  year         = {2025},
  hideurl          = {https://doi.org/10.1145/3735546.3735853},
  doi          = {10.1145/3735546.3735853}
}

@article{Erling12,
  author    = {Orri Erling},
  title     = {{Virtuoso, a Hybrid RDBMS/Graph Column Store}},
  journal   = {{IEEE} Data Eng. Bull.},
  volume    = {35},
  number    = {1},
  pages     = {3--8},
  year      = {2012},
  url       = {http://sites.computer.org/debull/A12mar/vicol.pdf},
  bibsource = {dblp computer science bibliography, https://dblp.org}
}

@inproceedings{FrancisGGLLMPRS18,
  author    = {Nadime Francis and
               Alastair Green and
               Paolo Guagliardo and
               Leonid Libkin and
               Tobias Lindaaker and
               Victor Marsault and
               Stefan Plantikow and
               Mats Rydberg and
               Petra Selmer and
               Andr{\'{e}}s Taylor},
  title     = {{Cypher: An Evolving Query Language for Property Graphs}},
  booktitle = {International Conference on Management of Data (SIGMOD)},
  year      = {2018},
  !url       = {https://doi.org/10.1145/3183713.3190657},
  doi       = {10.1145/3183713.3190657},
  timestamp = {Sat, 19 Oct 2019 20:14:55 +0200},
  biburl    = {https://dblp.org/rec/conf/sigmod/FrancisGGLLMPRS18.bib},
  bibsource = {dblp computer science bibliography, https://dblp.org}
}

@article{BonifatiMT20,
  author    = {Angela Bonifati and
               Wim Martens and
               Thomas Timm},
  title     = {{An analytical study of large SPARQL query logs}},
  journal   = {{VLDB} J.},
  volume    = {29},
  number    = {2-3},
  pages     = {655--679},
  year      = {2020},
  doi       = {10.1007/s00778-019-00558-9},
  bibsource = {dblp computer science bibliography, https://dblp.org}
}

@inproceedings{MartensP22,
  author       = {Wim Martens and
                  Tina Popp},
  title        = {The Complexity of Regular Trail and Simple Path Queries on Undirected
                  Graphs},
  booktitle    = {Principles of Database Systems (PODS)},
  pages        = {165--174},
  publisher    = {{ACM}},
  year         = {2022},
  url          = {https://doi.org/10.1145/3517804.3524149},
  doi          = {10.1145/3517804.3524149},
  bibsource    = {dblp computer science bibliography, https://dblp.org}
}

@inproceedings{FigueiraGKMNT20,
  author       = {Diego Figueira and
                  Adwait Godbole and
                  S. Krishna and
                  Wim Martens and
                  Matthias Niewerth and
                  Tina Trautner},
  editor       = {Diego Calvanese and
                  Esra Erdem and
                  Michael Thielscher},
  title        = {Containment of Simple Conjunctive Regular Path Queries},
  booktitle    = {Proceedings of the 17th International Conference on Principles of
                  Knowledge Representation and Reasoning, {KR} 2020, Rhodes, Greece,
                  September 12-18, 2020},
  pages        = {371--380},
  year         = {2020},
  url          = {https://doi.org/10.24963/kr.2020/38},
  doi          = {10.24963/KR.2020/38},
  bibsource    = {dblp computer science bibliography, https://dblp.org}
}

@inproceedings{MalyshevKGGB18,
  author    = {Stanislav Malyshev and
               Markus Kr{\"{o}}tzsch and
               Larry Gonz{\'{a}}lez and
               Julius Gonsior and
               Adrian Bielefeldt},
  title     = {{Getting the Most Out of Wikidata: Semantic Technology Usage in Wikipedia's
               Knowledge Graph}},
  booktitle = {International Semantic Web Conference (ISWC)},
  year      = {2018}
}

@misc{HarrisS13,
  author    = {Steve Harris and Andy Seaborne and Eric Prud'hommeaux},
  title     = {{SPARQL 1.1 Query Language}},
  url       = {https://www.w3.org/TR/sparql11-query/},
  year      = {2013},
  howpublished = {W3C Recommendation}
}

@article{HoganBC21,
  author       = {Aidan Hogan and
                  Eva Blomqvist and
                  Michael Cochez and
                  Claudia d'Amato and
                  Gerard de Melo and
                  Claudio Gutierrez and
                  Sabrina Kirrane and
                  Jos{\'{e}} Emilio Labra Gayo and
                  Roberto Navigli and
                  Sebastian Neumaier and
                  Axel{-}Cyrille Ngonga Ngomo and
                  Axel Polleres and
                  Sabbir M. Rashid and
                  Anisa Rula and
                  Lukas Schmelzeisen and
                  Juan F. Sequeda and
                  Steffen Staab and
                  Antoine Zimmermann},
  title        = {Knowledge Graphs},
  journal      = {{ACM} Comput. Surv.},
  volume       = {54},
  number       = {4},
  pages        = {71:1--71:37},
  year         = {2022},
  !url          = {https://doi.org/10.1145/3447772},
  doi          = {10.1145/3447772},
  timestamp    = {Wed, 23 Nov 2022 16:16:44 +0100},
  biburl       = {https://dblp.org/rec/journals/csur/HoganBCdMGKGNNN21.bib},
  bibsource    = {dblp computer science bibliography, https://dblp.org}
}

@inproceedings{MendelzonW89,
  author    = {Alberto O. Mendelzon and
               Peter T. Wood},
  title     = {Finding Regular Simple Paths in Graph Databases},
  booktitle = {Very Large Data Bases},
url          = {http://www.vldb.org/conf/1989/P185.PDF},
pages     = {185--193},
  year      = {1989}
}

@misc{JenaTDB,
  author    = {Jena Team},
  title     = {{Jena TDB}},
  url       = {https://jena.apache.org/documentation/tdb/},
  year      = {2021}
}

@misc{nebula,
  author    = {{Vesoft Inc/Nebula}},
  title     = {{NebulaGraph}},
  url       = {https://www.nebula-graph.io/},
  year      = {2023}
}

@misc{memgraph,
  author    = {Memgraph Team},
  title     = {{Memgraph}},
  url       = {https://memgraph.com/},
  year      = {2023}
}

@article{PerezAG09,
  author    = {Jorge P{\'{e}}rez and
               Marcelo Arenas and
               Claudio Guti{\'{e}}rrez},
  title     = {Semantics and complexity of {SPARQL}},
  journal   = {{ACM} Trans. Database Syst.},
  volume    = {34},
  number    = {3},
  pages     = {16:1--16:45},
  year      = {2009}
}

@book{ramakrishnan00,
  title={Database management systems},
  author={Ramakrishnan, Raghu and Gehrke, Johannes},
  year={2000},
  publisher={McGraw-Hill}
}

@incollection{ThompsonPC14,
  author    = {Bryan B. Thompson and
               Mike Personick and
               Martyn Cutcher},
  editor    = {Andreas Harth and
               Katja Hose and
               Ralf Schenkel},
  title     = {{The Bigdata{\textregistered} {RDF} Graph Database}},
  booktitle = {Linked Data Management},
  pages     = {193--237},
  publisher = {Chapman and Hall/CRC},
  year      = {2014},
  url       = {http://www.crcnetbase.com/doi/abs/10.1201/b16859-12},
  bibsource = {dblp computer science bibliography, https://dblp.org}
}

@misc{TigerGraph,
  author    = {TigerGraph Team},
  title     = {{TigerGraph Documentation -- version 3.1}},
  url       = {https://docs.tigergraph.com/},
  year      = {2021}
}

@article{VrandecicK14,
  author    = {Denny Vrandecic and
               Markus Kr{\"{o}}tzsch},
  title     = {Wikidata: a free collaborative knowledgebase},
  journal   = {Commun. {ACM}},
  volume    = {57},
  number    = {10},
  pages     = {78--85},
  year      = {2014},
  !url       = {https://doi.org/10.1145/2629489},
  doi       = {10.1145/2629489},
  timestamp = {Wed, 14 Nov 2018 10:22:37 +0100},
  biburl    = {https://dblp.org/rec/journals/cacm/VrandecicK14.bib},
  bibsource = {dblp computer science bibliography, https://dblp.org}
}

@inproceedings{Webber12,
  author    = {Jim Webber},
  !editor    = {Gary T. Leavens},
  title     = {{A programmatic introduction to Neo4j}},
  booktitle = {{SPLASH}},
  year      = {2012},
  !url       = {https://doi.org/10.1145/2384716.2384777},
  doi       = {10.1145/2384716.2384777},
  timestamp = {Tue, 06 Nov 2018 16:57:15 +0100},
  biburl    = {https://dblp.org/rec/conf/oopsla/Webber12.bib},
  bibsource = {dblp computer science bibliography, https://dblp.org}
}

@misc{WikiDataData,
editor = {The Wikimedia Foundation},
key = {Wikidata},
title = {Wikidata:Database download},
year = {2021},
url = {https://www.wikidata.org/wiki/Wikidata:Database_download}
}

@article{FiondaPG15,
  author    = {Valeria Fionda and
               Giuseppe Pirr{\`{o}} and
               Claudio Guti{\'{e}}rrez},
  title     = {NautiLOD: {A} Formal Language for the Web of Data Graph},
  journal   = {{ACM} Trans. Web},
  volume    = {9},
  number    = {1},
  pages     = {5:1--5:43},
  doi          = {10.1145/2697393},
year      = {2015}
}

@inproceedings{GQL,
  author    = {Alin Deutsch and
               Nadime Francis and
               Alastair Green and
               Keith Hare and
               Bei Li and
               Leonid Libkin and
               Tobias Lindaaker and
               Victor Marsault and
               Wim Martens and
               Jan Michels and
               Filip Murlak and
               Stefan Plantikow and
               Petra Selmer and
               Oskar van Rest and
               Hannes Voigt and
               Domagoj Vrgo\v{c} and
               Mingxi Wu and
               Fred Zemke},
  title     = {Graph Pattern Matching in {GQL} and {SQL/PGQ}},
  booktitle = {International Conference on Management of Data (SIGMOD)},
  year      = {2022},
  !url       = {https://doi.org/10.1145/3514221.3526057},
  doi       = {10.1145/3514221.3526057},
  timestamp = {Tue, 14 Jun 2022 18:31:24 +0200},
  biburl    = {https://dblp.org/rec/conf/sigmod/DeutschFGHLLLMM22.bib},
  bibsource = {dblp computer science bibliography, https://dblp.org}
}

@article{ReutterSV21,
  author    = {Juan L. Reutter and
               Adri{\'{a}}n Soto and
               Domagoj Vrgo\v{c}},
  title     = {Recursion in {SPARQL}},
  doi          = {10.3233/SW-200401},
  journal   = {Semantic Web},
  volume    = {12},
  number    = {5},
  pages     = {711--740},
  year      = {2021}
}

@article{CalvaneseGLV02,
  author    = {Diego Calvanese and
               Giuseppe De Giacomo and
               Maurizio Lenzerini and
               Moshe Y. Vardi},
  title     = {Rewriting of Regular Expressions and Regular Path Queries},
  journal   = {J. Comput. Syst. Sci.},
doi          = {10.1006/JCSS.2001.1805},
volume    = {64},
  number    = {3},
  pages     = {443--465},
  year      = {2002}
}

@inproceedings{CalvaneseGLV00,
  author       = {Diego Calvanese and
                  Giuseppe De Giacomo and
                  Maurizio Lenzerini and
                  Moshe Y. Vardi},
  editor       = {Anthony G. Cohn and
                  Fausto Giunchiglia and
                  Bart Selman},
  title        = {Containment of Conjunctive Regular Path Queries with Inverse},
  booktitle    = {{KR} 2000, Principles of Knowledge Representation and Reasoning Proceedings
                  of the Seventh International Conference, Breckenridge, Colorado, USA,
                  April 11-15, 2000},
  pages        = {176--185},
  publisher    = {Morgan Kaufmann},
  year         = {2000},
  bibsource    = {dblp computer science bibliography, https://dblp.org}
}

@article{FlorenzanoRUVV20,
  author    = {Fernando Florenzano and
               Cristian Riveros and
               Mart{\'{\i}}n Ugarte and
               Stijn Vansummeren and
               Domagoj Vrgoc},
  title     = {Efficient Enumeration Algorithms for Regular Document Spanners},
  journal   = {{ACM} Trans. Database Syst.},
  volume    = {45},
  number    = {1},
  pages     = {3:1--3:42},
  doi          = {10.1145/3351451},
year      = {2020}
}

@book{sakarovitch2009,
  title={Elements of automata theory},
  author={Sakarovitch, Jacques},
  year={2009},
  publisher={Cambridge University Press}
}

@article{MillDB,
  author       = {Domagoj Vrgoc and
                  Carlos Rojas and
                  Renzo Angles and
                  Marcelo Arenas and
                  Diego Arroyuelo and
                  Carlos Buil{-}Aranda and
                  Aidan Hogan and
                  Gonzalo Navarro and
                  Cristian Riveros and
                  Juan Romero},
  title        = {Millennium{DB}: An Open-Source Graph Database System},
  journal      = {Data Intell.},
  volume       = {5},
  number       = {3},
  pages        = {560--610},
  year         = {2023},
  url          = {https://doi.org/10.1162/dint_a_00229}
}

@inproceedings{Segoufin13,
  author    = {Luc Segoufin},
  editor    = {Wang{-}Chiew Tan and
               Giovanna Guerrini and
               Barbara Catania and
               Anastasios Gounaris},
  title     = {Enumerating with constant delay the answers to a query},
  booktitle = {Joint 2013 {EDBT/ICDT} Conferences, {ICDT} '13 Proceedings, Genoa,
               Italy, March 18-22, 2013},
  pages     = {10--20},
  publisher = {{ACM}},
  year      = {2013}
}

@inproceedings{bagan2006mso,
	title={MSO queries on tree decomposable structures are computable with linear delay},
	author={Bagan, Guillaume},
	booktitle={CSL},
	volume={4207},
	pages={167--181},
	year={2006},
	organization={Springer}
}

@inproceedings{GQLdigest,
  author       = {Nadime Francis and
                  Am{\'{e}}lie Gheerbrant and
                  Paolo Guagliardo and
                  Leonid Libkin and
                  Victor Marsault and
                  Wim Martens and
                  Filip Murlak and
                  Liat Peterfreund and
                  Alexandra Rogova and
                  Domagoj Vrgo\v{c}},
  title        = {A Researcher's Digest of {GQL} (Invited Talk)},
  booktitle    = {International Conference on Database Theory (ICDT)},
  year         = {2023},
  !url          = {https://doi.org/10.4230/LIPIcs.ICDT.2023.1},
  doi          = {10.4230/LIPIcs.ICDT.2023.1},
  timestamp    = {Fri, 17 Mar 2023 17:00:14 +0100},
  biburl       = {https://dblp.org/rec/conf/icdt/FrancisGGLMMMPR23.bib},
  bibsource    = {dblp computer science bibliography, https://dblp.org}
}

@article{pmr,
  author       = {Wim Martens and
                  Matthias Niewerth and
                  Tina Popp and
                  Carlos Rojas and
                  Stijn Vansummeren and
                  Domagoj Vrgo\v{c}},
  title        = {Representing Paths in Graph Database Pattern Matching},
  journal      = {Proc. {VLDB} Endow.},
  volume       = {16},
  number       = {7},
  pages        = {1790--1803},
  year         = {2023},
  url          = {https://www.vldb.org/pvldb/vol16/p1790-martens.pdf},
  bibsource    = {dblp computer science bibliography, https://dblp.org}
}

@inproceedings{csr,
  title={Parallel sparse matrix-vector and matrix-transpose-vector multiplication using compressed sparse blocks},
  author={Bulu{\c{c}}, Aydin and Fineman, Jeremy T and Frigo, Matteo and Gilbert, John R and Leiserson, Charles E},
  booktitle={Proceedings of the twenty-first annual symposium on Parallelism in algorithms and architectures},
  pages={233--244},
  year={2009}
}

@inproceedings{wdbench,
  author       = {Renzo Angles and
                  Carlos Buil Aranda and
                  Aidan Hogan and
                  Carlos Rojas and
                  Domagoj Vrgo\v{c}},
  title        = {{WDB}ench: {A} Wikidata Graph Query Benchmark},
  booktitle    = {The Semantic Web - {ISWC} 2022},
  year         = {2022},
  bibsource    = {dblp computer science bibliography, https://dblp.org}
}

@inproceedings{pathChallenge,
  title={Millennium{DB} Path Query Challenge},
  author={Benjam\'in Far\'ias and Carlos Rojas and Domagoj Vrgo\v{c}},
  booktitle={Alberto Mendelzon Workshop (AMW)},
url          = {https://ceur-ws.org/Vol-3409/paper13.pdf},
year={2023}
}

@online{wddata,
author = {Renzo Angles and Carlos Buil Aranda and Aidan Hogan and Carlos Rojas and Domagoj Vrgo\v{c}},
title = {{WDBench Dataset Download}},
year = {2022},
doi = {10.6084/m9.figshare.19599589}
}

@misc{repo,
  author    = {Benjam\'in Far\'ias and Wim Martens and  Carlos Rojas and Domagoj Vrgo\v{c}},
  title     = {{Regular path queries in MillenniumDB}},
  url       = {https://github.com/MillenniumDB/RPQPaper},
  year      = {2023}
}

@misc{ANONrepo,
  author    = {Benjam\'in Far\'ias and Wim Martens and  Carlos Rojas and Domagoj Vrgo\v{c}},
  title     = {{PathFinder: A unified approach for handling paths in graph query languages}},
  url       = {https://github.com/BFFV/PathFinder},
  year      = {2026}
}

@inproceedings{duckDB,
  author       = {Daniel ten Wolde and
                  Tavneet Singh and
                  G{\'{a}}bor Sz{\'{a}}rnyas and
                  Peter A. Boncz},
  title        = {Duck{PGQ}: Efficient Property Graph Queries in an analytical {RDBMS}},
  booktitle    = {Conference on Innovative Data Systems Research (CIDR)},
  !publisher    = {www.cidrdb.org},
  year         = {2023},
  url          = {https://www.cidrdb.org/cidr2023/papers/p66-wolde.pdf},
  timestamp    = {Wed, 19 Jul 2023 17:21:16 +0200},
  biburl       = {https://dblp.org/rec/conf/cidr/WoldeSSB23.bib},
  bibsource    = {dblp computer science bibliography, https://dblp.org}
}

@inproceedings{kuzu,
  author       = {Guodong Jin and
                  Xiyang Feng and
                  Ziyi Chen and
                  Chang Liu and
                  Semih Salihoglu},
  title        = {K{\`{U}}ZU Graph Database Management System},
  booktitle    = {13th Conference on Innovative Data Systems Research, {CIDR} 2023,
                  Amsterdam, The Netherlands, January 8-11, 2023},
  publisher    = {www.cidrdb.org},
  year         = {2023},
  url          = {https://www.cidrdb.org/cidr2023/papers/p48-jin.pdf},
  bibsource    = {dblp computer science bibliography, https://dblp.org}
}

@inproceedings{GubichevN11,
  author       = {Andrey Gubichev and
                  Thomas Neumann},
  title        = {Path Query Processing on Very Large {RDF} Graphs},
  booktitle    = {WebDB 2011},
  year         = {2011},
  url          = {http://webdb2011.rutgers.edu/papers/Paper21/pathwebdb.pdf},
  bibsource    = {dblp computer science bibliography, https://dblp.org}
}

@phdthesis{Gubichev15,
  author       = {Andrey Gubichev},
  title        = {Query Processing and Optimization in Graph Databases},
  school       = {Technical University Munich},
  year         = {2015},
  url          = {https://nbn-resolving.org/urn:nbn:de:bvb:91-diss-20150625-1238730-1-7},
  urn          = {urn:nbn:de:bvb:91-diss-20150625-1238730-1-7},
  bibsource    = {dblp computer science bibliography, https://dblp.org}
}

@inproceedings{SavenkovMUP17,
  author       = {Vadim Savenkov and
                  Qaiser Mehmood and
                  J{\"{u}}rgen Umbrich and
                  Axel Polleres},
  title        = {Counting to k or how {SPARQL1.1} Property Paths Can Be Extended to
                  Top-k Path Queries},
  booktitle    = {SEMANTiCS 2017},
  year         = {2017},
  !url          = {https://doi.org/10.1145/3132218.3132239},
  doi          = {10.1145/3132218.3132239},
  timestamp    = {Fri, 28 Jan 2022 12:12:02 +0100},
  biburl       = {https://dblp.org/rec/conf/i-semantics/SavenkovMUP17.bib},
  bibsource    = {dblp computer science bibliography, https://dblp.org}
}

@Misc{oracle,
  author =    {Oracle}, 
  key =       {Oracle},
  title =     {Oracle Graph Database},
  howpublished = {\url{https://www.oracle.com/database/graph/}}
}

@article{SakrBVIAAAABBDV-cacm21,
  author    = {Sherif Sakr and
               Angela Bonifati and
               Hannes Voigt and
               Alexandru Iosup and
               Khaled Ammar and
               Renzo Angles and
               Walid G. Aref and
               Marcelo Arenas and
               Maciej Besta and
               Peter A. Boncz and
               Khuzaima Daudjee and
               Emanuele Della Valle and
               Stefania Dumbrava and
               Olaf Hartig and
               Bernhard Haslhofer and
               Tim Hegeman and
               Jan Hidders and
               Katja Hose and
               Adriana Iamnitchi and
               Vasiliki Kalavri and
               Hugo Kapp and
               Wim Martens and
               M. Tamer {\"{O}}zsu and
               Eric Peukert and
               Stefan Plantikow and
               Mohamed Ragab and
               Matei Ripeanu and
               Semih Salihoglu and
               Christian Schulz and
               Petra Selmer and
               Juan F. Sequeda and
               Joshua Shinavier and
               G{\'{a}}bor Sz{\'{a}}rnyas and
               Riccardo Tommasini and
               Antonino Tumeo and
               Alexandru Uta and
               Ana Lucia Varbanescu and
               Hsiang{-}Yun Wu and
               Nikolay Yakovets and
               Da Yan and
               Eiko Yoneki},
  title     = {The future is big graphs: a community view on graph processing systems},
doi          = {10.1145/3434642},
journal   = {Commun. {ACM}},
  volume    = {64},
  number    = {9},
  pages     = {62--71},
  year      = {2021}
}

@ARTICLE{Alphafold,
  title    = "Highly accurate protein structure prediction with {AlphaFold}",
  author   = {Jumper, John and others},
  doi = {0.1038/s41586-021-03819-2},
  journal  = "Nature",
  volume   =  596,
  number   =  7873,
  pages    = "583--589",
  month    =  aug,
  year     =  2021
}

@article{katjaJedi,
  author       = {Christian Aebeloe and
                  Gabriela Montoya and
                  Vinay Setty and
                  Katja Hose},
  title        = {Discovering Diversified Paths in Knowledge Bases},
  journal      = {Proc. {VLDB} Endow.},
  volume       = {11},
  number       = {12},
  pages        = {2002--2005},
  year         = {2018},
  doi          = {10.14778/3229863.3236245}
}

@misc{snapnets,
  author       = {Jure Leskovec and Andrej Krevl},
  title        = {{SNAP Datasets}: {Stanford} Large Network Dataset Collection},
  howpublished = {\url{http://snap.stanford.edu/data}},
  month        = jun,
  year         = 2014
}

@inproceedings{YakovetsGG16,
  author       = {Nikolay Yakovets and
                  Parke Godfrey and
                  Jarek Gryz},
  title        = {Query Planning for Evaluating {SPARQL} Property Paths},
  booktitle    = {International Conference on Management of Data (SIGMOD)},
  pages        = {1875--1889},
  publisher    = {{ACM}},
  year         = {2016},
  !url          = {https://doi.org/10.1145/2882903.2882944},
  doi          = {10.1145/2882903.2882944},
  timestamp    = {Wed, 14 Nov 2018 10:56:20 +0100},
  biburl       = {https://dblp.org/rec/conf/sigmod/YakovetsGG16.bib},
  bibsource    = {dblp computer science bibliography, https://dblp.org}
}

@misc{nebulaIssue,
  author    = {{Vesoft Inc.}},
  title     = {{NebulaGraph Documentation}},
  url       = {https://docs.nebula-graph.io/3.5.0/20.appendix/0.FAQ/},
  year      = {2023}
}

@article{DavidFM24,
  author       = {Claire David and
                  Nadime Francis and
                  Victor Marsault},
  title        = {Distinct Shortest Walk Enumeration for RPQs},
  journal      = {Proc. {ACM} Manag. Data},
  volume       = {2},
  number       = {2},
  pages        = {100},
  year         = {2024},
  url          = {https://doi.org/10.1145/3651601},
  doi          = {10.1145/3651601},
  bibsource    = {dblp computer science bibliography, https://dblp.org}
}

@inproceedings{Pirro20,
  author       = {Giuseppe Pirr{\`{o}}},
  title        = {Fact-checking via Path Embedding and Aggregation},
  booktitle    = {Joint Proceedings of Workshops AI4LEGAL2020, NLIWOD, {PROFILES} 2020,
                  QuWeDa 2020 and {SEMIFORM2020} Colocated with the 19th International
                  Semantic Web Conference {(ISWC} 2020), Virtual Conference, November,
                  2020},
  series       = {{CEUR} Workshop Proceedings},
  volume       = {2722},
  pages        = {149--158},
  publisher    = {CEUR-WS.org},
  year         = {2020},
  url          = {https://ceur-ws.org/Vol-2722/semiform2020-paper-1.pdf},
  bibsource    = {dblp computer science bibliography, https://dblp.org}
}

@article{Eppstein98,
  author       = {David Eppstein},
  title        = {Finding the k Shortest Paths},
  journal      = {{SIAM} J. Comput.},
  volume       = {28},
  number       = {2},
  pages        = {652--673},
  year         = {1998},
  url          = {https://doi.org/10.1137/S0097539795290477},
  doi          = {10.1137/S0097539795290477},
  bibsource    = {dblp computer science bibliography, https://dblp.org}
}

@article{yen,
author = {Jin Y. Yen},
title = {Finding the K Shortest Loopless Paths in a Network},
journal = {Management Science},
volume = {17},
number = {11},
pages = {712--716},
year = {1971}
}

@techreport{sql-pgq-standard,
    author = {{ISO/IEC JTC 1/SC 32}},
    title = {ISO/IEC 9075-16:2023. Information technology -- Database languages SQL. Part 16: Property Graph Queries (SQL/PGQ)},
    institution = {ISO},
    howpublished = {\url{https://www.iso.org/standard/79473.html}},
    year = 2023
}

@techreport{gql-standard,
    author = {{ISO/IEC JTC 1/SC 32}},
    title = {ISO/IEC 39075:2024. Information technology -- Database languages GQL},
    institution = {ISO},
    howpublished = {\url{https://www.iso.org/standard/76120.html}},
    year = 2024
}

\end{document}